\documentclass[11pt]{article}

\usepackage{pstricks}
\usepackage{pst-plot}
\usepackage{pst-node}
\usepackage{amsmath}
\usepackage{pst-grad}
\usepackage{amssymb}
\usepackage{xspace}
\usepackage{pst-math}
\usepackage{pst-xkey}
\usepackage{pst-coil}
\usepackage{pst-3dplot}
\usepackage[T1]{fontenc}
\usepackage{ae}
\usepackage[ansinew]{inputenc}

\newrgbcolor{darkred}{0.8 0 0}
\newrgbcolor{darkerred}{0.7 0 0}
\newrgbcolor{darkestred}{0.5 0 0}
\newrgbcolor{darkgreen}{0 0.5 0.5}
\newrgbcolor{darkergreen}{0 0.6 0}
\newrgbcolor{darkestgreen}{0 0.5 0}
\newrgbcolor{darkblue}{0 0 0.6}
\newrgbcolor{darkestblue}{0 0 0.5}
\newrgbcolor{redgreen}{0.8 0.8 0}
\newrgbcolor{bluegreen}{0 0.4 0.7}
\newrgbcolor{bluewhite}{0 0.9 1}
\newrgbcolor{lightgreen}{0.6 1 0.6}
\newrgbcolor{lightred}{1 0.8 0.8}
\newrgbcolor{lightyred}{1 0.4 0.4}
\newrgbcolor{lightyblue}{0.5 0.5 1}
\newrgbcolor{lightblue}{0.8 1 1}
\newrgbcolor{lilas}{0.5 0.3 0.7}
\newrgbcolor{lilaswhite}{1 0.8 1}
\newrgbcolor{redy}{1 0.6 0.6}
\newrgbcolor{bluish}{0.6 0.6 1}
\newrgbcolor{sand}{0.9 0.9 0.7}
\newrgbcolor{brown}{0.4 0.4 0}
\newrgbcolor{blue1}{0 0.35 0.65}
\newrgbcolor{blue2}{0 0.4 0.6}
\newrgbcolor{blue3}{0 0.45 0.55}
\newrgbcolor{blue4}{0 0.5 0.5}
\newrgbcolor{blue5}{0 0.55 0.45}
\newrgbcolor{blue6}{0 0.6 0.4}
\newrgbcolor{blue7}{0 0.65 0.35}

\newcommand{\R }{\mathbb R}
\newcommand{\N }{\mathbb N}
\newcommand{\Z }{\mathbb Z}
\newcommand{\Q }{\mathbb Q}
\newcommand{\C }{\mathbb C}
\newcommand{\sen }{\ \! {\rm sen}\ \! }
\newcommand{\tg }{\ \! {\rm tg}\ \! }
\newcommand{\cosec }{\ \! {\rm cosec}\ \! }
\newcommand{\cotg }{\ \! {\rm cotg}\ \! }
\newcommand{\dx }{\ \! dx}
\newcommand{\e }{\ \! e}
\newcommand{\senh }{\ \! {\rm senh}\ \! }
\newcommand{\tgh }{\ \! {\rm tgh}\ \! }
\newcommand{\cotgh }{\ \! {\rm cotgh}\ \! }
\newcommand{\sech }{\ \! {\rm sech}\ \! }
\newcommand{\cosech }{\ \! {\rm cosech}\ \! }
\newcommand{\arcsen }{\ \! {\rm arcsen}\ \! }
\newcommand{\arctg }{\ \! {\rm arctg}\ \! }
\newcommand{\arccotg }{\ \! {\rm arccotg}\ \! }
\newcommand{\arcsenh }{\ \! {\rm arcsenh}\ \! }
\newcommand{\arccosh }{\ \! {\rm arccosh}\ \! }
\newcommand{\arctgh }{\ \! {\rm arctgh}\ \! }
\newcommand{\arccosech }{\ \! {\rm arccosech}\ \! }
\newcommand{\arcsech }{\ \! {\rm arcsech}\ \! }
\newcommand{\arcsec }{\ \! {\rm arcsec}\ \! }
\newcommand{\arccosec }{\ \! {\rm arccosec}\ \! }
\renewcommand{\arctgh }{\ \! {\rm arctgh}\ \! }
\newcommand{\arccotgh }{\ \! {\rm arccotgh}\ \! }
\newcommand{\nega }{\neg \ }
\newcommand{\eq }{\Leftrightarrow }

\begin{document}

\title{Cluster formation and evolution in networks of financial market indices}

\author{Leonidas Sandoval Junior \\ \\ Insper, Instituto de Ensino e Pesquisa}

\maketitle

\begin{abstract}
Using data from world stock exchange indices prior to and during periods of global financial crises, clusters and networks of indices are built for different thresholds and diverse periods of time, so that it is then possible to analyze how clusters are formed according to correlations among indices and how they evolve in time, particularly during times of financial crises. Further analysis is made on the eigenvectors corresponding to the second highest eigenvalues of the correlation matrices, revealing a structure peculiar to markets that operate in different time zones.
\end{abstract}

\section{Introduction}

This work uses a distance measure based on Spearman's rank correlation in order to build three dimensional networks based on hierarchies. This approach makes it possible to show how clusters form at different levels of correlations, and how the formations of those networks evolve in time.

The data used are the time series of international stock exchange indices taken from many countries in the world so as to represent both evolved and developing economies in a variety of continents. The series were taken both prior to and during some of the severest financial crises of the past decades, namely the 1987 Black Monday, the 1997 Asian Financial Crisis, the 1998 Russian Crisis, the burst of the dot-com bubble in 2001, and the crisis after September, 11, 2001, so that different regimes of volatility are represented. The idea is to be able to study how clusters behave when going from periods of low volatility to more turbulent times. The Subprime Mortgage crisis of 2008 is analyzed in a companion article, along with the years ranging from 2007 to 2010.

The data consists on some of the benchmark indices of a diversity of stock exchanges around the world. The number goes from 16 indices (1986) to 79 indices (2001). The choice of indices was mainly based on availability of data, since some of the stock exchanges being studied didn't develope or had no recorded indices until quite recently. As there are differences between some of the days certain stock markets operate, some of the operation days had to be deleted and others duplicated. The rule was the following: when more than 30\% of markets didn't operate on a certain day, that day was deleted. When that number was bellow 30\%, we repeated the value of the index for the previous day for the markets that didn't open. Special care had to be given to markets whose weekends didn't correspond to the usual ocidental days, like some Arab countries. For those markets, we adjusted the weekends in order to match those of the majority of markets. All that was done in order to certify that we minimized the number of days with missing data, so that we could measure the log-returns of two consecutive days of a market's operation whenever that was possible.

The correlation matrix of the time series of financial data encodes a large amount of information, and an even greater amount of noise. That information and noise must be filtered if one is to try to understand how the elements (in our case, indices) relate to each other and how that relation evolves in time. One of the most common filtering procedures is to represent those relations using a {\sl Minimum Spanning Tree} \cite{mst01}-\cite{mst15}, which is a graph containing all indices, connected by at least one edge, so that the sum of the edges is minimum, and which presents no loops. Another type of representation is that of {\sl Maximally Planar Filtered Graphs} \cite{pmfg01}-\cite{pmfg06}, which admits loops but must be representable in two dimensional graphs without crossings.

Yet another type of representation is obtained by establishing a number which defines how many connections (edges) are to be represented in a graph of the correlations between nodes. There is no limitation with respect to crossing of edges or to the formation of loops, and if the number is high enough, then one has a graph where all nodes are connected to one another. These are usualy called {\sl Asset Trees}, or {\sl Asset Graphs} \cite{asset01}-\cite{asset08}, for they are no trees in the network sense. Another way to build asset graphs is to establish a value (threshold) such that distances above it are not considered. This eliminates connections (edges) as well as indices (nodes), but also turn the diagrams more understandable by filtering both information and noise.

Some previous works using graphic representations of correlations between international assets (indices or otherwise) can be found in \cite{mst05}, \cite{pmfg05}, and \cite{asset07}.

In the present work, we shall build asset trees of indices based on threshold values for a distance measure between them that is based on the Spearman rank correlation among indices. By establishing increasing values for the thresholds, we shall be able to devise the formation of clusters among indices, and how they evolve in time. Details of the way that is done are given in Section 2.

A second topic is pursued in Section 3, where an analysis is made of the eigenvector corresponding to the second highest eigenvalue of the correlation matrix of each group of data being studied. It is well-known (\cite{leocorr} and references therein) that the eigenvector corresponding to the highest eigenvalue of a correlation matrix of financial assets corresponds to a ``market mode'' that resembles the common movement of the market as a whole, and that often the eigenvectors corresponding to the second and sometimes the third highest eigenvalues also carry some information on the internal structure of the market being studied. What is implied in Section 3 is that this internal structure reflects the fact that stock markets operate at different time zones, what is a characteristic that is unique to a global market.

\section{Depiction and evolution of clusters}

In this section, we use data from some key periods of time for financial markets in order to build clusters based on hierarchy. The procedure is to consider the daily log-return for each index, given by
\begin{equation}
\label{logreturn}
R_t=\ln P_t-\ln P_{t-1}\ ,
\end{equation}
where $P_t$ is the value of the index at day $t$ and $P_{t-1}$ is the value of the same index on day $t-1$. The log-returns are then used in order to create a correlation matrix $C$ based on Spearman's rank correlation, and then a distance is defined as
\begin{equation}
\label{distance}
d_{ij}=1-c_{ij}\ ,
\end{equation}
where $d_{ij}$ is the distance between indices $i$ and $j$ and $c_{ij}$ is the correlation between both indices.

Spearman rank correlation is being used instead of the usual Pearson correlation, for it better captures nonlinear relations between indices, and the distance measure was chosen as a linear realization of the correlation. The distance measure satisfies all conditions for an Euclidean measure.

In order to represent as best as possible the true distances of indices in a graph, we used three-dimensional maps based on principal component analysis, that minimizes the differences between the true distances and the distances represented in the graph. Then, those three-dimensional graphs were used in order to represent networks based on threshold values of the distances between nodes (indices). By using simulations (1000 for each period) with randomized data, where the log returns of each index were randomly shifted so as to eliminate any temporal correlation between them, but maintain their probability frequency distributions, we established thresholds above which random noise starts to interfere severely with the connections among indices. Those thresholds are discussed for each of the semesters depicted in this article.

A compromise had to be established between considering a good amount of data so as to minimize noise in the measurements of correlation and considering a small enough time interval so as to capture the relations between indices in a certain instant of time. Choosing to consider intervals of the size of a semester, this compromise was achieved, mainly because the crises being considered tended to happen on the second half of each year.

\subsection{1987 - Black Monday}

The first asset graphs to be built are based on data for the years 1986 and 1987, the year that preceded and the one that witnessed the crisis known as Black Monday, whose peak ocurred on Monday, October 19, 1987. Not yet completely explained, the crisis made stock markets worldwide drop up to 45\% in less than two weeks, and represented the worst financial crisis since 1929.

The networks for the two semesters of 1986 were built using the indices of 16 markets: S\&P 500 from the New York Stock Exchange (S\&P), and Nasdaq (Nasd), both from the USA, S\&P TSX from Canada (Cana), Ibovespa from Brazil (Braz), FTSE 100 from the United Kingdom (UK), DAX from Germany (Germ) - West Germany at the time - ATX from Austria (Autr), AEX from the Netherlands (Neth), SENSEX from India (Indi), Colombo All Share from Sri Lanka (SrLa), Nikkei 25 from Japan (Japa), Hang Seng from Hong Kong (HoKo), TAIEX from Taiwan (Taiw), Kospi from South Korea (SoKo), Kuala Lumpur Composite from Malaysia (Mala), and JCI from Indonesia (Indo). The number of countries is small, mainly due to lack of data, but offers a relatively wide variety of nations and cultures in three continents. The networks for the two semesters of 1987 were built using 23 indices, adding to the ones of 1986 indices ISEQ from Ireland (Irel), OMX from Sweden (Swed), OMX Helsinki from Finland (Finl), IBEX 35 from Spain (Spai), ASE General Index from Greece (Gree), and PSEi from the Philippines (Phil).

\subsubsection{First semester, 1986}

Figure 1 shows the three dimensional view of the asset trees for the first semester of 1986, with threshold ranging from $T=0.3$ to $T=0.8$. Since we are using only the indices that have connections bellow $T=0.8$, not all indices are displayed in the following graphs.

\vskip 0.3 cm

\begin{pspicture}(-4,-0.2)(1,1.9)
\rput(-3,1.3){T=0.3}
\psline(-3.9,-1.6)(3.7,-1.6)(3.7,1.7)(-3.9,1.7)(-3.9,-1.6)
\psset{xunit=5,yunit=5,Alpha=160,Beta=0} \scriptsize
\pstThreeDLine(0.2748,0.2635,0.1772)(0.3787,0.1646,0.1878) 
\pstThreeDDot[linecolor=orange,linewidth=1.2pt](0.2748,0.2635,0.1772) 
\pstThreeDDot[linecolor=orange,linewidth=1.2pt](0.3787,0.1646,0.1878) 
\pstThreeDPut(0.2748,0.2635,0.2272){S\&P}
\pstThreeDPut(0.3787,0.1646,0.2378){Nasd}
\end{pspicture}
\begin{pspicture}(-8,-0.2)(1,1.9)
\rput(-3,1.3){T=0.5}
\psline(-3.9,-1.6)(3.7,-1.6)(3.7,1.7)(-3.9,1.7)(-3.9,-1.6)
\psset{xunit=5,yunit=5,Alpha=160,Beta=0} \scriptsize
\pstThreeDLine(0.2748,0.2635,0.1772)(0.3787,0.1646,0.1878) 
\pstThreeDLine(0.3787,0.1646,0.1878)(0.4180,0.1087,-0.1400) 
\pstThreeDLine(0.0888,-0.5344,-0.0139)(0.2803,-0.4897,0.0337) 
\pstThreeDDot[linecolor=orange,linewidth=1.2pt](0.2748,0.2635,0.1772) 
\pstThreeDDot[linecolor=orange,linewidth=1.2pt](0.3787,0.1646,0.1878) 
\pstThreeDDot[linecolor=orange,linewidth=1.2pt](0.4180,0.1087,-0.1400) 
\pstThreeDDot[linecolor=blue,linewidth=1.2pt](0.0888,-0.5344,-0.0139) 
\pstThreeDDot[linecolor=blue,linewidth=1.2pt](0.2803,-0.4897,0.0337) 
\pstThreeDPut(0.2748,0.2635,0.2272){S\&P}
\pstThreeDPut(0.3787,0.1646,0.2378){Nasd}
\pstThreeDPut(0.4180,0.1087,-0.1900){Cana}
\pstThreeDPut(0.0888,-0.5344,-0.0639){Germ}
\pstThreeDPut(0.2803,-0.4897,0.0837){Neth}
\end{pspicture}

\vskip 1.8 cm

\begin{pspicture}(-4,-0.2)(1,1.9)
\rput(-3,1.3){T=0.7}
\psline(-3.9,-1.6)(3.7,-1.6)(3.7,1.7)(-3.9,1.7)(-3.9,-1.6)
\psset{xunit=5,yunit=5,Alpha=160,Beta=0} \scriptsize
\pstThreeDLine(0.2748,0.2635,0.1772)(0.3787,0.1646,0.1878) 
\pstThreeDLine(0.2748,0.2635,0.1772)(0.4180,0.1087,-0.1400) 
\pstThreeDLine(0.3787,0.1646,0.1878)(0.4180,0.1087,-0.1400) 
\pstThreeDLine(0.3787,0.1646,0.1878)(0.3638,0.0751,-0.1316) 
\pstThreeDLine(0.0888,-0.5344,-0.0139)(0.2803,-0.4897,0.0337) 
\pstThreeDDot[linecolor=orange,linewidth=1.2pt](0.2748,0.2635,0.1772) 
\pstThreeDDot[linecolor=orange,linewidth=1.2pt](0.3787,0.1646,0.1878) 
\pstThreeDDot[linecolor=orange,linewidth=1.2pt](0.4180,0.1087,-0.1400) 
\pstThreeDDot[linecolor=blue,linewidth=1.2pt](0.3638,0.0751,-0.1316) 
\pstThreeDDot[linecolor=blue,linewidth=1.2pt](0.0888,-0.5344,-0.0139) 
\pstThreeDDot[linecolor=blue,linewidth=1.2pt](0.2803,-0.4897,0.0337) 
\pstThreeDPut(0.2748,0.2635,0.2272){S\&P}
\pstThreeDPut(0.3787,0.1646,0.2378){Nasd}
\pstThreeDPut(0.4180,0.1087,-0.1900){Cana}
\pstThreeDPut(0.3638,0.0751,-0.1816){UK}
\pstThreeDPut(0.0888,-0.5344,-0.0639){Germ}
\pstThreeDPut(0.2803,-0.4897,0.0837){Neth}
\end{pspicture}
\begin{pspicture}(-8,-0.2)(1,1.9)
\rput(-3,1.3){T=0.8}
\psline(-3.9,-1.6)(3.7,-1.6)(3.7,1.7)(-3.9,1.7)(-3.9,-1.6)
\psset{xunit=5,yunit=5,Alpha=160,Beta=0} \scriptsize
\pstThreeDLine(0.2748,0.2635,0.1772)(0.3787,0.1646,0.1878) 
\pstThreeDLine(0.2748,0.2635,0.1772)(0.4180,0.1087,-0.1400) 
\pstThreeDLine(0.2748,0.2635,0.1772)(0.3638,0.0751,-0.1316) 
\pstThreeDLine(0.3787,0.1646,0.1878)(0.4180,0.1087,-0.1400) 
\pstThreeDLine(0.3787,0.1646,0.1878)(0.3638,0.0751,-0.1316) 
\pstThreeDLine(0.3787,0.1646,0.1878)(0.2803,-0.4897,0.0337) 
\pstThreeDLine(0.3787,0.1646,0.1878)(-0.1068,0.0420,-0.1631) 
\pstThreeDLine(0.4180,0.1087,-0.1400)(0.3638,0.0751,-0.1316) 
\pstThreeDLine(0.4180,0.1087,-0.1400)(0.2803,-0.4897,0.0337) 
\pstThreeDLine(-0.2655,-0.1135,0.0532)(-0.1068,0.0420,-0.1631) 
\pstThreeDLine(0.3638,0.0751,-0.1316)(0.2803,-0.4897,0.0337) 
\pstThreeDLine(0.0888,-0.5344,-0.0139)(0.2803,-0.4897,0.0337) 
\pstThreeDLine(-0.5160,0.3904,-0.1440)(-0.1068,0.0420,-0.1631) 
\pstThreeDLine(0.2803,-0.4897,0.0337)(-0.1293,-0.1766,-0.0161) 
\pstThreeDDot[linecolor=orange,linewidth=1.2pt](0.2748,0.2635,0.1772) 
\pstThreeDDot[linecolor=orange,linewidth=1.2pt](0.3787,0.1646,0.1878) 
\pstThreeDDot[linecolor=orange,linewidth=1.2pt](0.4180,0.1087,-0.1400) 
\pstThreeDDot[linecolor=green,linewidth=1.2pt](-0.2655,-0.1135,0.0532) 
\pstThreeDDot[linecolor=blue,linewidth=1.2pt](0.3638,0.0751,-0.1316) 
\pstThreeDDot[linecolor=blue,linewidth=1.2pt](0.0888,-0.5344,-0.0139) 
\pstThreeDDot[linecolor=blue,linewidth=1.2pt](-0.5160,0.3904,-0.1440) 
\pstThreeDDot[linecolor=blue,linewidth=1.2pt](0.2803,-0.4897,0.0337) 
\pstThreeDDot[linecolor=red,linewidth=1.2pt](-0.1068,0.0420,-0.1631) 
\pstThreeDDot[linecolor=red,linewidth=1.2pt](-0.1293,-0.1766,-0.0161) 
\pstThreeDPut(0.2748,0.2635,0.2272){S\&P}
\pstThreeDPut(0.3787,0.1646,0.2378){Nasd}
\pstThreeDPut(0.4180,0.1087,-0.1900){Cana}
\pstThreeDPut(-0.2655,-0.1135,0.1032){Braz}
\pstThreeDPut(0.3638,0.0751,-0.1816){UK}
\pstThreeDPut(0.0888,-0.5344,-0.0639){Germ}
\pstThreeDPut(-0.5160,0.3904,-0.2140){Autr}
\pstThreeDPut(0.2803,-0.4897,0.0837){Neth}
\pstThreeDPut(-0.1068,0.0420,-0.2131){Japa}
\pstThreeDPut(-0.1293,-0.1766,-0.0661){HoKo}
\end{pspicture}

\vskip 1.8 cm

\noindent Fig. 1. Three dimensional view of the asset trees for the first semester of 1986, with threshold ranging from $T=0.3$ to $T=0.8$.

\vskip 0.3 cm

At $T=0.3$, the first cluster is formed, with the connection between S\&P and Nasdaq, both indices from the USA. For $T=0.5$, Canada is added to the North American cluster and a new cluster is formed when Germany and Netherlands connect with each other. For $T=0.6$, a connection is formed between S\&P and Canada, strenghtening the North American cluster. At $T=0.7$, random noise starts to interfere with the results, but the connection formed between the UK and Nasdaq seems genuine. For $T=0.8$, the two clusters merge, and Austria, Brazil, Japan, and Hong Kong join the new single cluster. Noise overwhelms the results for $T>0.9$, and although many more connections are made for higher thresholds, they cannot be trusted. All the remaining indices connect with the resulting cluster within $T=0.9$ and $T=1.0$, and the last connections occur for $T=1.3$.

\subsubsection{Second semester, 1986}

Figure 2 shows the three dimensional view of the asset trees for the second semester of 1986, with threshold ranging from $T=0.3$ to $T=0.8$.

\vskip 0.3 cm

\begin{pspicture}(-4,-2.1)(1,1.7)
\rput(-3,1.1){T=0.3}
\psline(-3.9,-3.2)(3.7,-3.2)(3.7,1.5)(-3.9,1.5)(-3.9,-3.2)
\psset{xunit=5,yunit=5,Alpha=60,Beta=0} \scriptsize
\pstThreeDLine(-0.3183,0.2219,0.0548)(-0.2672,0.3351,-0.0145) 
\pstThreeDDot[linecolor=orange,linewidth=1.2pt](-0.3183,0.2219,0.0548) 
\pstThreeDDot[linecolor=orange,linewidth=1.2pt](-0.2672,0.3351,-0.0145) 
\pstThreeDPut(-0.3183,0.2219,0.1048){S\&P}
\pstThreeDPut(-0.2672,0.3351,-0.0645){Nasd}
\end{pspicture}
\begin{pspicture}(-8,-2.1)(1,1.7)
\rput(-3,1.1){T=0.5}
\psline(-3.9,-3.2)(3.7,-3.2)(3.7,1.5)(-3.9,1.5)(-3.9,-3.2)
\psset{xunit=5,yunit=5,Alpha=60,Beta=0} \scriptsize
\pstThreeDLine(-0.3183,0.2219,0.0548)(-0.2672,0.3351,-0.0145) 
\pstThreeDLine(-0.3183,0.2219,0.0548)(-0.3461,0.1142,0.1184) 
\pstThreeDDot[linecolor=orange,linewidth=1.2pt](-0.3183,0.2219,0.0548) 
\pstThreeDDot[linecolor=orange,linewidth=1.2pt](-0.2672,0.3351,-0.0145) 
\pstThreeDDot[linecolor=orange,linewidth=1.2pt](-0.3461,0.1142,0.1184) 
\pstThreeDPut(-0.3183,0.2219,0.1048){S\&P}
\pstThreeDPut(-0.2672,0.3351,-0.0645){Nasd}
\pstThreeDPut(-0.3461,0.1142,0.1684){Cana}
\end{pspicture}

\vskip 1.5 cm

\begin{pspicture}(-4,-2.1)(1,1.7)
\rput(-3,1.1){T=0.6}
\psline(-3.9,-3.2)(3.7,-3.2)(3.7,1.5)(-3.9,1.5)(-3.9,-3.2)
\psset{xunit=5,yunit=5,Alpha=60,Beta=0} \scriptsize
\pstThreeDLine(-0.3183,0.2219,0.0548)(-0.2672,0.3351,-0.0145) 
\pstThreeDLine(-0.3183,0.2219,0.0548)(-0.3461,0.1142,0.1184) 
\pstThreeDLine(-0.3183,0.2219,0.0548)(-0.2530,0.2321,0.1112) 
\pstThreeDLine(-0.2672,0.3351,-0.0145)(-0.3461,0.1142,0.1184) 
\pstThreeDLine(-0.2672,0.3351,-0.0145)(-0.2530,0.2321,0.1112) 
\pstThreeDLine(-0.3461,0.1142,0.1184)(-0.3106,0.0790,-0.0354) 
\pstThreeDLine(-0.2530,0.2321,0.1112)(-0.3106,0.0790,-0.0354) 
\pstThreeDLine(-0.0717,0.0271,-0.1439)(-0.3106,0.0790,-0.0354) 
\pstThreeDDot[linecolor=orange,linewidth=1.2pt](-0.3183,0.2219,0.0548) 
\pstThreeDDot[linecolor=orange,linewidth=1.2pt](-0.2672,0.3351,-0.0145) 
\pstThreeDDot[linecolor=orange,linewidth=1.2pt](-0.3461,0.1142,0.1184) 
\pstThreeDDot[linecolor=blue,linewidth=1.2pt](-0.2530,0.2321,0.1112) 
\pstThreeDDot[linecolor=blue,linewidth=1.2pt](-0.0717,0.0271,-0.1439) 
\pstThreeDDot[linecolor=blue,linewidth=1.2pt](-0.3106,0.0790,-0.0354) 
\pstThreeDPut(-0.3183,0.2219,0.1048){S\&P}
\pstThreeDPut(-0.2672,0.3351,-0.0645){Nasd}
\pstThreeDPut(-0.3461,0.1142,0.1684){Cana}
\pstThreeDPut(-0.2530,0.2321,0.1612){UK}
\pstThreeDPut(-0.0717,0.0271,-0.1939){Germ}
\pstThreeDPut(-0.3106,0.0790,-0.0854){Neth}
\end{pspicture}
\begin{pspicture}(-8,-2.1)(1,1.7)
\rput(-3,1.1){T=0.8}
\psline(-3.9,-3.2)(3.7,-3.2)(3.7,1.5)(-3.9,1.5)(-3.9,-3.2)
\psset{xunit=5,yunit=5,Alpha=60,Beta=0} \scriptsize
\pstThreeDLine(-0.3183,0.2219,0.0548)(-0.2672,0.3351,-0.0145) 
\pstThreeDLine(-0.3183,0.2219,0.0548)(-0.3461,0.1142,0.1184) 
\pstThreeDLine(-0.3183,0.2219,0.0548)(-0.2530,0.2321,0.1112) 
\pstThreeDLine(-0.3183,0.2219,0.0548)(-0.3106,0.0790,-0.0354) 
\pstThreeDLine(-0.2672,0.3351,-0.0145)(-0.3461,0.1142,0.1184) 
\pstThreeDLine(-0.2672,0.3351,-0.0145)(-0.2530,0.2321,0.1112) 
\pstThreeDLine(-0.2672,0.3351,-0.0145)(-0.3106,0.0790,-0.0354) 
\pstThreeDLine(-0.3461,0.1142,0.1184)(-0.2530,0.2321,0.1112) 
\pstThreeDLine(-0.3461,0.1142,0.1184)(-0.3106,0.0790,-0.0354) 
\pstThreeDLine(-0.2530,0.2321,0.1112)(-0.3106,0.0790,-0.0354) 
\pstThreeDLine(-0.0717,0.0271,-0.1439)(-0.3106,0.0790,-0.0354) 
\pstThreeDLine(-0.0717,0.0271,-0.1439)(-0.0851,-0.1377,-0.4672) 
\pstThreeDDot[linecolor=orange,linewidth=1.2pt](-0.3183,0.2219,0.0548) 
\pstThreeDDot[linecolor=orange,linewidth=1.2pt](-0.2672,0.3351,-0.0145) 
\pstThreeDDot[linecolor=orange,linewidth=1.2pt](-0.3461,0.1142,0.1184) 
\pstThreeDDot[linecolor=blue,linewidth=1.2pt](-0.2530,0.2321,0.1112) 
\pstThreeDDot[linecolor=blue,linewidth=1.2pt](-0.0717,0.0271,-0.1439) 
\pstThreeDDot[linecolor=blue,linewidth=1.2pt](-0.3106,0.0790,-0.0354) 
\pstThreeDDot[linecolor=red,linewidth=1.2pt](-0.0851,-0.1377,-0.4672) 
\pstThreeDPut(-0.3183,0.2219,0.1048){S\&P}
\pstThreeDPut(-0.2672,0.3351,-0.0645){Nasd}
\pstThreeDPut(-0.3461,0.1142,0.1684){Cana}
\pstThreeDPut(-0.2530,0.2321,0.1612){UK}
\pstThreeDPut(-0.0717,0.0271,-0.1939){Germ}
\pstThreeDPut(-0.3106,0.0790,-0.0854){Neth}
\pstThreeDPut(-0.0851,-0.1377,-0.5172){Japa}
\end{pspicture}

\vskip 1.6 cm

\noindent Fig. 2. Three dimensional view of the asset trees for the second semester of 1986, with threshold ranging from $T=0.3$ to $T=0.8$.

\vskip 0.3 cm

The first connection is formed at $T=0.3$, between S\&P and Nasdaq. This cluster grows only at $T=0.5$, with the addition of Canada. At $T=0.6$, the UK joins the North American cluster, which also connects with the Netherlands and, through it, with Germany. For $T=0.7$, Netherlands establishes itself as a major hub, and Japan joins, via Germany, at $T=0.8$. Random noise starts to have more effects at $T=0.7$, and by $T=0.9$, it becomes strong, so the many connections that are formed above $T=0.9$ cannot be trusted. The last connections occur at $T=1.3$.

\subsubsection{First semester, 1987}

Figure 3 shows the three dimensional view of the asset trees for the first semester of 1987, with threshold ranging from $T=0.3$ to $T=0.8$.

The first connection, again between S\&P and Nasdaq, forms at $T=0.3$. At $T=0.4$, Canada joins the emerging North American cluster. For $T=0.5$, a new cluster is formed, between Germany and Netherlands. At $T=0.6$, a connection is formed between Canada and S\&P, and for $T=0.7$, the UK connects with the North American cluster through Nasdaq. For $T=0.8$, random noise starts having strong effects. We have connections of the UK with Ireland and Hong Kong, and more connections among the members of the emerging European cluster, now comprised of Germany, Netherlands, Sweden, Finland, and Spain. South Korea connects with Ireland, and Sri Lanka connects with Malaysia. The last connections are probably due to random noise.

\vskip 0.8 cm

\begin{pspicture}(-3.5,-2.8)(1,2)
\rput(-3,1.7){T=0.3}
\psline(-3.9,-2.9)(3.7,-2.9)(3.7,2.1)(-3.9,2.1)(-3.9,-2.9)
\psset{xunit=5,yunit=5,Alpha=60,Beta=0} \scriptsize
\pstThreeDLine(-0.3548,0.3508,0.2828)(-0.4439,0.3895,0.1685) 
\pstThreeDDot[linecolor=orange,linewidth=1.2pt](-0.3548,0.3508,0.2828) 
\pstThreeDDot[linecolor=orange,linewidth=1.2pt](-0.4439,0.3895,0.1685) 
\pstThreeDPut(-0.3548,0.3508,0.3228){S\&P}
\pstThreeDPut(-0.4439,0.3895,0.2185){Nasd}
\end{pspicture}
\begin{pspicture}(-8,-2.8)(1,2)
\rput(-3,1.7){T=0.4}
\psline(-3.9,-2.9)(3.7,-2.9)(3.7,2.1)(-3.9,2.1)(-3.9,-2.9)
\psset{xunit=5,yunit=5,Alpha=60,Beta=0} \scriptsize
\pstThreeDLine(-0.3548,0.3508,0.2828)(-0.4439,0.3895,0.1685) 
\pstThreeDLine(-0.4439,0.3895,0.1685)(-0.2927,0.4073,0.1442) 
\pstThreeDDot[linecolor=orange,linewidth=1.2pt](-0.3548,0.3508,0.2828) 
\pstThreeDDot[linecolor=orange,linewidth=1.2pt](-0.4439,0.3895,0.1685) 
\pstThreeDDot[linecolor=orange,linewidth=1.2pt](-0.2927,0.4073,0.1442) 
\pstThreeDPut(-0.3548,0.3508,0.3228){S\&P}
\pstThreeDPut(-0.4439,0.3895,0.2185){Nasd}
\pstThreeDPut(-0.2927,0.4073,0.1942){Cana}
\end{pspicture}

\vskip 1 cm

\begin{pspicture}(-3.5,-2.8)(1,2)
\rput(-3,1.7){T=0.6}
\psline(-3.9,-2.9)(3.7,-2.9)(3.7,2.1)(-3.9,2.1)(-3.9,-2.9)
\psset{xunit=5,yunit=5,Alpha=60,Beta=0} \scriptsize
\pstThreeDLine(-0.3548,0.3508,0.2828)(-0.4439,0.3895,0.1685) 
\pstThreeDLine(-0.3548,0.3508,0.2828)(-0.2927,0.4073,0.1442) 
\pstThreeDLine(-0.4439,0.3895,0.1685)(-0.2927,0.4073,0.1442) 
\pstThreeDLine(-0.2182,-0.4967,0.1098)(-0.4761,-0.3290,0.1010) 
\pstThreeDDot[linecolor=orange,linewidth=1.2pt](-0.3548,0.3508,0.2828) 
\pstThreeDDot[linecolor=orange,linewidth=1.2pt](-0.4439,0.3895,0.1685) 
\pstThreeDDot[linecolor=orange,linewidth=1.2pt](-0.2927,0.4073,0.1442) 
\pstThreeDDot[linecolor=blue,linewidth=1.2pt](-0.2182,-0.4967,0.1098) 
\pstThreeDDot[linecolor=blue,linewidth=1.2pt](-0.4761,-0.3290,0.1010) 
\pstThreeDPut(-0.3548,0.3508,0.3228){S\&P}
\pstThreeDPut(-0.4439,0.3895,0.2185){Nasd}
\pstThreeDPut(-0.2927,0.4073,0.1942){Cana}
\pstThreeDPut(-0.2182,-0.4967,0.1598){Germ}
\pstThreeDPut(-0.4761,-0.3290,0.1510){Neth}
\end{pspicture}
\begin{pspicture}(-8,-2.8)(1,2)
\rput(-3,1.7){T=0.8}
\psline(-3.9,-2.9)(3.7,-2.9)(3.7,2.1)(-3.9,2.1)(-3.9,-2.9)
\psset{xunit=5,yunit=5,Alpha=60,Beta=0} \scriptsize
\pstThreeDLine(-0.3548,0.3508,0.2828)(-0.4439,0.3895,0.1685) 
\pstThreeDLine(-0.3548,0.3508,0.2828)(-0.2927,0.4073,0.1442) 
\pstThreeDLine(-0.4439,0.3895,0.1685)(-0.2927,0.4073,0.1442) 
\pstThreeDLine(-0.4439,0.3895,0.1685)(-0.3246,0.1166,-0.1551) 
\pstThreeDLine(-0.2927,0.4073,0.1442)(-0.3246,0.1166,-0.1551) 
\pstThreeDLine(-0.3246,0.1166,-0.1551)(-0.2024,-0.0006,-0.4309) 
\pstThreeDLine(-0.3246,0.1166,-0.1551)(-0.0297,0.1509,-0.2859) 
\pstThreeDLine(-0.2024,-0.0006,-0.4309)(-0.1040,0.0543,-0.3098) 
\pstThreeDLine(-0.2182,-0.4967,0.1098)(-0.4761,-0.3290,0.1010) 
\pstThreeDLine(-0.2182,-0.4967,0.1098)(-0.1440,-0.4232,0.0181) 
\pstThreeDLine(-0.4761,-0.3290,0.1010)(-0.1440,-0.4232,0.0181) 
\pstThreeDLine(-0.4761,-0.3290,0.1010)(-0.1865,-0.3497,0.1005) 
\pstThreeDLine(0.2693,0.2517,-0.2629)(0.4491,0.1496,-0.1378) 
\pstThreeDLine(0.3528,0.0400,0.1433)(0.2432,0.1585,0.2071) 
\pstThreeDDot[linecolor=orange,linewidth=1.2pt](-0.3548,0.3508,0.2828) 
\pstThreeDDot[linecolor=orange,linewidth=1.2pt](-0.4439,0.3895,0.1685) 
\pstThreeDDot[linecolor=orange,linewidth=1.2pt](-0.2927,0.4073,0.1442) 
\pstThreeDDot[linecolor=blue,linewidth=1.2pt](-0.3246,0.1166,-0.1551) 
\pstThreeDDot[linecolor=blue,linewidth=1.2pt](-0.2024,-0.0006,-0.4309) 
\pstThreeDDot[linecolor=blue,linewidth=1.2pt](-0.2182,-0.4967,0.1098) 
\pstThreeDDot[linecolor=blue,linewidth=1.2pt](-0.4761,-0.3290,0.1010) 
\pstThreeDDot[linecolor=blue,linewidth=1.2pt](-0.1440,-0.4232,0.0181) 
\pstThreeDDot[linecolor=blue,linewidth=1.2pt](0.2693,0.2517,-0.2629) 
\pstThreeDDot[linecolor=blue,linewidth=1.2pt](0.4491,0.1496,-0.1378) 
\pstThreeDDot[linecolor=brown,linewidth=1.2pt](0.3528,0.0400,0.1433) 
\pstThreeDDot[linecolor=red,linewidth=1.2pt](-0.1865,-0.3497,0.1005) 
\pstThreeDDot[linecolor=red,linewidth=1.2pt](-0.0297,0.1509,-0.2859) 
\pstThreeDDot[linecolor=red,linewidth=1.2pt](-0.1040,0.0543,-0.3098) 
\pstThreeDDot[linecolor=red,linewidth=1.2pt](0.2432,0.1585,0.2071) 
\pstThreeDPut(-0.3548,0.3508,0.3228){S\&P}
\pstThreeDPut(-0.4439,0.3895,0.2185){Nasd}
\pstThreeDPut(-0.2927,0.4073,0.1942){Cana}
\pstThreeDPut(-0.3246,0.1166,-0.2051){UK}
\pstThreeDPut(-0.2024,-0.0006,-0.4809){Irel}
\pstThreeDPut(-0.2182,-0.4967,0.1598){Germ}
\pstThreeDPut(-0.4761,-0.3290,0.1510){Neth}
\pstThreeDPut(-0.1440,-0.4232,0.0681){Swed}
\pstThreeDPut(0.2693,0.2517,-0.3129){Finl}
\pstThreeDPut(0.4491,0.1496,-0.1878){Spai}
\pstThreeDPut(0.3528,0.0400,0.1933){SrLa}
\pstThreeDPut(-0.1865,-0.3497,0.1505){Japa}
\pstThreeDPut(-0.0297,0.1509,-0.3359){HoKo}
\pstThreeDPut(-0.1040,0.0543,-0.3598){SoKo}
\pstThreeDPut(0.2432,0.1585,0.2571){Mala}
\end{pspicture}

\vskip 0.5 cm

\noindent Fig. 3. Three dimensional view of the asset trees for the first semester of 1987, with threshold ranging from $T=0.3$ to $T=0.8$.

\vskip 0.3 cm

\subsubsection{Second semester, 1987}

Figure 4 shows the three dimensional view of the asset trees for the second semester of 1987, with threshold ranging from $T=0.4$ to $T=0.7$.

At $T=0.3$, we have a cluster formed by the North American indices, S\&P, Nasdaq, and Canada. For $T=0.4$, the North American cluster gets fully connected, and a new cluster, composed by Germany and Netherlands, is formed. At $T=0.5$, many connections are established, the UK and Netherlands with the North American cluster, between the UK and Ireland, and between Sweden and other European indices. At $T=0.6$, Germany connects fully with the North American cluster, and Austria and Spain join the European cluster, while Japan connects with Sweden. Random noise starts to get much more influent for $T=0.7$. We now have connections of India, Hong Kong, and Malaysia with the North American and European clusters. More connections are formed for higher thresholds, until all indices are connected for $T=1.3$. Note that the size of the clusters shrank for the second semester of 1987, a consequence of the increase in correlation among the indices.

\vskip 0.5 cm

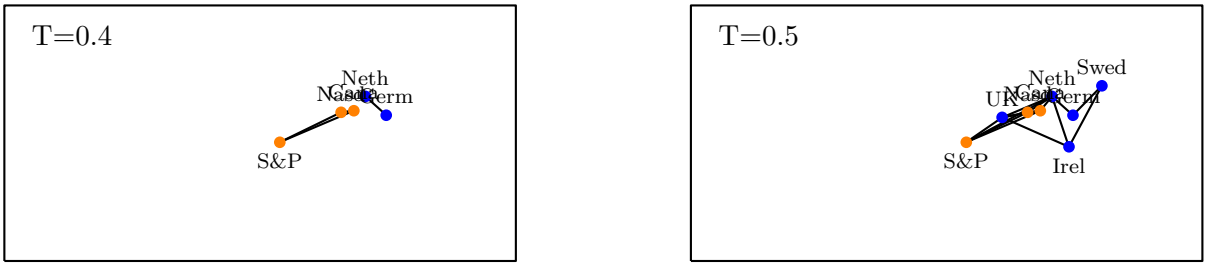
\begin{pspicture}(-4,-1.6)(1,1.5)
\rput(-3,1.1){T=0.4}
\psline(-3.9,-1.9)(2.9,-1.9)(2.9,1.5)(-3.9,1.5)(-3.9,-1.9)
\psset{xunit=5,yunit=5,Alpha=60,Beta=0} \scriptsize
\pstThreeDLine(-0.3962,-0.2840,-0.0640)(-0.3849,-0.0888,0.0158) 
\pstThreeDLine(-0.3962,-0.2840,-0.0640)(-0.3537,-0.0322,0.0199) 
\pstThreeDLine(-0.3849,-0.0888,0.0158)(-0.3537,-0.0322,0.0199) 
\pstThreeDLine(-0.2762,0.1126,0.0079)(-0.4088,-0.0281,0.0579) 
\pstThreeDDot[linecolor=orange,linewidth=1.2pt](-0.3962,-0.2840,-0.0640) 
\pstThreeDDot[linecolor=orange,linewidth=1.2pt](-0.3849,-0.0888,0.0158) 
\pstThreeDDot[linecolor=orange,linewidth=1.2pt](-0.3537,-0.0322,0.0199) 
\pstThreeDDot[linecolor=blue,linewidth=1.2pt](-0.2762,0.1126,0.0079) 
\pstThreeDDot[linecolor=blue,linewidth=1.2pt](-0.4088,-0.0281,0.0579) 
\pstThreeDPut(-0.3962,-0.2840,-0.1140){S\&P}
\pstThreeDPut(-0.3849,-0.0888,0.0658){Nasd}
\pstThreeDPut(-0.3537,-0.0322,0.0699){Cana}
\pstThreeDPut(-0.2762,0.1126,0.0579){Germ}
\pstThreeDPut(-0.4088,-0.0281,0.1079){Neth}
\end{pspicture}
\begin{pspicture}(-8,-1.6)(1,1.5)
\rput(-3,1.1){T=0.5}
\psline(-3.9,-1.9)(2.9,-1.9)(2.9,1.5)(-3.9,1.5)(-3.9,-1.9)
\psset{xunit=5,yunit=5,Alpha=60,Beta=0} \scriptsize
\pstThreeDLine(-0.3962,-0.2840,-0.0640)(-0.3849,-0.0888,0.0158) 
\pstThreeDLine(-0.3962,-0.2840,-0.0640)(-0.3537,-0.0322,0.0199) 
\pstThreeDLine(-0.3962,-0.2840,-0.0640)(-0.4076,-0.1801,0.0021) 
\pstThreeDLine(-0.3962,-0.2840,-0.0640)(-0.4088,-0.0281,0.0579) 
\pstThreeDLine(-0.3849,-0.0888,0.0158)(-0.3537,-0.0322,0.0199) 
\pstThreeDLine(-0.3849,-0.0888,0.0158)(-0.4076,-0.1801,0.0021) 
\pstThreeDLine(-0.3849,-0.0888,0.0158)(-0.4088,-0.0281,0.0579) 
\pstThreeDLine(-0.3537,-0.0322,0.0199)(-0.4076,-0.1801,0.0021) 
\pstThreeDLine(-0.3537,-0.0322,0.0199)(-0.4088,-0.0281,0.0579) 
\pstThreeDLine(-0.4076,-0.1801,0.0021)(-0.2572,0.1114,-0.0763) 
\pstThreeDLine(-0.4076,-0.1801,0.0021)(-0.4088,-0.0281,0.0579) 
\pstThreeDLine(-0.2572,0.1114,-0.0763)(-0.4088,-0.0281,0.0579) 
\pstThreeDLine(-0.2572,0.1114,-0.0763)(-0.2006,0.2453,0.0863) 
\pstThreeDLine(-0.2762,0.1126,0.0079)(-0.4088,-0.0281,0.0579) 
\pstThreeDLine(-0.2762,0.1126,0.0079)(-0.2006,0.2453,0.0863) 
\pstThreeDDot[linecolor=orange,linewidth=1.2pt](-0.3962,-0.2840,-0.0640) 
\pstThreeDDot[linecolor=orange,linewidth=1.2pt](-0.3849,-0.0888,0.0158) 
\pstThreeDDot[linecolor=orange,linewidth=1.2pt](-0.3537,-0.0322,0.0199) 
\pstThreeDDot[linecolor=blue,linewidth=1.2pt](-0.4076,-0.1801,0.0021) 
\pstThreeDDot[linecolor=blue,linewidth=1.2pt](-0.2572,0.1114,-0.0763) 
\pstThreeDDot[linecolor=blue,linewidth=1.2pt](-0.2762,0.1126,0.0079) 
\pstThreeDDot[linecolor=blue,linewidth=1.2pt](-0.4088,-0.0281,0.0579) 
\pstThreeDDot[linecolor=blue,linewidth=1.2pt](-0.2006,0.2453,0.0863) 
\pstThreeDPut(-0.3962,-0.2840,-0.1140){S\&P}
\pstThreeDPut(-0.3849,-0.0888,0.0658){Nasd}
\pstThreeDPut(-0.3537,-0.0322,0.0699){Cana}
\pstThreeDPut(-0.4076,-0.1801,0.0521){UK}
\pstThreeDPut(-0.2572,0.1114,-0.1263){Irel}
\pstThreeDPut(-0.2762,0.1126,0.0579){Germ}
\pstThreeDPut(-0.4088,-0.0281,0.1079){Neth}
\pstThreeDPut(-0.2006,0.2453,0.1363){Swed}
\end{pspicture}

\vskip 0.4 cm

\begin{pspicture}(-4,-1.6)(1,1.5)
\rput(-3,1.1){T=0.6}
\psline(-3.9,-1.9)(2.9,-1.9)(2.9,1.5)(-3.9,1.5)(-3.9,-1.9)
\psset{xunit=5,yunit=5,Alpha=60,Beta=0} \scriptsize
\pstThreeDLine(-0.3962,-0.2840,-0.0640)(-0.3849,-0.0888,0.0158) 
\pstThreeDLine(-0.3962,-0.2840,-0.0640)(-0.3537,-0.0322,0.0199) 
\pstThreeDLine(-0.3962,-0.2840,-0.0640)(-0.4076,-0.1801,0.0021) 
\pstThreeDLine(-0.3962,-0.2840,-0.0640)(-0.2762,0.1126,0.0079) 
\pstThreeDLine(-0.3962,-0.2840,-0.0640)(-0.4088,-0.0281,0.0579) 
\pstThreeDLine(-0.3849,-0.0888,0.0158)(-0.3537,-0.0322,0.0199) 
\pstThreeDLine(-0.3849,-0.0888,0.0158)(-0.4076,-0.1801,0.0021) 
\pstThreeDLine(-0.3849,-0.0888,0.0158)(-0.2572,0.1114,-0.0763) 
\pstThreeDLine(-0.3849,-0.0888,0.0158)(-0.2762,0.1126,0.0079) 
\pstThreeDLine(-0.3849,-0.0888,0.0158)(-0.4088,-0.0281,0.0579) 
\pstThreeDLine(-0.3537,-0.0322,0.0199)(-0.4076,-0.1801,0.0021) 
\pstThreeDLine(-0.3537,-0.0322,0.0199)(-0.2762,0.1126,0.0079) 
\pstThreeDLine(-0.3537,-0.0322,0.0199)(-0.4088,-0.0281,0.0579) 
\pstThreeDLine(-0.4076,-0.1801,0.0021)(-0.2572,0.1114,-0.0763) 
\pstThreeDLine(-0.4076,-0.1801,0.0021)(-0.2762,0.1126,0.0079) 
\pstThreeDLine(-0.4076,-0.1801,0.0021)(-0.4088,-0.0281,0.0579) 
\pstThreeDLine(-0.2572,0.1114,-0.0763)(-0.2762,0.1126,0.0079) 
\pstThreeDLine(-0.2572,0.1114,-0.0763)(-0.4088,-0.0281,0.0579) 
\pstThreeDLine(-0.2572,0.1114,-0.0763)(-0.2006,0.2453,0.0863) 
\pstThreeDLine(-0.2762,0.1126,0.0079)(-0.4088,-0.0281,0.0579) 
\pstThreeDLine(-0.2762,0.1126,0.0079)(-0.2006,0.2453,0.0863) 
\pstThreeDLine(0.2350,0.4744,-0.0619)(0.0839,0.4052,-0.2131) 
\pstThreeDLine(-0.4088,-0.0281,0.0579)(-0.2006,0.2453,0.0863) 
\pstThreeDLine(-0.2006,0.2453,0.0863)(-0.0684,0.2229,-0.1159) 
\pstThreeDDot[linecolor=orange,linewidth=1.2pt](-0.3962,-0.2840,-0.0640) 
\pstThreeDDot[linecolor=orange,linewidth=1.2pt](-0.3849,-0.0888,0.0158) 
\pstThreeDDot[linecolor=orange,linewidth=1.2pt](-0.3537,-0.0322,0.0199) 
\pstThreeDDot[linecolor=blue,linewidth=1.2pt](-0.4076,-0.1801,0.0021) 
\pstThreeDDot[linecolor=blue,linewidth=1.2pt](-0.2572,0.1114,-0.0763) 
\pstThreeDDot[linecolor=blue,linewidth=1.2pt](-0.2762,0.1126,0.0079) 
\pstThreeDDot[linecolor=blue,linewidth=1.2pt](0.2350,0.4744,-0.0619) 
\pstThreeDDot[linecolor=blue,linewidth=1.2pt](-0.4088,-0.0281,0.0579) 
\pstThreeDDot[linecolor=blue,linewidth=1.2pt](-0.2006,0.2453,0.0863) 
\pstThreeDDot[linecolor=blue,linewidth=1.2pt](0.0839,0.4052,-0.2131) 
\pstThreeDDot[linecolor=red,linewidth=1.2pt](-0.0684,0.2229,-0.1159) 
\pstThreeDPut(-0.3962,-0.2840,-0.1140){S\&P}
\pstThreeDPut(-0.3849,-0.0888,0.0658){Nasd}
\pstThreeDPut(-0.3537,-0.0322,0.0699){Cana}
\pstThreeDPut(-0.4076,-0.1801,0.0521){UK}
\pstThreeDPut(-0.2572,0.1114,-0.1263){Irel}
\pstThreeDPut(-0.2762,0.1126,0.0579){Germ}
\pstThreeDPut(0.2350,0.4744,-0.1119){Autr}
\pstThreeDPut(-0.4088,-0.0281,0.1079){Neth}
\pstThreeDPut(-0.2006,0.2453,0.1363){Swed}
\pstThreeDPut(0.0839,0.4052,-0.2631){Spai}
\pstThreeDPut(-0.0684,0.2229,-0.1659){Japa}
\end{pspicture}
\begin{pspicture}(-8,-1.6)(1,1.5)
\rput(-3,1.1){T=0.7}
\psline(-3.9,-1.9)(2.9,-1.9)(2.9,1.5)(-3.9,1.5)(-3.9,-1.9)
\psset{xunit=5,yunit=5,Alpha=60,Beta=0} \scriptsize
\pstThreeDLine(-0.3962,-0.2840,-0.0640)(-0.3849,-0.0888,0.0158) 
\pstThreeDLine(-0.3962,-0.2840,-0.0640)(-0.3537,-0.0322,0.0199) 
\pstThreeDLine(-0.3962,-0.2840,-0.0640)(-0.4076,-0.1801,0.0021) 
\pstThreeDLine(-0.3962,-0.2840,-0.0640)(-0.2762,0.1126,0.0079) 
\pstThreeDLine(-0.3962,-0.2840,-0.0640)(-0.4088,-0.0281,0.0579) 
\pstThreeDLine(-0.3849,-0.0888,0.0158)(-0.3537,-0.0322,0.0199) 
\pstThreeDLine(-0.3849,-0.0888,0.0158)(-0.4076,-0.1801,0.0021) 
\pstThreeDLine(-0.3849,-0.0888,0.0158)(-0.2572,0.1114,-0.0763) 
\pstThreeDLine(-0.3849,-0.0888,0.0158)(-0.2762,0.1126,0.0079) 
\pstThreeDLine(-0.3849,-0.0888,0.0158)(-0.4088,-0.0281,0.0579) 
\pstThreeDLine(-0.3849,-0.0888,0.0158)(-0.2006,0.2453,0.0863) 
\pstThreeDLine(-0.3849,-0.0888,0.0158)(-0.0506,-0.2656,0.0853) 
\pstThreeDLine(-0.3849,-0.0888,0.0158)(-0.0106,0.2446,0.0558) 
\pstThreeDLine(-0.3537,-0.0322,0.0199)(-0.4076,-0.1801,0.0021) 
\pstThreeDLine(-0.3537,-0.0322,0.0199)(-0.2762,0.1126,0.0079) 
\pstThreeDLine(-0.3537,-0.0322,0.0199)(-0.4088,-0.0281,0.0579) 
\pstThreeDLine(-0.3537,-0.0322,0.0199)(-0.2572,0.1114,-0.0763) 
\pstThreeDLine(-0.3537,-0.0322,0.0199)(-0.2006,0.2453,0.0863) 
\pstThreeDLine(-0.3537,-0.0322,0.0199)(-0.1158,0.2376,0.0565) 
\pstThreeDLine(-0.4076,-0.1801,0.0021)(-0.2572,0.1114,-0.0763) 
\pstThreeDLine(-0.4076,-0.1801,0.0021)(-0.2762,0.1126,0.0079) 
\pstThreeDLine(-0.4076,-0.1801,0.0021)(-0.4088,-0.0281,0.0579) 
\pstThreeDLine(-0.4076,-0.1801,0.0021)(-0.2006,0.2453,0.0863) 
\pstThreeDLine(-0.2572,0.1114,-0.0763)(-0.2762,0.1126,0.0079) 
\pstThreeDLine(-0.2572,0.1114,-0.0763)(-0.4088,-0.0281,0.0579) 
\pstThreeDLine(-0.2572,0.1114,-0.0763)(-0.2006,0.2453,0.0863) 
\pstThreeDLine(-0.2572,0.1114,-0.0763)(-0.1158,0.2376,0.0565) 
\pstThreeDLine(-0.2762,0.1126,0.0079)(-0.4088,-0.0281,0.0579) 
\pstThreeDLine(-0.2762,0.1126,0.0079)(-0.2006,0.2453,0.0863) 
\pstThreeDLine(-0.2762,0.1126,0.0079)(-0.0684,0.2229,-0.1159) 
\pstThreeDLine(-0.2762,0.1126,0.0079)(-0.1158,0.2376,0.0565) 
\pstThreeDLine(0.2350,0.4744,-0.0619)(0.0839,0.4052,-0.2131) 
\pstThreeDLine(0.2350,0.4744,-0.0619)(-0.2006,0.2453,0.0863) 
\pstThreeDLine(0.2350,0.4744,-0.0619)(-0.0106,0.2446,0.0558) 
\pstThreeDLine(-0.4088,-0.0281,0.0579)(-0.2006,0.2453,0.0863) 
\pstThreeDLine(-0.4088,-0.0281,0.0579)(-0.1158,0.2376,0.0565) 
\pstThreeDLine(-0.2006,0.2453,0.0863)(-0.0684,0.2229,-0.1159) 
\pstThreeDLine(-0.2006,0.2453,0.0863)(-0.1158,0.2376,0.0565) 
\pstThreeDLine(0.0839,0.4052,-0.2131)(-0.1158,0.2376,0.0565) 
\pstThreeDLine(0.0839,0.4052,-0.2131)(-0.0106,0.2446,0.0558) 
\pstThreeDLine(-0.0684,0.2229,-0.1159)(-0.0106,0.2446,0.0558) 
\pstThreeDDot[linecolor=orange,linewidth=1.2pt](-0.3962,-0.2840,-0.0640) 
\pstThreeDDot[linecolor=orange,linewidth=1.2pt](-0.3849,-0.0888,0.0158) 
\pstThreeDDot[linecolor=orange,linewidth=1.2pt](-0.3537,-0.0322,0.0199) 
\pstThreeDDot[linecolor=blue,linewidth=1.2pt](-0.4076,-0.1801,0.0021) 
\pstThreeDDot[linecolor=blue,linewidth=1.2pt](-0.2572,0.1114,-0.0763) 
\pstThreeDDot[linecolor=blue,linewidth=1.2pt](-0.2762,0.1126,0.0079) 
\pstThreeDDot[linecolor=blue,linewidth=1.2pt](0.2350,0.4744,-0.0619) 
\pstThreeDDot[linecolor=blue,linewidth=1.2pt](-0.4088,-0.0281,0.0579) 
\pstThreeDDot[linecolor=blue,linewidth=1.2pt](-0.2006,0.2453,0.0863) 
\pstThreeDDot[linecolor=blue,linewidth=1.2pt](0.0839,0.4052,-0.2131) 
\pstThreeDDot[linecolor=brown,linewidth=1.2pt](-0.0506,-0.2656,0.0853) 
\pstThreeDDot[linecolor=red,linewidth=1.2pt](-0.0684,0.2229,-0.1159) 
\pstThreeDDot[linecolor=red,linewidth=1.2pt](-0.1158,0.2376,0.0565) 
\pstThreeDDot[linecolor=red,linewidth=1.2pt](-0.0106,0.2446,0.0558) 
\pstThreeDPut(-0.3962,-0.2840,-0.1140){S\&P}
\pstThreeDPut(-0.3849,-0.0888,0.0658){Nasd}
\pstThreeDPut(-0.3537,-0.0322,0.0699){Cana}
\pstThreeDPut(-0.4076,-0.1801,0.0521){UK}
\pstThreeDPut(-0.2572,0.1114,-0.1263){Irel}
\pstThreeDPut(-0.2762,0.1126,0.0579){Germ}
\pstThreeDPut(0.2350,0.4744,-0.1119){Autr}
\pstThreeDPut(-0.4088,-0.0281,0.1079){Neth}
\pstThreeDPut(-0.2006,0.2453,0.1363){Swed}
\pstThreeDPut(0.0839,0.4052,-0.2631){Spai}
\pstThreeDPut(-0.0506,-0.2656,0.1353){Indi}
\pstThreeDPut(-0.0684,0.2229,-0.1659){Japa}
\pstThreeDPut(-0.1158,0.2376,0.1065){HoKo}
\pstThreeDPut(-0.0106,0.2446,0.1058){Mala}
\end{pspicture}

\vskip 0.7 cm

\noindent Fig. 4. Three dimensional view of the asset trees for the second semester of 1987, with threshold ranging from $T=0.4$ to $T=0.7$.

\vskip 0.3 cm

\subsection{1997 and 1998 - Asian Financial Crisis and Russian Crisis}

We now jump some years ahead to the next two crises we are going to analyze. The first one is the Asian Financial Crisis of 1997, which began with the devaluation of the Thailandese currency and spread to other Pacific Asian markets. The second was almost a consequence of the first one, as prices of commodities fell worldwide, affecting the Russian economy with particular acuteness. The networks for 1997 are built using 57 indices, adding the indices IPC from Mexico (Mexi), BCT Corp from Costa Rica (CoRi), Bermuda SX Index from Bermuda (Bermu), Jamaica SX from Jamaica (Jama), Merval from Argentina (Arge), IPSA from Chile (Chil), IBC from Venezuela (Vene), IGBVL from Peru (Peru), CAC 40 from France (Fran), SMI from Switzerland (Swit), BEL 20 from Belgium (Belg), OMX Copenhagen 20 from Denmark (Denm), OBX from Norway (Norw), OMX Iceland from Iceland (Icel), PSI 20 from Portugal (Port), PX (or PX50) from the Czech Republic (CzRe), SAX from Slovakia (Slok), Budapest SX from Hungary (Hung), WIG from Poland (Pola), OMXT from Estonia (Esto), ISE National 100 from Turkey (Turk), Tel Aviv 25 from Israel (Isra), BLOM from Lebanon (Leba), TASI from Saudi Arabia (SaAr), MSM 30 from Ohman (Ohma), Karachi 100 from Pakistan (Paki), Shanghai SE Composite from China (Chin), SET from Thailand (Thai), S\&P/ASX 200 from Australia (Aust), CFG25 from Morocco (Moro), Ghana All Share from Ghana (Ghan), NSE 20 from Kenya (Keny), FTSE/JSE Africa All Share from South Africa (SoAf), and SEMDEX from Mauritius (Maur). For 1997, we have no data about Russia, which is added as the index MICEX from Russia (Russ) to the data used for building networks for 1998 (which then has 58 indices).

\subsubsection{First semester, 1997}

Figure 5 shows the three dimensional view of the asset trees for the first semester of 1997, with threshold ranging from $T=0.4$ to $T=0.7$.

At threshold $T=0.3$, the only connection is between S\&P ans Nasdaq. For $T=0.4$, Canada joins the North American cluster, and an European cluster appears, formed by France, Germany, Switzerland, Belgium, Netherlands, Sweden, Finland, and Norway. At $T=0.5$, nothing changes in the North Amercian cluster, but the European cluster becomes denser, with more connections formed among its indices. The UK, Ireland, Austria, Denmark, Spain, and Portugal join that cluster. Something to notice is that there are strong ties formed among the Scandinavian indices, and among the Central European ones. For $T=0.6$, Brazil and Argentina connect with the now American cluster. There is also further consolidation in the European cluster. Strange connections are formed, like two connections of Peru, with Denmark and Portugal. Less strange are the connections of Australia with Germany, and of Hong Kong with Austria. Noise starts to become important at this threshold, so one should be careful not to consider all connections now as true information. At $T=0.7$, noise becomes a strong effect, but we analize this case, anyway, because it also has some important information. We now have connections between the American and the European clusters, formed via Canada and Chile. Mexico connects with Argentina, and thus to the American cluster. Poland joins the European cluster, and Hong Kong, Australia, and South Africa establish more connections with Europe (South Africa, for the first time). There is also a small, separate cluster, formed by Malaysia and Indonesia. Peru connects with Norway, and Iceland forms a cluster with Mauritius, the latter almost clearly an effect of random noise. More connections are formed for higher thresholds, but many of them are the effect of random connections, and are indistinguishable from connections containing true information.

\vskip 0.3 cm



\vskip 2.2 cm

\noindent Fig. 5. Three dimensional view of the asset trees for the first semester of 1997, with threshold ranging from $T=0.4$ to $T=0.7$.

\vskip 0.3 cm

\subsubsection{Second semester, 1997}

Figure 6 shows the three dimensional view of the asset trees for the second semester of 1997, with threshold ranging from $T=0.3$ to $T=0.6$.

\vskip -0.2 cm

%

\vskip 2 cm

\noindent Fig. 6. Three dimensional view of the asset trees for the second semester of 1997, with threshold ranging from $T=0.3$ to $T=0.6$.

\vskip 0.3 cm

At $T=0.2$, two seeds of clusters appear: one comprised of S\&P and Nasdaq, and the other made of Switzerland and Netherlands. The European cluster grows at $T=0.3$, with the addition of the UK, France, Germany, Austria, Belgium, Sweden, Finland, and Spain. At $T=0.4$, an American cluster is formed, with Canada, Mexico, Brazil, and Argentina joining S\&P and Nasdaq. The European cluster is joined by Ireland and Denmark, and becomes denser. At $T=0.5$, connections are established between the two existing clusters, and the European cluster is joined by Chile, which does not connect with the American cluster. Norway and Portugal also join the European cluster. At $T=0.6$, there is a massive number of connections between the American and the European clusters. Chile connects further with Europe and also with America, Peru also connects with both clusters, and Hungary and South Africa get fully integrated with the European cluster, with Hungary also connecting with the American cluster. Poland, Israel, Hong Kong, and Australia also make connections with the European cluster. Japan connects with Australia, and Hong Kong connects with Malaysia. For increasing values of the threshold, many more connections are made, but random noise becomes so strong we cannot separate meaningful connections from random ones.

\subsubsection{First semester, 1998}

Figure 7 shows the three dimensional view of the asset trees for the first semester of 1998, with threshold ranging from $T=0.3$ to $T=0.6$.

Connections are not formed until $T=0.3$. At this threshold, we have connections between S\&P and Nasdaq, and between the UK, France, Switzerland, Belgium, Netherlands, Sweden, Finland, and Spain, forming the North American and the European clusters. At $T=0.4$, the North American cluster stays unaltered, while the European cluster becomes denser, with more connections among its members, and with the joining of Germany and Portugal. For $T=0.5$, Canada, Mexico, Argentina and Brazil, join the American cluster, which connects with the European one via Canada. Ireland, Denmark, and Norway join the European cluster, which becomes even denser, and establishes connections with Hong Kong and South Africa. Hungary connects with Hong Kong and we see the emergence of a Pacific Asian cluster, comprised of Hong Kong, Thailand, and Malaysia. At $T=0.6$, just bellow the area dominated by noise, we have an American cluster very much integrated with Europe, and Chile also connected with the European cluster. South Africa establishes a connection with America and more connections with Europe, along with a connection with the Pacific Asian cluster via Hong Kong. The European cluster, connected now with Hungary, now also has connections with Israel, India, and Australia. Hong Kong has now more connections with Europe, and also forms connections with Taiwan, Thailand, Philipines, the latter also forming connections with Thailand and Malaysia, strenghtening the Pacific Asian cluster. Above this threshold, noise starts to dominate, although we have the formation of many more meaningful connections, which are indistiguishable from the randomly made ones. The last connection occurs at $T=1.4$, between Iceland and Portugal.

\vskip 0.8 cm



\vskip 2 cm

\noindent Fig. 7. Three dimensional view of the asset trees for the first semester of 1998, with threshold ranging from $T=0.3$ to $T=0.6$.

\vskip 0.3 cm

\subsubsection{Second semester, 1998}

Figure 8 shows the three dimensional view of the asset trees for the second semester of 1998, with threshold ranging from $T=0.3$ to $T=0.6$.

\vskip 0.8 cm

%

\vskip 2.3 cm

\noindent Fig. 8. Three dimensional view of the asset trees for the second semester of 1998, with threshold ranging from $T=0.3$ to $T=0.6$.

\vskip 0.3 cm

At $T=0.2$, two clusters appear: the North American one, formed by S\&P and Nasdaq, and the European cluster, comprised of the UK, France, Germany, Switzerland, Belgium, Netherlands, Sweden, Finland, and Spain. At $T=0.3$, a third cluster, comprised of Brazil and Argentina, is formed. The European cluster strenghtens its ties, and gains Denmark, Norway, Portugal, the Czech Republic, and Hungary. For $T=0.4$, Canada joins the North American cluster, and Mexico and Chile join the South American one. The European cluster becomes even denser and absorbs Ireland, Austria, and Greece. South Africa makes connections with most of the European indices, effectively joining the European cluster. For $T=0.5$, the North and South American clusters merge, and Canada makes many connections with Europe. Poland joins the European cluster, which now makes connections with Turkey, Israel, Hong Kong, and Australia. Hong Kong connects with South Africa, and Australia connects with Hong Kong, Philippines, and South Africa. At $T=0.6$, there is a strong integration between the American and the European clusters, now with the addition of Venezuela and Peru. Turkey, Israel, Hong Kong, and Australia fully integrate with Europe, and Japan connects both with the European cluster and with a Pacific Asian one, comprised of Japan, Hong Kong, Thailand, Indonesia, and Australia. South Korea and Malaysia connect with Europe and not with the Pacific Asian cluster. For $T>0.6$, the connections between clusters strenghten, and the Pacific Asian cluster becomes more self-connected, but noise becomes strong from here onwards, and so connections cannot be trusted anymore. The last connections occur at $T=1.3$.

\subsection{2001 - Burst of the dot-com Bubble and September 11}

Two very distinct crises arose in 2001: the first was the burst of a bubble due to the overvaluing of the new internet-based companies, and the second was the fear caused by the severest terrorist attack a country has ever faced, when the USA was targeted by commercial airplanes overtaken by terrorists. Both crises were very different in origin and in duration, and their analysis might shed some light on how crises may develope.

For the networks concerning the year 2000, we add now the indices Bolsa de Panama General from Panama (Pana), FTSE MIB (or MIB-30) from Italy (Ital), Malta SX from Malta (Malt), LUxX from Luxembourg (Luxe), BET from Romania (Roma), OMXR from Latvia (Latv), OMXV from Lithuania (Lith), PFTS from Ukraine (Ukra), Al Quds from Palestine (Pale), ASE General from Jordan (Jorda), QE (or DSM 20) from Qatar (Qata), MSE Top 20 from Mongolia (Mongo), Straits Times from Singapore (Sing), TUNINDEX from Tunisia (Tuni), EGX 30 from Egypt (Egyp), and Nigeria SX All Share from Nigeria (Nige), for a total of 74 indices. For the networks of 2001, we also add the indices SOFIX from Bulgaria (Bulga), KASE from Kazakhstan (Kaza), VN-Index from Vietnam (Viet), NZX 50 from New Zealand (NeZe), and Gaborone from Botswana (Bots), for a total of 79 indices.

\subsubsection{First semester, 2000}

Figure 9 shows the three dimensional view of the asset trees for the first semester of 2000, with threshold ranging from $T=0.3$ to $T=0.6$.

\vskip 0.8 cm

\begin{pspicture}(-4,-1.9)(1,1.8)
\rput(-3,1.7){T=0.3}
\psline(-3.9,-2.9)(3.7,-2.9)(3.7,2.1)(-3.9,2.1)(-3.9,-2.9)
\psset{xunit=7,yunit=7,Alpha=40,Beta=0} \scriptsize
\pstThreeDLine(-0.2591,-0.4498,-0.1778)(-0.2765,-0.3585,-0.1471) 
\pstThreeDLine(-0.4405,-0.0111,0.0425)(-0.4254,-0.0865,0.0298) 
\pstThreeDLine(-0.4405,-0.0111,0.0425)(-0.3999,-0.0837,0.1406) 
\pstThreeDLine(-0.4405,-0.0111,0.0425)(-0.4286,0.0333,-0.0189) 
\pstThreeDLine(-0.4405,-0.0111,0.0425)(-0.3927,-0.0353,0.1513) 
\pstThreeDLine(-0.4254,-0.0865,0.0298)(-0.3999,-0.0837,0.1406) 
\pstThreeDLine(-0.4254,-0.0865,0.0298)(-0.4286,0.0333,-0.0189) 
\pstThreeDLine(-0.3999,-0.0837,0.1406)(-0.3999,-0.0837,0.1406) 
\pstThreeDLine(-0.3855,0.0439,0.0641)(-0.3508,0.1151,0.0570) 
\pstThreeDLine(-0.3927,-0.0353,0.1513)(-0.3622,-0.0730,0.1113) 
\pstThreeDDot[linecolor=orange,linewidth=1.2pt](-0.2591,-0.4498,-0.1778) 
\pstThreeDDot[linecolor=orange,linewidth=1.2pt](-0.2765,-0.3585,-0.1471) 
\pstThreeDDot[linecolor=blue,linewidth=1.2pt](-0.4405,-0.0111,0.0425) 
\pstThreeDDot[linecolor=blue,linewidth=1.2pt](-0.4254,-0.0865,0.0298) 
\pstThreeDDot[linecolor=blue,linewidth=1.2pt](-0.3999,-0.0837,0.1406) 
\pstThreeDDot[linecolor=blue,linewidth=1.2pt](-0.4286,0.0333,-0.0189) 
\pstThreeDDot[linecolor=blue,linewidth=1.2pt](-0.3855,0.0439,0.0641) 
\pstThreeDDot[linecolor=blue,linewidth=1.2pt](-0.3508,0.1151,0.0570) 
\pstThreeDDot[linecolor=blue,linewidth=1.2pt](-0.3927,-0.0353,0.1513) 
\pstThreeDDot[linecolor=blue,linewidth=1.2pt](-0.3622,-0.0730,0.1113) 
\pstThreeDPut(-0.2591,-0.4498,-0.2078){S\&P}
\pstThreeDPut(-0.2765,-0.3585,-0.1771){Nasd}
\pstThreeDPut(-0.4405,-0.0111,0.0725){Fran}
\pstThreeDPut(-0.4254,-0.0865,0.0598){Germ}
\pstThreeDPut(-0.3999,-0.0837,0.1706){Ital}
\pstThreeDPut(-0.4286,0.0333,-0.0489){Neth}
\pstThreeDPut(-0.3855,0.0439,0.0941){Swed}
\pstThreeDPut(-0.3508,0.1151,0.0870){Finl}
\pstThreeDPut(-0.3927,-0.0353,0.1813){Spai}
\pstThreeDPut(-0.3622,-0.0730,0.1413){Port}
\end{pspicture}
\begin{pspicture}(-8,-1.9)(1,1.8)
\rput(-3,1.7){T=0.4}
\psline(-3.9,-2.9)(3.7,-2.9)(3.7,2.1)(-3.9,2.1)(-3.9,-2.9)
\psset{xunit=7,yunit=7,Alpha=40,Beta=0} \scriptsize
\pstThreeDLine(-0.2591,-0.4498,-0.1778)(-0.2765,-0.3585,-0.1471) 
\pstThreeDLine(-0.2591,-0.4498,-0.1778)(-0.2965,-0.2159,-0.1210) 
\pstThreeDLine(-0.2591,-0.4498,-0.1778)(-0.2411,-0.3198,-0.1859) 
\pstThreeDLine(-0.2765,-0.3585,-0.1471)(-0.2965,-0.2159,-0.1210) 
\pstThreeDLine(-0.2765,-0.3585,-0.1471)(-0.2411,-0.3198,-0.1859) 
\pstThreeDLine(-0.4256,-0.0475,0.0253)(-0.4405,-0.0111,0.0425) 
\pstThreeDLine(-0.4256,-0.0475,0.0253)(-0.4254,-0.0865,0.0298) 
\pstThreeDLine(-0.4256,-0.0475,0.0253)(-0.3999,-0.0837,0.1406) 
\pstThreeDLine(-0.4256,-0.0475,0.0253)(-0.4286,0.0333,-0.0189) 
\pstThreeDLine(-0.4256,-0.0475,0.0253)(-0.3927,-0.0353,0.1513) 
\pstThreeDLine(-0.4405,-0.0111,0.0425)(-0.4254,-0.0865,0.0298) 
\pstThreeDLine(-0.4405,-0.0111,0.0425)(-0.3999,-0.0837,0.1406) 
\pstThreeDLine(-0.4405,-0.0111,0.0425)(-0.4286,0.0333,-0.0189) 
\pstThreeDLine(-0.4405,-0.0111,0.0425)(-0.3855,0.0439,0.0641) 
\pstThreeDLine(-0.4405,-0.0111,0.0425)(-0.3508,0.1151,0.0570) 
\pstThreeDLine(-0.4405,-0.0111,0.0425)(-0.3927,-0.0353,0.1513) 
\pstThreeDLine(-0.4405,-0.0111,0.0425)(-0.3622,-0.0730,0.1113) 
\pstThreeDLine(-0.4254,-0.0865,0.0298)(-0.3999,-0.0837,0.1406) 
\pstThreeDLine(-0.4254,-0.0865,0.0298)(-0.4286,0.0333,-0.0189) 
\pstThreeDLine(-0.4254,-0.0865,0.0298)(-0.3855,0.0439,0.0641) 
\pstThreeDLine(-0.4254,-0.0865,0.0298)(-0.3927,-0.0353,0.1513) 
\pstThreeDLine(-0.4254,-0.0865,0.0298)(-0.3622,-0.0730,0.1113) 
\pstThreeDLine(-0.3999,-0.0837,0.1406)(-0.4286,0.0333,-0.0189) 
\pstThreeDLine(-0.3999,-0.0837,0.1406)(-0.3999,-0.0837,0.1406) 
\pstThreeDLine(-0.4286,0.0333,-0.0189)(-0.3855,0.0439,0.0641) 
\pstThreeDLine(-0.4286,0.0333,-0.0189)(-0.3508,0.1151,0.0570) 
\pstThreeDLine(-0.4286,0.0333,-0.0189)(-0.3927,-0.0353,0.1513) 
\pstThreeDLine(-0.3855,0.0439,0.0641)(-0.3508,0.1151,0.0570) 
\pstThreeDLine(-0.3855,0.0439,0.0641)(-0.3927,-0.0353,0.1513) 
\pstThreeDLine(-0.3855,0.0439,0.0641)(-0.3622,-0.0730,0.1113) 
\pstThreeDLine(-0.3508,0.1151,0.0570)(-0.3927,-0.0353,0.1513) 
\pstThreeDLine(-0.3508,0.1151,0.0570)(-0.3622,-0.0730,0.1113) 
\pstThreeDLine(-0.3927,-0.0353,0.1513)(-0.3622,-0.0730,0.1113) 
\pstThreeDDot[linecolor=orange,linewidth=1.2pt](-0.2591,-0.4498,-0.1778) 
\pstThreeDDot[linecolor=orange,linewidth=1.2pt](-0.2765,-0.3585,-0.1471) 
\pstThreeDDot[linecolor=orange,linewidth=1.2pt](-0.2965,-0.2159,-0.1210) 
\pstThreeDDot[linecolor=orange,linewidth=1.2pt](-0.2411,-0.3198,-0.1859) 
\pstThreeDDot[linecolor=blue,linewidth=1.2pt](-0.4256,-0.0475,0.0253) 
\pstThreeDDot[linecolor=blue,linewidth=1.2pt](-0.4405,-0.0111,0.0425) 
\pstThreeDDot[linecolor=blue,linewidth=1.2pt](-0.4254,-0.0865,0.0298) 
\pstThreeDDot[linecolor=blue,linewidth=1.2pt](-0.3999,-0.0837,0.1406) 
\pstThreeDDot[linecolor=blue,linewidth=1.2pt](-0.4286,0.0333,-0.0189) 
\pstThreeDDot[linecolor=blue,linewidth=1.2pt](-0.3855,0.0439,0.0641) 
\pstThreeDDot[linecolor=blue,linewidth=1.2pt](-0.3508,0.1151,0.0570) 
\pstThreeDDot[linecolor=blue,linewidth=1.2pt](-0.3927,-0.0353,0.1513) 
\pstThreeDDot[linecolor=blue,linewidth=1.2pt](-0.3622,-0.0730,0.1113) 
\pstThreeDPut(-0.2591,-0.4498,-0.2078){S\&P}
\pstThreeDPut(-0.2765,-0.3585,-0.1771){Nasd}
\pstThreeDPut(-0.2965,-0.2159,-0.1510){Cana}
\pstThreeDPut(-0.2411,-0.3198,-0.2159){Mexi}
\pstThreeDPut(-0.4256,-0.0475,0.0553){UK}
\pstThreeDPut(-0.4405,-0.0111,0.0725){Fran}
\pstThreeDPut(-0.4254,-0.0865,0.0598){Germ}
\pstThreeDPut(-0.3999,-0.0837,0.1706){Ital}
\pstThreeDPut(-0.4286,0.0333,-0.0489){Neth}
\pstThreeDPut(-0.3855,0.0439,0.0941){Swed}
\pstThreeDPut(-0.3508,0.1151,0.0870){Finl}
\pstThreeDPut(-0.3927,-0.0353,0.1813){Spai}
\pstThreeDPut(-0.3622,-0.0730,0.1413){Port}
\end{pspicture}

\vskip 1.7 cm

\begin{pspicture}(-4,-1.9)(1,1.8)
\rput(-3,1.7){T=0.5}
\psline(-3.9,-2.9)(3.7,-2.9)(3.7,2.1)(-3.9,2.1)(-3.9,-2.9)
\psset{xunit=7,yunit=7,Alpha=40,Beta=0} \scriptsize
\pstThreeDLine(-0.2591,-0.4498,-0.1778)(-0.2765,-0.3585,-0.1471) 
\pstThreeDLine(-0.2591,-0.4498,-0.1778)(-0.2965,-0.2159,-0.1210) 
\pstThreeDLine(-0.2591,-0.4498,-0.1778)(-0.2411,-0.3198,-0.1859) 
\pstThreeDLine(-0.2591,-0.4498,-0.1778)(-0.2339,-0.3289,-0.1857) 
\pstThreeDLine(-0.2591,-0.4498,-0.1778)(-0.2311,-0.4492,-0.0349) 
\pstThreeDLine(-0.2765,-0.3585,-0.1471)(-0.2965,-0.2159,-0.1210) 
\pstThreeDLine(-0.2765,-0.3585,-0.1471)(-0.2411,-0.3198,-0.1859) 
\pstThreeDLine(-0.2765,-0.3585,-0.1471)(-0.2339,-0.3289,-0.1857) 
\pstThreeDLine(-0.2765,-0.3585,-0.1471)(-0.2311,-0.4492,-0.0349) 
\pstThreeDLine(-0.2965,-0.2159,-0.1210)(-0.4254,-0.0865,0.0298) 
\pstThreeDLine(-0.2411,-0.3198,-0.1859)(-0.2339,-0.3289,-0.1857) 
\pstThreeDLine(-0.2411,-0.3198,-0.1859)(-0.2311,-0.4492,-0.0349) 
\pstThreeDLine(-0.2339,-0.3289,-0.1857)(-0.2311,-0.4492,-0.0349) 
\pstThreeDLine(-0.4256,-0.0475,0.0253)(-0.4405,-0.0111,0.0425) 
\pstThreeDLine(-0.4256,-0.0475,0.0253)(-0.4254,-0.0865,0.0298) 
\pstThreeDLine(-0.4256,-0.0475,0.0253)(-0.3999,-0.0837,0.1406) 
\pstThreeDLine(-0.4256,-0.0475,0.0253)(-0.4286,0.0333,-0.0189) 
\pstThreeDLine(-0.4256,-0.0475,0.0253)(-0.3855,0.0439,0.0641) 
\pstThreeDLine(-0.4256,-0.0475,0.0253)(-0.3508,0.1151,0.0570) 
\pstThreeDLine(-0.4256,-0.0475,0.0253)(-0.3047,0.0054,0.1790) 
\pstThreeDLine(-0.4256,-0.0475,0.0253)(-0.3927,-0.0353,0.1513) 
\pstThreeDLine(-0.4256,-0.0475,0.0253)(-0.3622,-0.0730,0.1113) 
\pstThreeDLine(-0.4405,-0.0111,0.0425)(-0.4254,-0.0865,0.0298) 
\pstThreeDLine(-0.4405,-0.0111,0.0425)(-0.3999,-0.0837,0.1406) 
\pstThreeDLine(-0.4405,-0.0111,0.0425)(-0.4286,0.0333,-0.0189) 
\pstThreeDLine(-0.4405,-0.0111,0.0425)(-0.3855,0.0439,0.0641) 
\pstThreeDLine(-0.4405,-0.0111,0.0425)(-0.3508,0.1151,0.0570) 
\pstThreeDLine(-0.4405,-0.0111,0.0425)(-0.3927,-0.0353,0.1513) 
\pstThreeDLine(-0.4405,-0.0111,0.0425)(-0.3622,-0.0730,0.1113) 
\pstThreeDLine(-0.4254,-0.0865,0.0298)(-0.3999,-0.0837,0.1406) 
\pstThreeDLine(-0.4254,-0.0865,0.0298)(-0.4286,0.0333,-0.0189) 
\pstThreeDLine(-0.4254,-0.0865,0.0298)(-0.3855,0.0439,0.0641) 
\pstThreeDLine(-0.4254,-0.0865,0.0298)(-0.3508,0.1151,0.0570) 
\pstThreeDLine(-0.4254,-0.0865,0.0298)(-0.3927,-0.0353,0.1513) 
\pstThreeDLine(-0.4254,-0.0865,0.0298)(-0.3622,-0.0730,0.1113) 
\pstThreeDLine(-0.3999,-0.0837,0.1406)(-0.4286,0.0333,-0.0189) 
\pstThreeDLine(-0.3999,-0.0837,0.1406)(-0.3855,0.0439,0.0641) 
\pstThreeDLine(-0.3999,-0.0837,0.1406)(-0.3508,0.1151,0.0570) 
\pstThreeDLine(-0.3999,-0.0837,0.1406)(-0.3999,-0.0837,0.1406) 
\pstThreeDLine(-0.3999,-0.0837,0.1406)(-0.3622,-0.0730,0.1113) 
\pstThreeDLine(-0.4286,0.0333,-0.0189)(-0.3855,0.0439,0.0641) 
\pstThreeDLine(-0.4286,0.0333,-0.0189)(-0.2527,0.1271,0.0834) 
\pstThreeDLine(-0.4286,0.0333,-0.0189)(-0.3508,0.1151,0.0570) 
\pstThreeDLine(-0.4286,0.0333,-0.0189)(-0.3047,0.0054,0.1790) 
\pstThreeDLine(-0.4286,0.0333,-0.0189)(-0.3927,-0.0353,0.1513) 
\pstThreeDLine(-0.4286,0.0333,-0.0189)(-0.3622,-0.0730,0.1113) 
\pstThreeDLine(-0.3855,0.0439,0.0641)(-0.2527,0.1271,0.0834) 
\pstThreeDLine(-0.3855,0.0439,0.0641)(-0.3508,0.1151,0.0570) 
\pstThreeDLine(-0.3855,0.0439,0.0641)(-0.3047,0.0054,0.1790) 
\pstThreeDLine(-0.3855,0.0439,0.0641)(-0.3927,-0.0353,0.1513) 
\pstThreeDLine(-0.3855,0.0439,0.0641)(-0.3622,-0.0730,0.1113) 
\pstThreeDLine(-0.3508,0.1151,0.0570)(-0.3047,0.0054,0.1790) 
\pstThreeDLine(-0.3508,0.1151,0.0570)(-0.3927,-0.0353,0.1513) 
\pstThreeDLine(-0.3508,0.1151,0.0570)(-0.3622,-0.0730,0.1113) 
\pstThreeDLine(-0.3047,0.0054,0.1790)(-0.3927,-0.0353,0.1513) 
\pstThreeDLine(-0.3927,-0.0353,0.1513)(-0.3622,-0.0730,0.1113) 
\pstThreeDDot[linecolor=orange,linewidth=1.2pt](-0.2591,-0.4498,-0.1778) 
\pstThreeDDot[linecolor=orange,linewidth=1.2pt](-0.2765,-0.3585,-0.1471) 
\pstThreeDDot[linecolor=orange,linewidth=1.2pt](-0.2965,-0.2159,-0.1210) 
\pstThreeDDot[linecolor=orange,linewidth=1.2pt](-0.2411,-0.3198,-0.1859) 
\pstThreeDDot[linecolor=green,linewidth=1.2pt](-0.2339,-0.3289,-0.1857) 
\pstThreeDDot[linecolor=green,linewidth=1.2pt](-0.2311,-0.4492,-0.0349) 
\pstThreeDDot[linecolor=blue,linewidth=1.2pt](-0.4256,-0.0475,0.0253) 
\pstThreeDDot[linecolor=blue,linewidth=1.2pt](-0.4405,-0.0111,0.0425) 
\pstThreeDDot[linecolor=blue,linewidth=1.2pt](-0.4254,-0.0865,0.0298) 
\pstThreeDDot[linecolor=blue,linewidth=1.2pt](-0.3999,-0.0837,0.1406) 
\pstThreeDDot[linecolor=blue,linewidth=1.2pt](-0.4286,0.0333,-0.0189) 
\pstThreeDDot[linecolor=blue,linewidth=1.2pt](-0.3855,0.0439,0.0641) 
\pstThreeDDot[linecolor=blue,linewidth=1.2pt](-0.2527,0.1271,0.0834) 
\pstThreeDDot[linecolor=blue,linewidth=1.2pt](-0.3508,0.1151,0.0570) 
\pstThreeDDot[linecolor=blue,linewidth=1.2pt](-0.3047,0.0054,0.1790) 
\pstThreeDDot[linecolor=blue,linewidth=1.2pt](-0.3927,-0.0353,0.1513) 
\pstThreeDDot[linecolor=blue,linewidth=1.2pt](-0.3622,-0.0730,0.1113) 
\pstThreeDPut(-0.2591,-0.4498,-0.2078){S\&P}
\pstThreeDPut(-0.2765,-0.3585,-0.1771){Nasd}
\pstThreeDPut(-0.2965,-0.2159,-0.1510){Cana}
\pstThreeDPut(-0.2411,-0.3198,-0.2159){Mexi}
\pstThreeDPut(-0.2339,-0.3289,-0.2157){Braz}
\pstThreeDPut(-0.2311,-0.4492,-0.0649){Arge}
\pstThreeDPut(-0.4256,-0.0475,0.0553){UK}
\pstThreeDPut(-0.4405,-0.0111,0.0725){Fran}
\pstThreeDPut(-0.4254,-0.0865,0.0598){Germ}
\pstThreeDPut(-0.3999,-0.0837,0.1706){Ital}
\pstThreeDPut(-0.4286,0.0333,-0.0489){Neth}
\pstThreeDPut(-0.3855,0.0439,0.0941){Swed}
\pstThreeDPut(-0.2527,0.1271,0.1134){Denm}
\pstThreeDPut(-0.3508,0.1151,0.0870){Finl}
\pstThreeDPut(-0.3047,0.0054,0.2090){Norw}
\pstThreeDPut(-0.3927,-0.0353,0.1813){Spai}
\pstThreeDPut(-0.3622,-0.0730,0.1413){Port}
\end{pspicture}
\begin{pspicture}(-8,-1.9)(1,1.8)
\rput(-3,1.7){T=0.6}
\psline(-3.9,-2.9)(3.7,-2.9)(3.7,2.1)(-3.9,2.1)(-3.9,-2.9)
\psset{xunit=7,yunit=7,Alpha=40,Beta=0} \scriptsize
\pstThreeDLine(-0.2591,-0.4498,-0.1778)(-0.2765,-0.3585,-0.1471) 
\pstThreeDLine(-0.2591,-0.4498,-0.1778)(-0.2965,-0.2159,-0.1210) 
\pstThreeDLine(-0.2591,-0.4498,-0.1778)(-0.2411,-0.3198,-0.1859) 
\pstThreeDLine(-0.2591,-0.4498,-0.1778)(-0.2339,-0.3289,-0.1857) 
\pstThreeDLine(-0.2591,-0.4498,-0.1778)(-0.2311,-0.4492,-0.0349) 
\pstThreeDLine(-0.2591,-0.4498,-0.1778)(-0.2190,-0.1776,-0.2638) 
\pstThreeDLine(-0.2591,-0.4498,-0.1778)(-0.4254,-0.0865,0.0298) 
\pstThreeDLine(-0.2765,-0.3585,-0.1471)(-0.2965,-0.2159,-0.1210) 
\pstThreeDLine(-0.2765,-0.3585,-0.1471)(-0.2411,-0.3198,-0.1859) 
\pstThreeDLine(-0.2765,-0.3585,-0.1471)(-0.2339,-0.3289,-0.1857) 
\pstThreeDLine(-0.2765,-0.3585,-0.1471)(-0.2311,-0.4492,-0.0349) 
\pstThreeDLine(-0.2765,-0.3585,-0.1471)(-0.4254,-0.0865,0.0298) 
\pstThreeDLine(-0.2765,-0.3585,-0.1471)(-0.4286,0.0333,-0.0189) 
\pstThreeDLine(-0.2765,-0.3585,-0.1471)(-0.3141,-0.0763,0.0303) 
\pstThreeDLine(-0.2965,-0.2159,-0.1210)(-0.2411,-0.3198,-0.1859) 
\pstThreeDLine(-0.2965,-0.2159,-0.1210)(-0.2339,-0.3289,-0.1857) 
\pstThreeDLine(-0.2965,-0.2159,-0.1210)(-0.2311,-0.4492,-0.0349) 
\pstThreeDLine(-0.2965,-0.2159,-0.1210)(-0.4256,-0.0475,0.0253) 
\pstThreeDLine(-0.2965,-0.2159,-0.1210)(-0.4405,-0.0111,0.0425) 
\pstThreeDLine(-0.2965,-0.2159,-0.1210)(-0.4254,-0.0865,0.0298) 
\pstThreeDLine(-0.2965,-0.2159,-0.1210)(-0.4286,0.0333,-0.0189) 
\pstThreeDLine(-0.2965,-0.2159,-0.1210)(-0.3855,0.0439,0.0641) 
\pstThreeDLine(-0.2411,-0.3198,-0.1859)(-0.2339,-0.3289,-0.1857) 
\pstThreeDLine(-0.2411,-0.3198,-0.1859)(-0.2311,-0.4492,-0.0349) 
\pstThreeDLine(-0.2339,-0.3289,-0.1857)(-0.2311,-0.4492,-0.0349) 
\pstThreeDLine(-0.2339,-0.3289,-0.1857)(-0.2190,-0.1776,-0.2638) 
\pstThreeDLine(-0.2339,-0.3289,-0.1857)(-0.4254,-0.0865,0.0298) 
\pstThreeDLine(-0.2311,-0.4492,-0.0349)(-0.4254,-0.0865,0.0298) 
\pstThreeDLine(-0.2311,-0.4492,-0.0349)(-0.3927,-0.0353,0.1513) 
\pstThreeDLine(-0.2190,-0.1776,-0.2638)(-0.4256,-0.0475,0.0253) 
\pstThreeDLine(-0.2190,-0.1776,-0.2638)(-0.4286,0.0333,-0.0189) 
\pstThreeDLine(-0.1883,-0.0771,-0.0206)(-0.3622,-0.0730,0.1113) 
\pstThreeDLine(-0.4256,-0.0475,0.0253)(-0.1587,0.1543,0.1406) 
\pstThreeDLine(-0.4256,-0.0475,0.0253)(-0.4405,-0.0111,0.0425) 
\pstThreeDLine(-0.4256,-0.0475,0.0253)(-0.4254,-0.0865,0.0298) 
\pstThreeDLine(-0.4256,-0.0475,0.0253)(-0.2335,-0.1044,0.1696) 
\pstThreeDLine(-0.4256,-0.0475,0.0253)(-0.3999,-0.0837,0.1406) 
\pstThreeDLine(-0.4256,-0.0475,0.0253)(-0.4286,0.0333,-0.0189) 
\pstThreeDLine(-0.4256,-0.0475,0.0253)(-0.3855,0.0439,0.0641) 
\pstThreeDLine(-0.4256,-0.0475,0.0253)(-0.2527,0.1271,0.0834) 
\pstThreeDLine(-0.4256,-0.0475,0.0253)(-0.3508,0.1151,0.0570) 
\pstThreeDLine(-0.4256,-0.0475,0.0253)(-0.3047,0.0054,0.1790) 
\pstThreeDLine(-0.4256,-0.0475,0.0253)(-0.3927,-0.0353,0.1513) 
\pstThreeDLine(-0.4256,-0.0475,0.0253)(-0.3622,-0.0730,0.1113) 
\pstThreeDLine(-0.4256,-0.0475,0.0253)(-0.2360,0.0914,0.1646) 
\pstThreeDLine(-0.4256,-0.0475,0.0253)(-0.2335,0.0939,0.0524) 
\pstThreeDLine(-0.4256,-0.0475,0.0253)(-0.3141,-0.0763,0.0303) 
\pstThreeDLine(-0.4256,-0.0475,0.0253)(-0.2270,0.1120,-0.0735) 
\pstThreeDLine(-0.4405,-0.0111,0.0425)(-0.4254,-0.0865,0.0298) 
\pstThreeDLine(-0.4405,-0.0111,0.0425)(-0.2335,-0.1044,0.1696) 
\pstThreeDLine(-0.4405,-0.0111,0.0425)(-0.3999,-0.0837,0.1406) 
\pstThreeDLine(-0.4405,-0.0111,0.0425)(-0.4286,0.0333,-0.0189) 
\pstThreeDLine(-0.4405,-0.0111,0.0425)(-0.3855,0.0439,0.0641) 
\pstThreeDLine(-0.4405,-0.0111,0.0425)(-0.2527,0.1271,0.0834) 
\pstThreeDLine(-0.4405,-0.0111,0.0425)(-0.3508,0.1151,0.0570) 
\pstThreeDLine(-0.4405,-0.0111,0.0425)(-0.3047,0.0054,0.1790) 
\pstThreeDLine(-0.4405,-0.0111,0.0425)(-0.3927,-0.0353,0.1513) 
\pstThreeDLine(-0.4405,-0.0111,0.0425)(-0.3622,-0.0730,0.1113) 
\pstThreeDLine(-0.4405,-0.0111,0.0425)(-0.2360,0.0914,0.1646) 
\pstThreeDLine(-0.4405,-0.0111,0.0425)(-0.2335,0.0939,0.0524) 
\pstThreeDLine(-0.4405,-0.0111,0.0425)(-0.3141,-0.0763,0.0303) 
\pstThreeDLine(-0.4405,-0.0111,0.0425)(-0.1730,0.2247,-0.1478) 
\pstThreeDLine(-0.4254,-0.0865,0.0298)(-0.2335,-0.1044,0.1696) 
\pstThreeDLine(-0.4254,-0.0865,0.0298)(-0.3999,-0.0837,0.1406) 
\pstThreeDLine(-0.4254,-0.0865,0.0298)(-0.4286,0.0333,-0.0189) 
\pstThreeDLine(-0.4254,-0.0865,0.0298)(-0.3855,0.0439,0.0641) 
\pstThreeDLine(-0.4254,-0.0865,0.0298)(-0.2527,0.1271,0.0834) 
\pstThreeDLine(-0.4254,-0.0865,0.0298)(-0.3508,0.1151,0.0570) 
\pstThreeDLine(-0.4254,-0.0865,0.0298)(-0.3047,0.0054,0.1790) 
\pstThreeDLine(-0.4254,-0.0865,0.0298)(-0.3927,-0.0353,0.1513) 
\pstThreeDLine(-0.4254,-0.0865,0.0298)(-0.3622,-0.0730,0.1113) 
\pstThreeDLine(-0.4254,-0.0865,0.0298)(-0.2360,0.0914,0.1646) 
\pstThreeDLine(-0.4254,-0.0865,0.0298)(-0.2335,0.0939,0.0524) 
\pstThreeDLine(-0.4254,-0.0865,0.0298)(-0.3141,-0.0763,0.0303) 
\pstThreeDLine(-0.2335,-0.1044,0.1696)(-0.3999,-0.0837,0.1406) 
\pstThreeDLine(-0.2335,-0.1044,0.1696)(-0.4286,0.0333,-0.0189) 
\pstThreeDLine(-0.2335,-0.1044,0.1696)(-0.3047,0.0054,0.1790) 
\pstThreeDLine(-0.2335,-0.1044,0.1696)(-0.3927,-0.0353,0.1513) 
\pstThreeDLine(-0.3999,-0.0837,0.1406)(-0.4286,0.0333,-0.0189) 
\pstThreeDLine(-0.3999,-0.0837,0.1406)(-0.3855,0.0439,0.0641) 
\pstThreeDLine(-0.3999,-0.0837,0.1406)(-0.2527,0.1271,0.0834) 
\pstThreeDLine(-0.3999,-0.0837,0.1406)(-0.3508,0.1151,0.0570) 
\pstThreeDLine(-0.3999,-0.0837,0.1406)(-0.3047,0.0054,0.1790) 
\pstThreeDLine(-0.3999,-0.0837,0.1406)(-0.3999,-0.0837,0.1406) 
\pstThreeDLine(-0.3999,-0.0837,0.1406)(-0.3622,-0.0730,0.1113) 
\pstThreeDLine(-0.3999,-0.0837,0.1406)(-0.3141,-0.0763,0.0303) 
\pstThreeDLine(-0.4286,0.0333,-0.0189)(-0.3855,0.0439,0.0641) 
\pstThreeDLine(-0.4286,0.0333,-0.0189)(-0.2527,0.1271,0.0834) 
\pstThreeDLine(-0.4286,0.0333,-0.0189)(-0.3508,0.1151,0.0570) 
\pstThreeDLine(-0.4286,0.0333,-0.0189)(-0.3047,0.0054,0.1790) 
\pstThreeDLine(-0.4286,0.0333,-0.0189)(-0.3927,-0.0353,0.1513) 
\pstThreeDLine(-0.4286,0.0333,-0.0189)(-0.3622,-0.0730,0.1113) 
\pstThreeDLine(-0.4286,0.0333,-0.0189)(-0.2360,0.0914,0.1646) 
\pstThreeDLine(-0.4286,0.0333,-0.0189)(-0.2335,0.0939,0.0524) 
\pstThreeDLine(-0.4286,0.0333,-0.0189)(-0.3141,-0.0763,0.0303) 
\pstThreeDLine(-0.4286,0.0333,-0.0189)(-0.1922,0.3397,-0.1574) 
\pstThreeDLine(-0.4286,0.0333,-0.0189)(-0.1730,0.2247,-0.1478) 
\pstThreeDLine(-0.4286,0.0333,-0.0189)(-0.1219,0.2300,-0.1623) 
\pstThreeDLine(-0.4286,0.0333,-0.0189)(-0.2270,0.1120,-0.0735) 
\pstThreeDLine(-0.3855,0.0439,0.0641)(-0.2527,0.1271,0.0834) 
\pstThreeDLine(-0.3855,0.0439,0.0641)(-0.3508,0.1151,0.0570) 
\pstThreeDLine(-0.3855,0.0439,0.0641)(-0.3047,0.0054,0.1790) 
\pstThreeDLine(-0.3855,0.0439,0.0641)(-0.3927,-0.0353,0.1513) 
\pstThreeDLine(-0.3855,0.0439,0.0641)(-0.3622,-0.0730,0.1113) 
\pstThreeDLine(-0.3855,0.0439,0.0641)(-0.2360,0.0914,0.1646) 
\pstThreeDLine(-0.3855,0.0439,0.0641)(-0.3141,-0.0763,0.0303) 
\pstThreeDLine(-0.3855,0.0439,0.0641)(-0.2270,0.1120,-0.0735) 
\pstThreeDLine(-0.2527,0.1271,0.0834)(-0.3508,0.1151,0.0570) 
\pstThreeDLine(-0.2527,0.1271,0.0834)(-0.3047,0.0054,0.1790) 
\pstThreeDLine(-0.2527,0.1271,0.0834)(-0.3999,-0.0837,0.1406) 
\pstThreeDLine(-0.2527,0.1271,0.0834)(-0.3622,-0.0730,0.1113) 
\pstThreeDLine(-0.3508,0.1151,0.0570)(-0.3047,0.0054,0.1790) 
\pstThreeDLine(-0.3508,0.1151,0.0570)(-0.3927,-0.0353,0.1513) 
\pstThreeDLine(-0.3508,0.1151,0.0570)(-0.3622,-0.0730,0.1113) 
\pstThreeDLine(-0.3508,0.1151,0.0570)(-0.2360,0.0914,0.1646) 
\pstThreeDLine(-0.3508,0.1151,0.0570)(-0.3141,-0.0763,0.0303) 
\pstThreeDLine(-0.3508,0.1151,0.0570)(-0.1922,0.3397,-0.1574) 
\pstThreeDLine(-0.3508,0.1151,0.0570)(-0.2270,0.1120,-0.0735) 
\pstThreeDLine(-0.3047,0.0054,0.1790)(-0.3927,-0.0353,0.1513) 
\pstThreeDLine(-0.3047,0.0054,0.1790)(-0.3622,-0.0730,0.1113) 
\pstThreeDLine(-0.3047,0.0054,0.1790)(-0.2360,0.0914,0.1646) 
\pstThreeDLine(-0.3047,0.0054,0.1790)(-0.2335,0.0939,0.0524) 
\pstThreeDLine(-0.3047,0.0054,0.1790)(-0.1593,0.2363,0.0988) 
\pstThreeDLine(-0.3047,0.0054,0.1790)(-0.3141,-0.0763,0.0303) 
\pstThreeDLine(-0.3047,0.0054,0.1790)(-0.2270,0.1120,-0.0735) 
\pstThreeDLine(-0.3927,-0.0353,0.1513)(-0.3622,-0.0730,0.1113) 
\pstThreeDLine(-0.3927,-0.0353,0.1513)(-0.2360,0.0914,0.1646) 
\pstThreeDLine(-0.3927,-0.0353,0.1513)(-0.2335,0.0939,0.0524) 
\pstThreeDLine(-0.3622,-0.0730,0.1113)(-0.3141,-0.0763,0.0303) 
\pstThreeDLine(-0.2360,0.0914,0.1646)(-0.2335,0.0939,0.0524) 
\pstThreeDLine(-0.2360,0.0914,0.1646)(-0.1593,0.2363,0.0988) 
\pstThreeDLine(-0.2335,0.0939,0.0524)(-0.1593,0.2363,0.0988) 
\pstThreeDLine(-0.2335,0.0939,0.0524)(-0.2270,0.1120,-0.0735) 
\pstThreeDLine(-0.1406,0.3275,0.0449)(-0.1922,0.3397,-0.1574) 
\pstThreeDLine(-0.1593,0.2363,0.0988)(-0.2270,0.1120,-0.0735) 
\pstThreeDLine(-0.0781,0.3082,-0.0955)(-0.1922,0.3397,-0.1574) 
\pstThreeDLine(-0.0781,0.3082,-0.0955)(-0.1437,0.3074,-0.1581) 
\pstThreeDLine(-0.0781,0.3082,-0.0955)(-0.1219,0.2300,-0.1623) 
\pstThreeDLine(-0.1922,0.3397,-0.1574)(-0.1437,0.3074,-0.1581) 
\pstThreeDLine(-0.1922,0.3397,-0.1574)(-0.1730,0.2247,-0.1478) 
\pstThreeDLine(-0.1922,0.3397,-0.1574)(-0.1219,0.2300,-0.1623) 
\pstThreeDDot[linecolor=orange,linewidth=1.2pt](-0.2591,-0.4498,-0.1778) 
\pstThreeDDot[linecolor=orange,linewidth=1.2pt](-0.2765,-0.3585,-0.1471) 
\pstThreeDDot[linecolor=orange,linewidth=1.2pt](-0.2965,-0.2159,-0.1210) 
\pstThreeDDot[linecolor=orange,linewidth=1.2pt](-0.2411,-0.3198,-0.1859) 
\pstThreeDDot[linecolor=green,linewidth=1.2pt](-0.2339,-0.3289,-0.1857) 
\pstThreeDDot[linecolor=green,linewidth=1.2pt](-0.2311,-0.4492,-0.0349) 
\pstThreeDDot[linecolor=green,linewidth=1.2pt](-0.2190,-0.1776,-0.2638) 
\pstThreeDDot[linecolor=green,linewidth=1.2pt](-0.1883,-0.0771,-0.0206) 
\pstThreeDDot[linecolor=blue,linewidth=1.2pt](-0.4256,-0.0475,0.0253) 
\pstThreeDDot[linecolor=blue,linewidth=1.2pt](-0.1587,0.1543,0.1406) 
\pstThreeDDot[linecolor=blue,linewidth=1.2pt](-0.4405,-0.0111,0.0425) 
\pstThreeDDot[linecolor=blue,linewidth=1.2pt](-0.4254,-0.0865,0.0298) 
\pstThreeDDot[linecolor=blue,linewidth=1.2pt](-0.2335,-0.1044,0.1696) 
\pstThreeDDot[linecolor=blue,linewidth=1.2pt](-0.3999,-0.0837,0.1406) 
\pstThreeDDot[linecolor=blue,linewidth=1.2pt](-0.4286,0.0333,-0.0189) 
\pstThreeDDot[linecolor=blue,linewidth=1.2pt](-0.3855,0.0439,0.0641) 
\pstThreeDDot[linecolor=blue,linewidth=1.2pt](-0.2527,0.1271,0.0834) 
\pstThreeDDot[linecolor=blue,linewidth=1.2pt](-0.3508,0.1151,0.0570) 
\pstThreeDDot[linecolor=blue,linewidth=1.2pt](-0.3047,0.0054,0.1790) 
\pstThreeDDot[linecolor=blue,linewidth=1.2pt](-0.3927,-0.0353,0.1513) 
\pstThreeDDot[linecolor=blue,linewidth=1.2pt](-0.3622,-0.0730,0.1113) 
\pstThreeDDot[linecolor=blue,linewidth=1.2pt](-0.2360,0.0914,0.1646) 
\pstThreeDDot[linecolor=blue,linewidth=1.2pt](-0.2335,0.0939,0.0524) 
\pstThreeDDot[linecolor=blue,linewidth=1.2pt](-0.1406,0.3275,0.0449) 
\pstThreeDDot[linecolor=bluish,linewidth=1.2pt](-0.1593,0.2363,0.0988) 
\pstThreeDDot[linecolor=brown,linewidth=1.2pt](-0.3141,-0.0763,0.0303) 
\pstThreeDDot[linecolor=red,linewidth=1.2pt](-0.0781,0.3082,-0.0955) 
\pstThreeDDot[linecolor=red,linewidth=1.2pt](-0.1922,0.3397,-0.1574) 
\pstThreeDDot[linecolor=red,linewidth=1.2pt](-0.1437,0.3074,-0.1581) 
\pstThreeDDot[linecolor=red,linewidth=1.2pt](-0.1730,0.2247,-0.1478) 
\pstThreeDDot[linecolor=black,linewidth=1.2pt](-0.1219,0.2300,-0.1623) 
\pstThreeDDot[linecolor=magenta,linewidth=1.2pt](-0.2270,0.1120,-0.0735) 
\pstThreeDPut(-0.2591,-0.4498,-0.2078){S\&P}
\pstThreeDPut(-0.2765,-0.3585,-0.1771){Nasd}
\pstThreeDPut(-0.2965,-0.2159,-0.1510){Cana}
\pstThreeDPut(-0.2411,-0.3198,-0.2159){Mexi}
\pstThreeDPut(-0.2339,-0.3289,-0.2157){Braz}
\pstThreeDPut(-0.2311,-0.4492,-0.0649){Arge}
\pstThreeDPut(-0.2190,-0.1776,-0.2938){Chil}
\pstThreeDPut(-0.1883,-0.0771,-0.0506){Peru}
\pstThreeDPut(-0.4256,-0.0475,0.0553){UK}
\pstThreeDPut(-0.1587,0.1543,0.1706){Irel}
\pstThreeDPut(-0.4405,-0.0111,0.0725){Fran}
\pstThreeDPut(-0.4254,-0.0865,0.0598){Germ}
\pstThreeDPut(-0.2335,-0.1044,0.1996){Swit}
\pstThreeDPut(-0.3999,-0.0837,0.1706){Ital}
\pstThreeDPut(-0.4286,0.0333,-0.0489){Neth}
\pstThreeDPut(-0.3855,0.0439,0.0941){Swed}
\pstThreeDPut(-0.2527,0.1271,0.1134){Denm}
\pstThreeDPut(-0.3508,0.1151,0.0870){Finl}
\pstThreeDPut(-0.3047,0.0054,0.2090){Norw}
\pstThreeDPut(-0.3927,-0.0353,0.1813){Spai}
\pstThreeDPut(-0.3622,-0.0730,0.1413){Port}
\pstThreeDPut(-0.2360,0.0914,0.1946){CzRe}
\pstThreeDPut(-0.2335,0.0939,0.0824){Hung}
\pstThreeDPut(-0.1406,0.3275,0.0749){Pola}
\pstThreeDPut(-0.1593,0.2363,0.1288){Russ}
\pstThreeDPut(-0.3141,-0.0763,0.0603){Isra}
\pstThreeDPut(-0.0781,0.3082,-0.1255){Japa}
\pstThreeDPut(-0.1922,0.3397,-0.1874){HoKo}
\pstThreeDPut(-0.1437,0.3074,-0.1881){SoKo}
\pstThreeDPut(-0.1730,0.2247,-0.1778){Sing}
\pstThreeDPut(-0.1219,0.2300,-0.1923){Aust}
\pstThreeDPut(-0.2270,0.1120,-0.1035){SoAf}
\end{pspicture}

\vskip 1.2 cm

\noindent Fig. 9. Three dimensional view of the asset trees for the first semester of 2000, with threshold ranging from $T=0.3$ to $T=0.6$.

\vskip 0.3 cm

For $T=0.2$, the only connection is between S\&P and Nasdaq. For $T=0.3$, a second cluster, comprised of European indices, is formed, with France, Germany, Italy, Netherlands, Sweden, Finland, Spain, and Portugal. At $T=0.4$, the North American cluster is fully connected, with the addition of Canada and Mexico. The UK joins the European cluster, which also becomes denser, with the establishment of more connections between its members. For $T=0.5$, Brazil and Argentina join the American cluster, and also connect with one another. Canada connects with Germany, in the European cluster, which now grows even denser, and has Denmark added to it. At $T=0.6$, many more connections are formed, inside clusters and between clusters. Israel connects with both America and Europe, Hong Kong, Singapore, Australia and South Africa connect with Europe. Switzerland, Norway, the Czech Republic, Hungary, and Russia also join the European cluster. A Pacific Asian cluster is formed, comprised of Japan, Hong Kong, South Korea, and Singapore, strongly connected with Australia, and lightly connected with Europe. For larger thresholds, noise starts to dominate, and many connections are made, some of them apparently at random. At $T=1.3$, all indices are connected, with the last connection being between Jamaica and Lebanon.

\subsubsection{Second semester, 2000}

Figure 10 shows the three dimensional view of the asset trees for the second semester of 2000, with threshold ranging from $T=0.3$ to $T=0.6$.

Once more, at $T=0.2$, a single connection is formed between S\&P and Nasdaq. For $T=0.3$, we also have an European cluster comprised of the UK, France, Germany, Italy, Netherlands, Sweden, Finland, and Spain. At $T=0.4$, the North American cluster, now comprised of S\&P, Nasdaq, Canada, and Mexico, joins the European cluster via Germany. The European cluster has now only the addition of Portugal, but it becomes denser, with more connections formed among its members. At $T=0.5$, Argentina joins the American cluster, which makes stronger connections with Europe. Switzerland, Norway, and Russia join the European cluster. At $T=0.6$, a myriad of connections is made, with Brazil and Chile joining the American cluster, and Ireland, Austria, Belgium, the Czech Republic, Hungary, and Poland joining the European Cluster. The connections between both clusters grow stronger, and Israel, Hong Kong, South Korea, Singapore, Australia, and South Africa establish connections with some or both clusters. There is also a new cluster, of Pacific Asian indices, comprised of Japan, Hong Kong, South Korea, and Singapore, with connections to Australia and to South Africa. At larger thresholds, new connections are made, subject to a lot of random noise, and the last connection is formed at $T=1.4$, between Ireland and Lebanon.

\vskip 0.6 cm

\begin{pspicture}(-4,-1.1)(1,3.5)
\rput(-3.4,2.9){T=0.3}
\psline(-4.3,-2.1)(4.1,-2.1)(4.1,3.3)(-4.3,3.3)(-4.3,-2.1)
\psset{xunit=7,yunit=7,Alpha=40,Beta=30} \scriptsize
\pstThreeDLine(-0.3293,0.3307,-0.1048)(-0.2767,0.3087,-0.0883) 
\pstThreeDLine(-0.4497,-0.0240,-0.0150)(-0.4143,0.1411,-0.0345) 
\pstThreeDLine(-0.4143,0.1411,-0.0345)(-0.4211,0.1692,-0.0737) 
\pstThreeDLine(-0.4143,0.1411,-0.0345)(-0.4098,0.1700,-0.0147) 
\pstThreeDLine(-0.4143,0.1411,-0.0345)(-0.4483,-0.0002,0.0071) 
\pstThreeDLine(-0.4143,0.1411,-0.0345)(-0.4327,0.1368,0.0159) 
\pstThreeDLine(-0.4143,0.1411,-0.0345)(-0.3578,0.1557,0.0036) 
\pstThreeDLine(-0.4211,0.1692,-0.0737)(-0.4098,0.1700,-0.0147) 
\pstThreeDLine(-0.4211,0.1692,-0.0737)(-0.4483,-0.0002,0.0071) 
\pstThreeDLine(-0.4211,0.1692,-0.0737)(-0.4327,0.1368,0.0159) 
\pstThreeDLine(-0.4098,0.1700,-0.0147)(-0.4483,-0.0002,0.0071) 
\pstThreeDLine(-0.4327,0.1368,0.0159)(-0.3743,0.0417,0.0536) 
\pstThreeDDot[linecolor=orange,linewidth=1.2pt](-0.3293,0.3307,-0.1048) 
\pstThreeDDot[linecolor=orange,linewidth=1.2pt](-0.2767,0.3087,-0.0883) 
\pstThreeDDot[linecolor=blue,linewidth=1.2pt](-0.4497,-0.0240,-0.0150) 
\pstThreeDDot[linecolor=blue,linewidth=1.2pt](-0.4143,0.1411,-0.0345) 
\pstThreeDDot[linecolor=blue,linewidth=1.2pt](-0.4211,0.1692,-0.0737) 
\pstThreeDDot[linecolor=blue,linewidth=1.2pt](-0.4098,0.1700,-0.0147) 
\pstThreeDDot[linecolor=blue,linewidth=1.2pt](-0.4483,-0.0002,0.0071) 
\pstThreeDDot[linecolor=blue,linewidth=1.2pt](-0.4327,0.1368,0.0159) 
\pstThreeDDot[linecolor=blue,linewidth=1.2pt](-0.3743,0.0417,0.0536) 
\pstThreeDDot[linecolor=blue,linewidth=1.2pt](-0.3578,0.1557,0.0036) 
\pstThreeDPut(-0.3293,0.3307,-0.1348){S\&P}
\pstThreeDPut(-0.2767,0.3087,-0.1183){Nasd}
\pstThreeDPut(-0.4497,-0.0240,-0.0450){UK}
\pstThreeDPut(-0.4143,0.1411,-0.0645){Fran}
\pstThreeDPut(-0.4211,0.1692,-0.1037){Germ}
\pstThreeDPut(-0.4098,0.1700,-0.0447){Ital}
\pstThreeDPut(-0.4483,-0.0002,0.0371){Neth}
\pstThreeDPut(-0.4327,0.1368,0.0459){Swed}
\pstThreeDPut(-0.3743,0.0417,0.0836){Finl}
\pstThreeDPut(-0.3578,0.1557,0.0336){Spai}
\end{pspicture}
\begin{pspicture}(-8,-1.1)(1,3.5)
\rput(-3.4,2.9){T=0.4}
\psline(-4.3,-2.1)(4.1,-2.1)(4.1,3.3)(-4.3,3.3)(-4.3,-2.1)
\psset{xunit=7,yunit=7,Alpha=40,Beta=30} \scriptsize
\pstThreeDLine(-0.3293,0.3307,-0.1048)(-0.2767,0.3087,-0.0883) 
\pstThreeDLine(-0.3293,0.3307,-0.1048)(-0.3053,0.1377,-0.1069) 
\pstThreeDLine(-0.3293,0.3307,-0.1048)(-0.2434,0.2850,0.1640) 
\pstThreeDLine(-0.3293,0.3307,-0.1048)(-0.4211,0.1692,-0.0737) 
\pstThreeDLine(-0.2767,0.3087,-0.0883)(-0.3053,0.1377,-0.1069) 
\pstThreeDLine(-0.2767,0.3087,-0.0883)(-0.2434,0.2850,0.1640) 
\pstThreeDLine(-0.4497,-0.0240,-0.0150)(-0.4143,0.1411,-0.0345) 
\pstThreeDLine(-0.4497,-0.0240,-0.0150)(-0.4211,0.1692,-0.0737) 
\pstThreeDLine(-0.4497,-0.0240,-0.0150)(-0.4098,0.1700,-0.0147) 
\pstThreeDLine(-0.4497,-0.0240,-0.0150)(-0.4483,-0.0002,0.0071) 
\pstThreeDLine(-0.4497,-0.0240,-0.0150)(-0.4327,0.1368,0.0159) 
\pstThreeDLine(-0.4497,-0.0240,-0.0150)(-0.3743,0.0417,0.0536) 
\pstThreeDLine(-0.4143,0.1411,-0.0345)(-0.4211,0.1692,-0.0737) 
\pstThreeDLine(-0.4143,0.1411,-0.0345)(-0.4098,0.1700,-0.0147) 
\pstThreeDLine(-0.4143,0.1411,-0.0345)(-0.4483,-0.0002,0.0071) 
\pstThreeDLine(-0.4143,0.1411,-0.0345)(-0.4327,0.1368,0.0159) 
\pstThreeDLine(-0.4143,0.1411,-0.0345)(-0.3743,0.0417,0.0536) 
\pstThreeDLine(-0.4143,0.1411,-0.0345)(-0.3578,0.1557,0.0036) 
\pstThreeDLine(-0.4143,0.1411,-0.0345)(-0.3423,0.1394,-0.0176) 
\pstThreeDLine(-0.4211,0.1692,-0.0737)(-0.4098,0.1700,-0.0147) 
\pstThreeDLine(-0.4211,0.1692,-0.0737)(-0.4483,-0.0002,0.0071) 
\pstThreeDLine(-0.4211,0.1692,-0.0737)(-0.4327,0.1368,0.0159) 
\pstThreeDLine(-0.4211,0.1692,-0.0737)(-0.3743,0.0417,0.0536) 
\pstThreeDLine(-0.4211,0.1692,-0.0737)(-0.3578,0.1557,0.0036) 
\pstThreeDLine(-0.4098,0.1700,-0.0147)(-0.4483,-0.0002,0.0071) 
\pstThreeDLine(-0.4098,0.1700,-0.0147)(-0.4327,0.1368,0.0159) 
\pstThreeDLine(-0.4098,0.1700,-0.0147)(-0.3578,0.1557,0.0036) 
\pstThreeDLine(-0.4483,-0.0002,0.0071)(-0.4327,0.1368,0.0159) 
\pstThreeDLine(-0.4483,-0.0002,0.0071)(-0.3743,0.0417,0.0536) 
\pstThreeDLine(-0.4483,-0.0002,0.0071)(-0.3578,0.1557,0.0036) 
\pstThreeDLine(-0.4327,0.1368,0.0159)(-0.3743,0.0417,0.0536) 
\pstThreeDLine(-0.4327,0.1368,0.0159)(-0.3578,0.1557,0.0036) 
\pstThreeDLine(-0.4327,0.1368,0.0159)(-0.3423,0.1394,-0.0176) 
\pstThreeDLine(-0.3743,0.0417,0.0536)(-0.3423,0.1394,-0.0176) 
\pstThreeDLine(-0.3578,0.1557,0.0036)(-0.3423,0.1394,-0.0176) 
\pstThreeDDot[linecolor=orange,linewidth=1.2pt](-0.3293,0.3307,-0.1048) 
\pstThreeDDot[linecolor=orange,linewidth=1.2pt](-0.2767,0.3087,-0.0883) 
\pstThreeDDot[linecolor=orange,linewidth=1.2pt](-0.3053,0.1377,-0.1069) 
\pstThreeDDot[linecolor=orange,linewidth=1.2pt](-0.2434,0.2850,0.1640) 
\pstThreeDDot[linecolor=blue,linewidth=1.2pt](-0.4497,-0.0240,-0.0150) 
\pstThreeDDot[linecolor=blue,linewidth=1.2pt](-0.4143,0.1411,-0.0345) 
\pstThreeDDot[linecolor=blue,linewidth=1.2pt](-0.4211,0.1692,-0.0737) 
\pstThreeDDot[linecolor=blue,linewidth=1.2pt](-0.2692,0.1314,0.0996) 
\pstThreeDDot[linecolor=blue,linewidth=1.2pt](-0.4098,0.1700,-0.0147) 
\pstThreeDDot[linecolor=blue,linewidth=1.2pt](-0.4483,-0.0002,0.0071) 
\pstThreeDDot[linecolor=blue,linewidth=1.2pt](-0.4327,0.1368,0.0159) 
\pstThreeDDot[linecolor=blue,linewidth=1.2pt](-0.3743,0.0417,0.0536) 
\pstThreeDDot[linecolor=blue,linewidth=1.2pt](-0.3578,0.1557,0.0036) 
\pstThreeDDot[linecolor=blue,linewidth=1.2pt](-0.3423,0.1394,-0.0176) 
\pstThreeDPut(-0.3293,0.3307,-0.1348){S\&P}
\pstThreeDPut(-0.2767,0.3087,-0.1183){Nasd}
\pstThreeDPut(-0.3053,0.1377,-0.1369){Cana}
\pstThreeDPut(-0.2434,0.2850,0.1940){Mexi}
\pstThreeDPut(-0.4497,-0.0240,-0.0450){UK}
\pstThreeDPut(-0.4143,0.1411,-0.0645){Fran}
\pstThreeDPut(-0.4211,0.1692,-0.1037){Germ}
\pstThreeDPut(-0.2692,0.1314,0.1296){Swit}
\pstThreeDPut(-0.4098,0.1700,-0.0447){Ital}
\pstThreeDPut(-0.4483,-0.0002,0.0371){Neth}
\pstThreeDPut(-0.4327,0.1368,0.0459){Swed}
\pstThreeDPut(-0.3743,0.0417,0.0836){Finl}
\pstThreeDPut(-0.3578,0.1557,0.0336){Spai}
\pstThreeDPut(-0.3423,0.1394,-0.0476){Port}
\end{pspicture}

\vskip 1.5 cm

\begin{pspicture}(-4,-1.1)(1,3.5)
\rput(-3.4,2.9){T=0.5}
\psline(-4.3,-2.1)(4.1,-2.1)(4.1,3.3)(-4.3,3.3)(-4.3,-2.1)
\psset{xunit=7,yunit=7,Alpha=40,Beta=30} \scriptsize
\pstThreeDLine(-0.3293,0.3307,-0.1048)(-0.2767,0.3087,-0.0883) 
\pstThreeDLine(-0.3293,0.3307,-0.1048)(-0.3053,0.1377,-0.1069) 
\pstThreeDLine(-0.3293,0.3307,-0.1048)(-0.2434,0.2850,0.1640) 
\pstThreeDLine(-0.3293,0.3307,-0.1048)(-0.2008,0.5136,0.0513) 
\pstThreeDLine(-0.3293,0.3307,-0.1048)(-0.4143,0.1411,-0.0345) 
\pstThreeDLine(-0.3293,0.3307,-0.1048)(-0.4211,0.1692,-0.0737) 
\pstThreeDLine(-0.3293,0.3307,-0.1048)(-0.4098,0.1700,-0.0147) 
\pstThreeDLine(-0.3293,0.3307,-0.1048)(-0.4483,-0.0002,0.0071) 
\pstThreeDLine(-0.2767,0.3087,-0.0883)(-0.3053,0.1377,-0.1069) 
\pstThreeDLine(-0.2767,0.3087,-0.0883)(-0.2434,0.2850,0.1640) 
\pstThreeDLine(-0.2767,0.3087,-0.0883)(-0.2008,0.5136,0.0513) 
\pstThreeDLine(-0.2767,0.3087,-0.0883)(-0.4211,0.1692,-0.0737) 
\pstThreeDLine(-0.2767,0.3087,-0.0883)(-0.4098,0.1700,-0.0147) 
\pstThreeDLine(-0.3053,0.1377,-0.1069)(-0.4211,0.1692,-0.0737) 
\pstThreeDLine(-0.3053,0.1377,-0.1069)(-0.4483,-0.0002,0.0071) 
\pstThreeDLine(-0.3053,0.1377,-0.1069)(-0.4327,0.1368,0.0159) 
\pstThreeDLine(-0.3053,0.1377,-0.1069)(-0.3743,0.0417,0.0536) 
\pstThreeDLine(-0.4497,-0.0240,-0.0150)(-0.4143,0.1411,-0.0345) 
\pstThreeDLine(-0.4497,-0.0240,-0.0150)(-0.4211,0.1692,-0.0737) 
\pstThreeDLine(-0.4497,-0.0240,-0.0150)(-0.4098,0.1700,-0.0147) 
\pstThreeDLine(-0.4497,-0.0240,-0.0150)(-0.4483,-0.0002,0.0071) 
\pstThreeDLine(-0.4497,-0.0240,-0.0150)(-0.4327,0.1368,0.0159) 
\pstThreeDLine(-0.4497,-0.0240,-0.0150)(-0.3743,0.0417,0.0536) 
\pstThreeDLine(-0.4497,-0.0240,-0.0150)(-0.3331,-0.0481,-0.1432) 
\pstThreeDLine(-0.4497,-0.0240,-0.0150)(-0.3578,0.1557,0.0036) 
\pstThreeDLine(-0.4497,-0.0240,-0.0150)(-0.3423,0.1394,-0.0176) 
\pstThreeDLine(-0.4497,-0.0240,-0.0150)(-0.2380,-0.1616,-0.0303) 
\pstThreeDLine(-0.4143,0.1411,-0.0345)(-0.4211,0.1692,-0.0737) 
\pstThreeDLine(-0.4143,0.1411,-0.0345)(-0.4098,0.1700,-0.0147) 
\pstThreeDLine(-0.4143,0.1411,-0.0345)(-0.4483,-0.0002,0.0071) 
\pstThreeDLine(-0.4143,0.1411,-0.0345)(-0.4327,0.1368,0.0159) 
\pstThreeDLine(-0.4143,0.1411,-0.0345)(-0.3743,0.0417,0.0536) 
\pstThreeDLine(-0.4143,0.1411,-0.0345)(-0.3331,-0.0481,-0.1432) 
\pstThreeDLine(-0.4143,0.1411,-0.0345)(-0.3578,0.1557,0.0036) 
\pstThreeDLine(-0.4143,0.1411,-0.0345)(-0.3423,0.1394,-0.0176) 
\pstThreeDLine(-0.4211,0.1692,-0.0737)(-0.4098,0.1700,-0.0147) 
\pstThreeDLine(-0.4211,0.1692,-0.0737)(-0.4483,-0.0002,0.0071) 
\pstThreeDLine(-0.4211,0.1692,-0.0737)(-0.4327,0.1368,0.0159) 
\pstThreeDLine(-0.4211,0.1692,-0.0737)(-0.3743,0.0417,0.0536) 
\pstThreeDLine(-0.4211,0.1692,-0.0737)(-0.3331,-0.0481,-0.1432) 
\pstThreeDLine(-0.4211,0.1692,-0.0737)(-0.3578,0.1557,0.0036) 
\pstThreeDLine(-0.4211,0.1692,-0.0737)(-0.3423,0.1394,-0.0176) 
\pstThreeDLine(-0.2692,0.1314,0.0996)(-0.4098,0.1700,-0.0147) 
\pstThreeDLine(-0.2692,0.1314,0.0996)(-0.4483,-0.0002,0.0071) 
\pstThreeDLine(-0.2457,0.0421,-0.0528)(-0.4483,-0.0002,0.0071) 
\pstThreeDLine(-0.4098,0.1700,-0.0147)(-0.4483,-0.0002,0.0071) 
\pstThreeDLine(-0.4098,0.1700,-0.0147)(-0.4327,0.1368,0.0159) 
\pstThreeDLine(-0.4098,0.1700,-0.0147)(-0.3743,0.0417,0.0536) 
\pstThreeDLine(-0.4098,0.1700,-0.0147)(-0.3331,-0.0481,-0.1432) 
\pstThreeDLine(-0.4098,0.1700,-0.0147)(-0.3578,0.1557,0.0036) 
\pstThreeDLine(-0.4098,0.1700,-0.0147)(-0.3423,0.1394,-0.0176) 
\pstThreeDLine(-0.4483,-0.0002,0.0071)(-0.4327,0.1368,0.0159) 
\pstThreeDLine(-0.4483,-0.0002,0.0071)(-0.3743,0.0417,0.0536) 
\pstThreeDLine(-0.4483,-0.0002,0.0071)(-0.3331,-0.0481,-0.1432) 
\pstThreeDLine(-0.4483,-0.0002,0.0071)(-0.3578,0.1557,0.0036) 
\pstThreeDLine(-0.4483,-0.0002,0.0071)(-0.3423,0.1394,-0.0176) 
\pstThreeDLine(-0.4483,-0.0002,0.0071)(-0.2602,-0.0687,-0.0884) 
\pstThreeDLine(-0.4327,0.1368,0.0159)(-0.3743,0.0417,0.0536) 
\pstThreeDLine(-0.4327,0.1368,0.0159)(-0.3331,-0.0481,-0.1432) 
\pstThreeDLine(-0.4327,0.1368,0.0159)(-0.3578,0.1557,0.0036) 
\pstThreeDLine(-0.4327,0.1368,0.0159)(-0.3423,0.1394,-0.0176) 
\pstThreeDLine(-0.3743,0.0417,0.0536)(-0.3423,0.1394,-0.0176) 
\pstThreeDLine(-0.3743,0.0417,0.0536)(-0.3331,-0.0481,-0.1432) 
\pstThreeDLine(-0.3743,0.0417,0.0536)(-0.3578,0.1557,0.0036) 
\pstThreeDLine(-0.3331,-0.0481,-0.1432)(-0.3578,0.1557,0.0036) 
\pstThreeDLine(-0.3331,-0.0481,-0.1432)(-0.2441,-0.3293,-0.0370) 
\pstThreeDLine(-0.3578,0.1557,0.0036)(-0.3423,0.1394,-0.0176) 
\pstThreeDDot[linecolor=orange,linewidth=1.2pt](-0.3293,0.3307,-0.1048) 
\pstThreeDDot[linecolor=orange,linewidth=1.2pt](-0.2767,0.3087,-0.0883) 
\pstThreeDDot[linecolor=orange,linewidth=1.2pt](-0.3053,0.1377,-0.1069) 
\pstThreeDDot[linecolor=orange,linewidth=1.2pt](-0.2434,0.2850,0.1640) 
\pstThreeDDot[linecolor=green,linewidth=1.2pt](-0.2008,0.5136,0.0513) 
\pstThreeDDot[linecolor=blue,linewidth=1.2pt](-0.4497,-0.0240,-0.0150) 
\pstThreeDDot[linecolor=blue,linewidth=1.2pt](-0.4143,0.1411,-0.0345) 
\pstThreeDDot[linecolor=blue,linewidth=1.2pt](-0.4211,0.1692,-0.0737) 
\pstThreeDDot[linecolor=blue,linewidth=1.2pt](-0.2692,0.1314,0.0996) 
\pstThreeDDot[linecolor=blue,linewidth=1.2pt](-0.2457,0.0421,-0.0528) 
\pstThreeDDot[linecolor=blue,linewidth=1.2pt](-0.4098,0.1700,-0.0147) 
\pstThreeDDot[linecolor=blue,linewidth=1.2pt](-0.4483,-0.0002,0.0071) 
\pstThreeDDot[linecolor=blue,linewidth=1.2pt](-0.4327,0.1368,0.0159) 
\pstThreeDDot[linecolor=blue,linewidth=1.2pt](-0.3743,0.0417,0.0536) 
\pstThreeDDot[linecolor=blue,linewidth=1.2pt](-0.3331,-0.0481,-0.1432) 
\pstThreeDDot[linecolor=blue,linewidth=1.2pt](-0.3578,0.1557,0.0036) 
\pstThreeDDot[linecolor=blue,linewidth=1.2pt](-0.3423,0.1394,-0.0176) 
\pstThreeDDot[linecolor=bluish,linewidth=1.2pt](-0.2441,-0.3293,-0.0370) 
\pstThreeDDot[linecolor=red,linewidth=1.2pt](-0.2380,-0.1616,-0.0303) 
\pstThreeDDot[linecolor=magenta,linewidth=1.2pt](-0.2602,-0.0687,-0.0884) 
\pstThreeDPut(-0.3293,0.3307,-0.1348){S\&P}
\pstThreeDPut(-0.2767,0.3087,-0.1183){Nasd}
\pstThreeDPut(-0.3053,0.1377,-0.1369){Cana}
\pstThreeDPut(-0.2434,0.2850,0.1940){Mexi}
\pstThreeDPut(-0.2008,0.5136,0.0813){Arge}
\pstThreeDPut(-0.4497,-0.0240,-0.0450){UK}
\pstThreeDPut(-0.4143,0.1411,-0.0645){Fran}
\pstThreeDPut(-0.4211,0.1692,-0.1037){Germ}
\pstThreeDPut(-0.2692,0.1314,0.1296){Swit}
\pstThreeDPut(-0.2457,0.0421,-0.0828){Autr}
\pstThreeDPut(-0.4098,0.1700,-0.0447){Ital}
\pstThreeDPut(-0.4483,-0.0002,0.0371){Neth}
\pstThreeDPut(-0.4327,0.1368,0.0459){Swed}
\pstThreeDPut(-0.2500,-0.0978,0.0509){Denm}
\pstThreeDPut(-0.3743,0.0417,0.0836){Finl}
\pstThreeDPut(-0.3331,-0.0481,-0.1732){Norw}
\pstThreeDPut(-0.3578,0.1557,0.0336){Spai}
\pstThreeDPut(-0.3423,0.1394,-0.0476){Port}
\pstThreeDPut(-0.2441,-0.3293,-0.0670){Russ}
\pstThreeDPut(-0.2380,-0.1616,-0.0603){Sing}
\pstThreeDPut(-0.2602,-0.0687,-0.1184){SoAf}
\end{pspicture}
\begin{pspicture}(-8,-1.1)(1,3.5)
\rput(-3.4,2.9){T=0.6}
\psline(-4.3,-2.1)(4.1,-2.1)(4.1,3.3)(-4.3,3.3)(-4.3,-2.1)
\psset{xunit=7,yunit=7,Alpha=40,Beta=30} \scriptsize
\pstThreeDLine(-0.3293,0.3307,-0.1048)(-0.2767,0.3087,-0.0883) 
\pstThreeDLine(-0.3293,0.3307,-0.1048)(-0.3053,0.1377,-0.1069) 
\pstThreeDLine(-0.3293,0.3307,-0.1048)(-0.2434,0.2850,0.1640) 
\pstThreeDLine(-0.3293,0.3307,-0.1048)(-0.1590,0.2962,0.0355) 
\pstThreeDLine(-0.3293,0.3307,-0.1048)(-0.2008,0.5136,0.0513) 
\pstThreeDLine(-0.3293,0.3307,-0.1048)(-0.4497,-0.0240,-0.0150) 
\pstThreeDLine(-0.3293,0.3307,-0.1048)(-0.4143,0.1411,-0.0345) 
\pstThreeDLine(-0.3293,0.3307,-0.1048)(-0.4211,0.1692,-0.0737) 
\pstThreeDLine(-0.3293,0.3307,-0.1048)(-0.4098,0.1700,-0.0147) 
\pstThreeDLine(-0.3293,0.3307,-0.1048)(-0.4483,-0.0002,0.0071) 
\pstThreeDLine(-0.3293,0.3307,-0.1048)(-0.4327,0.1368,0.0159) 
\pstThreeDLine(-0.3293,0.3307,-0.1048)(-0.3578,0.1557,0.0036) 
\pstThreeDLine(-0.3293,0.3307,-0.1048)(-0.2602,-0.0687,-0.0884) 
\pstThreeDLine(-0.2767,0.3087,-0.0883)(-0.3053,0.1377,-0.1069) 
\pstThreeDLine(-0.2767,0.3087,-0.0883)(-0.2434,0.2850,0.1640) 
\pstThreeDLine(-0.2767,0.3087,-0.0883)(-0.2008,0.5136,0.0513) 
\pstThreeDLine(-0.2767,0.3087,-0.0883)(-0.4497,-0.0240,-0.0150) 
\pstThreeDLine(-0.2767,0.3087,-0.0883)(-0.4143,0.1411,-0.0345) 
\pstThreeDLine(-0.2767,0.3087,-0.0883)(-0.4211,0.1692,-0.0737) 
\pstThreeDLine(-0.2767,0.3087,-0.0883)(-0.4098,0.1700,-0.0147) 
\pstThreeDLine(-0.2767,0.3087,-0.0883)(-0.4483,-0.0002,0.0071) 
\pstThreeDLine(-0.2767,0.3087,-0.0883)(-0.3743,0.0417,0.0536) 
\pstThreeDLine(-0.2767,0.3087,-0.0883)(-0.2824,-0.0553,-0.1019) 
\pstThreeDLine(-0.3053,0.1377,-0.1069)(-0.2434,0.2850,0.1640) 
\pstThreeDLine(-0.3053,0.1377,-0.1069)(-0.4497,-0.0240,-0.0150) 
\pstThreeDLine(-0.3053,0.1377,-0.1069)(-0.4143,0.1411,-0.0345) 
\pstThreeDLine(-0.3053,0.1377,-0.1069)(-0.4211,0.1692,-0.0737) 
\pstThreeDLine(-0.3053,0.1377,-0.1069)(-0.4098,0.1700,-0.0147) 
\pstThreeDLine(-0.3053,0.1377,-0.1069)(-0.4483,-0.0002,0.0071) 
\pstThreeDLine(-0.3053,0.1377,-0.1069)(-0.4327,0.1368,0.0159) 
\pstThreeDLine(-0.3053,0.1377,-0.1069)(-0.3743,0.0417,0.0536) 
\pstThreeDLine(-0.3053,0.1377,-0.1069)(-0.3578,0.1557,0.0036) 
\pstThreeDLine(-0.3053,0.1377,-0.1069)(-0.2602,-0.0687,-0.0884) 
\pstThreeDLine(-0.2434,0.2850,0.1640)(-0.1590,0.2962,0.0355) 
\pstThreeDLine(-0.2434,0.2850,0.1640)(-0.2008,0.5136,0.0513) 
\pstThreeDLine(-0.2434,0.2850,0.1640)(-0.4211,0.1692,-0.0737) 
\pstThreeDLine(-0.2434,0.2850,0.1640)(-0.4098,0.1700,-0.0147) 
\pstThreeDLine(-0.2434,0.2850,0.1640)(-0.4483,-0.0002,0.0071) 
\pstThreeDLine(-0.2434,0.2850,0.1640)(-0.4327,0.1368,0.0159) 
\pstThreeDLine(-0.1590,0.2962,0.0355)(-0.2008,0.5136,0.0513) 
\pstThreeDLine(-0.2008,0.5136,0.0513)(-0.4211,0.1692,-0.0737) 
\pstThreeDLine(-0.2008,0.5136,0.0513)(-0.4098,0.1700,-0.0147) 
\pstThreeDLine(-0.2008,0.5136,0.0513)(-0.4327,0.1368,0.0159) 
\pstThreeDLine(-0.2008,0.5136,0.0513)(-0.3578,0.1557,0.0036) 
\pstThreeDLine(-0.2137,0.0764,-0.0031)(-0.4497,-0.0240,-0.0150) 
\pstThreeDLine(-0.4497,-0.0240,-0.0150)(-0.4143,0.1411,-0.0345) 
\pstThreeDLine(-0.4497,-0.0240,-0.0150)(-0.4211,0.1692,-0.0737) 
\pstThreeDLine(-0.4497,-0.0240,-0.0150)(-0.4098,0.1700,-0.0147) 
\pstThreeDLine(-0.4497,-0.0240,-0.0150)(-0.4483,-0.0002,0.0071) 
\pstThreeDLine(-0.4497,-0.0240,-0.0150)(-0.4327,0.1368,0.0159) 
\pstThreeDLine(-0.4497,-0.0240,-0.0150)(-0.2500,-0.0978,0.0209) 
\pstThreeDLine(-0.4497,-0.0240,-0.0150)(-0.3743,0.0417,0.0536) 
\pstThreeDLine(-0.4497,-0.0240,-0.0150)(-0.3331,-0.0481,-0.1432) 
\pstThreeDLine(-0.4497,-0.0240,-0.0150)(-0.3578,0.1557,0.0036) 
\pstThreeDLine(-0.4497,-0.0240,-0.0150)(-0.3423,0.1394,-0.0176) 
\pstThreeDLine(-0.4497,-0.0240,-0.0150)(-0.2300,-0.1152,0.1511) 
\pstThreeDLine(-0.4497,-0.0240,-0.0150)(-0.1998,-0.1644,-0.0280) 
\pstThreeDLine(-0.4497,-0.0240,-0.0150)(-0.2441,-0.3293,-0.0370) 
\pstThreeDLine(-0.4497,-0.0240,-0.0150)(-0.1606,-0.3229,-0.0570) 
\pstThreeDLine(-0.4497,-0.0240,-0.0150)(-0.2380,-0.1616,-0.0303) 
\pstThreeDLine(-0.4497,-0.0240,-0.0150)(-0.2602,-0.0687,-0.0884) 
\pstThreeDLine(-0.1861,-0.2205,0.1529)(-0.4483,-0.0002,0.0071) 
\pstThreeDLine(-0.1861,-0.2205,0.1529)(-0.1998,-0.1644,-0.0280) 
\pstThreeDLine(-0.1861,-0.2205,0.1529)(-0.1725,-0.3789,0.0721) 
\pstThreeDLine(-0.4143,0.1411,-0.0345)(-0.4211,0.1692,-0.0737) 
\pstThreeDLine(-0.4143,0.1411,-0.0345)(-0.2692,0.1314,0.0996) 
\pstThreeDLine(-0.4143,0.1411,-0.0345)(-0.2457,0.0421,-0.0528) 
\pstThreeDLine(-0.4143,0.1411,-0.0345)(-0.4098,0.1700,-0.0147) 
\pstThreeDLine(-0.4143,0.1411,-0.0345)(-0.4483,-0.0002,0.0071) 
\pstThreeDLine(-0.4143,0.1411,-0.0345)(-0.4327,0.1368,0.0159) 
\pstThreeDLine(-0.4143,0.1411,-0.0345)(-0.2500,-0.0978,0.0209) 
\pstThreeDLine(-0.4143,0.1411,-0.0345)(-0.3743,0.0417,0.0536) 
\pstThreeDLine(-0.4143,0.1411,-0.0345)(-0.3331,-0.0481,-0.1432) 
\pstThreeDLine(-0.4143,0.1411,-0.0345)(-0.3578,0.1557,0.0036) 
\pstThreeDLine(-0.4143,0.1411,-0.0345)(-0.3423,0.1394,-0.0176) 
\pstThreeDLine(-0.4143,0.1411,-0.0345)(-0.2300,-0.1152,0.1511) 
\pstThreeDLine(-0.4143,0.1411,-0.0345)(-0.2441,-0.3293,-0.0370) 
\pstThreeDLine(-0.4143,0.1411,-0.0345)(-0.2380,-0.1616,-0.0303) 
\pstThreeDLine(-0.4143,0.1411,-0.0345)(-0.2602,-0.0687,-0.0884) 
\pstThreeDLine(-0.4211,0.1692,-0.0737)(-0.2692,0.1314,0.0996) 
\pstThreeDLine(-0.4211,0.1692,-0.0737)(-0.4098,0.1700,-0.0147) 
\pstThreeDLine(-0.4211,0.1692,-0.0737)(-0.4483,-0.0002,0.0071) 
\pstThreeDLine(-0.4211,0.1692,-0.0737)(-0.4327,0.1368,0.0159) 
\pstThreeDLine(-0.4211,0.1692,-0.0737)(-0.3743,0.0417,0.0536) 
\pstThreeDLine(-0.4211,0.1692,-0.0737)(-0.3331,-0.0481,-0.1432) 
\pstThreeDLine(-0.4211,0.1692,-0.0737)(-0.3578,0.1557,0.0036) 
\pstThreeDLine(-0.4211,0.1692,-0.0737)(-0.3423,0.1394,-0.0176) 
\pstThreeDLine(-0.4211,0.1692,-0.0737)(-0.2300,-0.1152,0.1511) 
\pstThreeDLine(-0.4211,0.1692,-0.0737)(-0.2380,-0.1616,-0.0303) 
\pstThreeDLine(-0.4211,0.1692,-0.0737)(-0.2602,-0.0687,-0.0884) 
\pstThreeDLine(-0.2692,0.1314,0.0996)(-0.4098,0.1700,-0.0147) 
\pstThreeDLine(-0.2692,0.1314,0.0996)(-0.4483,-0.0002,0.0071) 
\pstThreeDLine(-0.2692,0.1314,0.0996)(-0.4327,0.1368,0.0159) 
\pstThreeDLine(-0.2692,0.1314,0.0996)(-0.3331,-0.0481,-0.1432) 
\pstThreeDLine(-0.2692,0.1314,0.0996)(-0.3578,0.1557,0.0036) 
\pstThreeDLine(-0.2457,0.0421,-0.0528)(-0.4483,-0.0002,0.0071) 
\pstThreeDLine(-0.2457,0.0421,-0.0528)(-0.3743,0.0417,0.0536) 
\pstThreeDLine(-0.2457,0.0421,-0.0528)(-0.3331,-0.0481,-0.1432) 
\pstThreeDLine(-0.2457,0.0421,-0.0528)(-0.3578,0.1557,0.0036) 
\pstThreeDLine(-0.2457,0.0421,-0.0528)(-0.2602,-0.0687,-0.0884) 
\pstThreeDLine(-0.4098,0.1700,-0.0147)(-0.4483,-0.0002,0.0071) 
\pstThreeDLine(-0.4098,0.1700,-0.0147)(-0.4327,0.1368,0.0159) 
\pstThreeDLine(-0.4098,0.1700,-0.0147)(-0.2500,-0.0978,0.0209) 
\pstThreeDLine(-0.4098,0.1700,-0.0147)(-0.3743,0.0417,0.0536) 
\pstThreeDLine(-0.4098,0.1700,-0.0147)(-0.3331,-0.0481,-0.1432) 
\pstThreeDLine(-0.4098,0.1700,-0.0147)(-0.3578,0.1557,0.0036) 
\pstThreeDLine(-0.4098,0.1700,-0.0147)(-0.3423,0.1394,-0.0176) 
\pstThreeDLine(-0.4098,0.1700,-0.0147)(-0.2380,-0.1616,-0.0303) 
\pstThreeDLine(-0.4098,0.1700,-0.0147)(-0.2602,-0.0687,-0.0884) 
\pstThreeDLine(-0.1193,0.0304,-0.1252)(-0.4483,-0.0002,0.0071) 
\pstThreeDLine(-0.4483,-0.0002,0.0071)(-0.4327,0.1368,0.0159) 
\pstThreeDLine(-0.4483,-0.0002,0.0071)(-0.2500,-0.0978,0.0209) 
\pstThreeDLine(-0.4483,-0.0002,0.0071)(-0.3743,0.0417,0.0536) 
\pstThreeDLine(-0.4483,-0.0002,0.0071)(-0.3331,-0.0481,-0.1432) 
\pstThreeDLine(-0.4483,-0.0002,0.0071)(-0.3578,0.1557,0.0036) 
\pstThreeDLine(-0.4483,-0.0002,0.0071)(-0.3423,0.1394,-0.0176) 
\pstThreeDLine(-0.4483,-0.0002,0.0071)(-0.2300,-0.1152,0.1511) 
\pstThreeDLine(-0.4483,-0.0002,0.0071)(-0.1998,-0.1644,-0.0280) 
\pstThreeDLine(-0.4483,-0.0002,0.0071)(-0.2441,-0.3293,-0.0370) 
\pstThreeDLine(-0.4483,-0.0002,0.0071)(-0.1725,-0.3789,0.0721) 
\pstThreeDLine(-0.4483,-0.0002,0.0071)(-0.2380,-0.1616,-0.0303) 
\pstThreeDLine(-0.4483,-0.0002,0.0071)(-0.2602,-0.0687,-0.0884) 
\pstThreeDLine(-0.4327,0.1368,0.0159)(-0.3743,0.0417,0.0536) 
\pstThreeDLine(-0.4327,0.1368,0.0159)(-0.3331,-0.0481,-0.1432) 
\pstThreeDLine(-0.4327,0.1368,0.0159)(-0.3578,0.1557,0.0036) 
\pstThreeDLine(-0.4327,0.1368,0.0159)(-0.3423,0.1394,-0.0176) 
\pstThreeDLine(-0.4327,0.1368,0.0159)(-0.2824,-0.0553,-0.1019) 
\pstThreeDLine(-0.4327,0.1368,0.0159)(-0.2602,-0.0687,-0.0884) 
\pstThreeDLine(-0.2500,-0.0978,0.0209)(-0.3331,-0.0481,-0.1432) 
\pstThreeDLine(-0.2500,-0.0978,0.0209)(-0.2441,-0.3293,-0.0370) 
\pstThreeDLine(-0.3743,0.0417,0.0536)(-0.3423,0.1394,-0.0176) 
\pstThreeDLine(-0.3743,0.0417,0.0536)(-0.3331,-0.0481,-0.1432) 
\pstThreeDLine(-0.3743,0.0417,0.0536)(-0.3578,0.1557,0.0036) 
\pstThreeDLine(-0.3743,0.0417,0.0536)(-0.2300,-0.1152,0.1511) 
\pstThreeDLine(-0.3743,0.0417,0.0536)(-0.2602,-0.0687,-0.0884) 
\pstThreeDLine(-0.3331,-0.0481,-0.1432)(-0.3578,0.1557,0.0036) 
\pstThreeDLine(-0.3331,-0.0481,-0.1432)(-0.3423,0.1394,-0.0176) 
\pstThreeDLine(-0.3331,-0.0481,-0.1432)(-0.2441,-0.3293,-0.0370) 
\pstThreeDLine(-0.3331,-0.0481,-0.1432)(-0.2300,-0.1152,0.1511) 
\pstThreeDLine(-0.3331,-0.0481,-0.1432)(-0.1998,-0.1644,-0.0280) 
\pstThreeDLine(-0.3331,-0.0481,-0.1432)(-0.2824,-0.0553,-0.1019) 
\pstThreeDLine(-0.3331,-0.0481,-0.1432)(-0.2602,-0.0687,-0.0884) 
\pstThreeDLine(-0.3578,0.1557,0.0036)(-0.3423,0.1394,-0.0176) 
\pstThreeDLine(-0.3578,0.1557,0.0036)(-0.2602,-0.0687,-0.0884) 
\pstThreeDLine(-0.2300,-0.1152,0.1511)(-0.1998,-0.1644,-0.0280) 
\pstThreeDLine(-0.2300,-0.1152,0.1511)(-0.1905,-0.2449,0.2364) 
\pstThreeDLine(-0.2300,-0.1152,0.1511)(-0.2441,-0.3293,-0.0370) 
\pstThreeDLine(-0.2300,-0.1152,0.1511)(-0.2602,-0.0687,-0.0884) 
\pstThreeDLine(-0.1998,-0.1644,-0.0280)(-0.2441,-0.3293,-0.0370) 
\pstThreeDLine(-0.1998,-0.1644,-0.0280)(-0.1606,-0.3229,-0.0570) 
\pstThreeDLine(-0.1998,-0.1644,-0.0280)(-0.2602,-0.0687,-0.0884) 
\pstThreeDLine(-0.1905,-0.2449,0.2364)(-0.1725,-0.3789,0.0721) 
\pstThreeDLine(-0.1905,-0.2449,0.2364)(-0.1073,-0.3379,0.0185) 
\pstThreeDLine(-0.2441,-0.3293,-0.0370)(-0.1725,-0.3789,0.0721) 
\pstThreeDLine(-0.2441,-0.3293,-0.0370)(-0.1606,-0.3229,-0.0570) 
\pstThreeDLine(-0.2441,-0.3293,-0.0370)(-0.1073,-0.3379,0.0185) 
\pstThreeDLine(-0.1149,-0.2784,0.0225)(-0.1725,-0.3789,0.0721) 
\pstThreeDLine(-0.1149,-0.2784,0.0225)(-0.1073,-0.3379,0.0185) 
\pstThreeDLine(-0.1725,-0.3789,0.0721)(-0.1606,-0.3229,-0.0570) 
\pstThreeDLine(-0.1725,-0.3789,0.0721)(-0.2380,-0.1616,-0.0303) 
\pstThreeDLine(-0.1725,-0.3789,0.0721)(-0.1073,-0.3379,0.0185) 
\pstThreeDLine(-0.1725,-0.3789,0.0721)(-0.2602,-0.0687,-0.0884) 
\pstThreeDLine(-0.1606,-0.3229,-0.0570)(-0.2380,-0.1616,-0.0303) 
\pstThreeDLine(-0.1073,-0.3379,0.0185)(-0.2602,-0.0687,-0.0884) 
\pstThreeDDot[linecolor=orange,linewidth=1.2pt](-0.3293,0.3307,-0.1048) 
\pstThreeDDot[linecolor=orange,linewidth=1.2pt](-0.2767,0.3087,-0.0883) 
\pstThreeDDot[linecolor=orange,linewidth=1.2pt](-0.3053,0.1377,-0.1069) 
\pstThreeDDot[linecolor=orange,linewidth=1.2pt](-0.2434,0.2850,0.1640) 
\pstThreeDDot[linecolor=green,linewidth=1.2pt](-0.1590,0.2962,0.0355) 
\pstThreeDDot[linecolor=green,linewidth=1.2pt](-0.2008,0.5136,0.0513) 
\pstThreeDDot[linecolor=green,linewidth=1.2pt](-0.2137,0.0764,-0.0031) 
\pstThreeDDot[linecolor=blue,linewidth=1.2pt](-0.4497,-0.0240,-0.0150) 
\pstThreeDDot[linecolor=blue,linewidth=1.2pt](-0.1861,-0.2205,0.1529) 
\pstThreeDDot[linecolor=blue,linewidth=1.2pt](-0.4143,0.1411,-0.0345) 
\pstThreeDDot[linecolor=blue,linewidth=1.2pt](-0.4211,0.1692,-0.0737) 
\pstThreeDDot[linecolor=blue,linewidth=1.2pt](-0.2692,0.1314,0.0996) 
\pstThreeDDot[linecolor=blue,linewidth=1.2pt](-0.2457,0.0421,-0.0528) 
\pstThreeDDot[linecolor=blue,linewidth=1.2pt](-0.4098,0.1700,-0.0147) 
\pstThreeDDot[linecolor=blue,linewidth=1.2pt](-0.1193,0.0304,-0.1252) 
\pstThreeDDot[linecolor=blue,linewidth=1.2pt](-0.4483,-0.0002,0.0071) 
\pstThreeDDot[linecolor=blue,linewidth=1.2pt](-0.4327,0.1368,0.0159) 
\pstThreeDDot[linecolor=blue,linewidth=1.2pt](-0.2500,-0.0978,0.0209) 
\pstThreeDDot[linecolor=blue,linewidth=1.2pt](-0.3743,0.0417,0.0536) 
\pstThreeDDot[linecolor=blue,linewidth=1.2pt](-0.3331,-0.0481,-0.1432) 
\pstThreeDDot[linecolor=blue,linewidth=1.2pt](-0.3578,0.1557,0.0036) 
\pstThreeDDot[linecolor=blue,linewidth=1.2pt](-0.3423,0.1394,-0.0176) 
\pstThreeDDot[linecolor=blue,linewidth=1.2pt](-0.2300,-0.1152,0.1511) 
\pstThreeDDot[linecolor=blue,linewidth=1.2pt](-0.1998,-0.1644,-0.0280) 
\pstThreeDDot[linecolor=blue,linewidth=1.2pt](-0.1905,-0.2449,0.2364) 
\pstThreeDDot[linecolor=bluish,linewidth=1.2pt](-0.2441,-0.3293,-0.0370) 
\pstThreeDDot[linecolor=brown,linewidth=1.2pt](-0.2824,-0.0553,-0.1019) 
\pstThreeDDot[linecolor=red,linewidth=1.2pt](-0.1149,-0.2784,0.0225) 
\pstThreeDDot[linecolor=red,linewidth=1.2pt](-0.1725,-0.3789,0.0721) 
\pstThreeDDot[linecolor=red,linewidth=1.2pt](-0.1606,-0.3229,-0.0570) 
\pstThreeDDot[linecolor=red,linewidth=1.2pt](-0.2380,-0.1616,-0.0303) 
\pstThreeDDot[linecolor=black,linewidth=1.2pt](-0.1073,-0.3379,0.0185) 
\pstThreeDDot[linecolor=magenta,linewidth=1.2pt](-0.2602,-0.0687,-0.0884) 
\pstThreeDPut(-0.3293,0.3307,-0.1348){S\&P}
\pstThreeDPut(-0.2767,0.3087,-0.1183){Nasd}
\pstThreeDPut(-0.3053,0.1377,-0.1369){Cana}
\pstThreeDPut(-0.2434,0.2850,0.1940){Mexi}
\pstThreeDPut(-0.1590,0.2962,0.0655){Braz}
\pstThreeDPut(-0.2008,0.5136,0.0813){Arge}
\pstThreeDPut(-0.2137,0.0764,-0.0331){Chil}
\pstThreeDPut(-0.4497,-0.0240,-0.0450){UK}
\pstThreeDPut(-0.1861,-0.2205,0.1829){Irel}
\pstThreeDPut(-0.4143,0.1411,-0.0645){Fran}
\pstThreeDPut(-0.4211,0.1692,-0.1037){Germ}
\pstThreeDPut(-0.2692,0.1314,0.1296){Swit}
\pstThreeDPut(-0.2457,0.0421,-0.0828){Autr}
\pstThreeDPut(-0.4098,0.1700,-0.0447){Ital}
\pstThreeDPut(-0.1193,0.0304,-0.1552){Belg}
\pstThreeDPut(-0.4483,-0.0002,0.0371){Neth}
\pstThreeDPut(-0.4327,0.1368,0.0459){Swed}
\pstThreeDPut(-0.2500,-0.0978,0.0509){Denm}
\pstThreeDPut(-0.3743,0.0417,0.0836){Finl}
\pstThreeDPut(-0.3331,-0.0481,-0.1732){Norw}
\pstThreeDPut(-0.3578,0.1557,0.0336){Spai}
\pstThreeDPut(-0.3423,0.1394,-0.0476){Port}
\pstThreeDPut(-0.2300,-0.1152,0.1811){CzRe}
\pstThreeDPut(-0.1998,-0.1644,-0.0580){Hung}
\pstThreeDPut(-0.1905,-0.2449,0.2664){Pola}
\pstThreeDPut(-0.2441,-0.3293,-0.0670){Russ}
\pstThreeDPut(-0.2824,-0.0553,-0.1319){Isra}
\pstThreeDPut(-0.1149,-0.2784,0.0525){Japa}
\pstThreeDPut(-0.1725,-0.3789,0.1021){HoKo}
\pstThreeDPut(-0.1606,-0.3229,-0.0870){SoKo}
\pstThreeDPut(-0.2380,-0.1616,-0.0603){Sing}
\pstThreeDPut(-0.1073,-0.3379,0.0485){Aust}
\pstThreeDPut(-0.2602,-0.0687,-0.1184){SoAf}
\end{pspicture}

\vskip 1.5 cm

\noindent Fig. 10. Three dimensional view of the asset trees for the second semester of 2000, with threshold ranging from $T=0.3$ to $T=0.6$.

\vskip 0.3 cm

\subsubsection{First semester, 2001}

Figure 11 shows the three dimensional view of the asset trees for the first semester of 2001, with threshold ranging from $T=0.3$ to $T=0.6$.

At $T=0.2$, there are two small clusters: one formed by $S\&P$ and Nasdaq, and the other formed by France, Italy, and Netherlands. At $T=0.3$, Canada joins the North American cluster, and the UK, Germany, Sweden, and Spain join the European cluster. At $T=0.4$, the European cluster is more interwoven, and have the addition of Switzerland, Belgium, Finland, Norway, and Portugal. At $T=0.5$, the North American cluster is fully formed, with the addition of Mexico, and a South American cluster forms, composed of Brazil and Argentina. Europe becomes even more densily connected, and receives the addition of Ireland, Denmark, the Czech Republic, and Hungary, the last two considered as part of Eastern Europe. We also have Israel and South Africa connecting with Europe. A new cluster is born, made of Pacific Asian countries, namely Japan, Hong Kong, and South Korea. The North American and European clusters start to get connected. At $T=0.6$, North and South America are integrated with each other and with Europe. Newcommers are Chile to South America, Austria, Luxembourg, Greece, Poland, and Russia to the European cluster, Singapore to the Pacific Asian cluster, and New Zealand, which also connects with the Pacific Asian cluster. Through Singapore and Hong Kong, the Pacific Asian cluster connects with the European one. For $T>0.6$, we see further integration and more markets joining the fold, but we then enter a region where noise dominates, and connections cannot be fully trusted. By $T=1.3$, all indices are connected with one another.

\begin{pspicture}(-3,-2.3)(1,3.9)
\rput(-2.5,2.7){T=0.3}
\psline(-3.4,-2.7)(4.5,-2.7)(4.5,3.2)(-3.4,3.2)(-3.4,-2.7)
\psset{xunit=7,yunit=7,Alpha=40,Beta=0} \scriptsize
\pstThreeDLine(-0.3058,0.4366,-0.2130)(-0.2370,0.3880,-0.2352) 
\pstThreeDLine(-0.4573,0.0026,0.0890)(-0.4374,0.0304,0.0586) 
\pstThreeDLine(-0.4573,0.0026,0.0890)(-0.4068,0.0142,0.1478) 
\pstThreeDLine(-0.3058,0.4366,-0.2130)(-0.3105,0.3550,-0.2430) 
\pstThreeDLine(-0.2370,0.3880,-0.2352)(-0.3105,0.3550,-0.2430) 
\pstThreeDLine(-0.4188,0.1322,0.0751)(-0.4573,0.0026,0.0890) 
\pstThreeDLine(-0.4573,0.0026,0.0890)(-0.4269,0.1338,0.0221) 
\pstThreeDLine(-0.4573,0.0026,0.0890)(-0.4438,0.0797,0.0084) 
\pstThreeDLine(-0.4573,0.0026,0.0890)(-0.3813,-0.0473,0.0646) 
\pstThreeDLine(-0.4573,0.0026,0.0890)(-0.4188,0.0934,0.0241) 
\pstThreeDLine(-0.4269,0.1338,0.0221)(-0.4374,0.0304,0.0586) 
\pstThreeDLine(-0.4269,0.1338,0.0221)(-0.4068,0.0142,0.1478) 
\pstThreeDLine(-0.4269,0.1338,0.0221)(-0.4438,0.0797,0.0084) 
\pstThreeDLine(-0.4269,0.1338,0.0221)(-0.4188,0.0934,0.0241) 
\pstThreeDLine(-0.4374,0.0304,0.0586)(-0.4068,0.0142,0.1478) 
\pstThreeDLine(-0.4374,0.0304,0.0586)(-0.4438,0.0797,0.0084) 
\pstThreeDLine(-0.4374,0.0304,0.0586)(-0.4188,0.0934,0.0241) 
\pstThreeDLine(-0.4438,0.0797,0.0084)(-0.3813,-0.0473,0.0646) 
\pstThreeDDot[linecolor=orange,linewidth=1.2pt](-0.3058,0.4366,-0.2130) 
\pstThreeDDot[linecolor=orange,linewidth=1.2pt](-0.2370,0.3880,-0.2352) 
\pstThreeDDot[linecolor=orange,linewidth=1.2pt](-0.3105,0.3550,-0.2430) 
\pstThreeDDot[linecolor=blue,linewidth=1.2pt](-0.4188,0.1322,0.0751) 
\pstThreeDDot[linecolor=blue,linewidth=1.2pt](-0.4573,0.0026,0.0890) 
\pstThreeDDot[linecolor=blue,linewidth=1.2pt](-0.4269,0.1338,0.0221) 
\pstThreeDDot[linecolor=blue,linewidth=1.2pt](-0.4374,0.0304,0.0586) 
\pstThreeDDot[linecolor=blue,linewidth=1.2pt](-0.4068,0.0142,0.1478) 
\pstThreeDDot[linecolor=blue,linewidth=1.2pt](-0.4438,0.0797,0.0084) 
\pstThreeDDot[linecolor=blue,linewidth=1.2pt](-0.3813,-0.0473,0.0646) 
\pstThreeDDot[linecolor=blue,linewidth=1.2pt](-0.4188,0.0934,0.0241) 
\pstThreeDPut(-0.3058,0.4366,-0.2430){S\&P}
\pstThreeDPut(-0.2370,0.3880,-0.2652){Nasd}
\pstThreeDPut(-0.3105,0.3550,-0.2730){Cana}
\pstThreeDPut(-0.4188,0.1322,0.1051){UK}
\pstThreeDPut(-0.4573,0.0026,0.1190){Fran}
\pstThreeDPut(-0.4269,0.1338,0.0521){Germ}
\pstThreeDPut(-0.4374,0.0304,0.0886){Ital}
\pstThreeDPut(-0.4068,0.0142,0.1778){Neth}
\pstThreeDPut(-0.4438,0.0797,0.0384){Swed}
\pstThreeDPut(-0.3813,-0.0473,0.0946){Finl}
\pstThreeDPut(-0.4188,0.0934,0.0541){Spai}
\end{pspicture}
\begin{pspicture}(-8,-2.3)(1,3.9)
\rput(-2.5,2.7){T=0.4}
\psline(-3.4,-2.7)(4.5,-2.7)(4.5,3.2)(-3.4,3.2)(-3.4,-2.7)
\psset{xunit=7,yunit=7,Alpha=40,Beta=0} \scriptsize
\pstThreeDLine(-0.3058,0.4366,-0.2130)(-0.2370,0.3880,-0.2352) 
\pstThreeDLine(-0.4573,0.0026,0.0890)(-0.4374,0.0304,0.0586) 
\pstThreeDLine(-0.4573,0.0026,0.0890)(-0.4068,0.0142,0.1478) 
\pstThreeDLine(-0.3058,0.4366,-0.2130)(-0.3105,0.3550,-0.2430) 
\pstThreeDLine(-0.2370,0.3880,-0.2352)(-0.3105,0.3550,-0.2430) 
\pstThreeDLine(-0.4188,0.1322,0.0751)(-0.4573,0.0026,0.0890) 
\pstThreeDLine(-0.4573,0.0026,0.0890)(-0.4269,0.1338,0.0221) 
\pstThreeDLine(-0.4573,0.0026,0.0890)(-0.4438,0.0797,0.0084) 
\pstThreeDLine(-0.4573,0.0026,0.0890)(-0.3813,-0.0473,0.0646) 
\pstThreeDLine(-0.4573,0.0026,0.0890)(-0.4188,0.0934,0.0241) 
\pstThreeDLine(-0.4269,0.1338,0.0221)(-0.4374,0.0304,0.0586) 
\pstThreeDLine(-0.4269,0.1338,0.0221)(-0.4068,0.0142,0.1478) 
\pstThreeDLine(-0.4269,0.1338,0.0221)(-0.4438,0.0797,0.0084) 
\pstThreeDLine(-0.4269,0.1338,0.0221)(-0.4188,0.0934,0.0241) 
\pstThreeDLine(-0.4374,0.0304,0.0586)(-0.4068,0.0142,0.1478) 
\pstThreeDLine(-0.4374,0.0304,0.0586)(-0.4438,0.0797,0.0084) 
\pstThreeDLine(-0.4374,0.0304,0.0586)(-0.4188,0.0934,0.0241) 
\pstThreeDLine(-0.4438,0.0797,0.0084)(-0.3813,-0.0473,0.0646) 
\pstThreeDLine(-0.4188,0.1322,0.0751)(-0.4269,0.1338,0.0221) 
\pstThreeDLine(-0.4188,0.1322,0.0751)(-0.4374,0.0304,0.0586) 
\pstThreeDLine(-0.4188,0.1322,0.0751)(-0.4068,0.0142,0.1478) 
\pstThreeDLine(-0.4188,0.1322,0.0751)(-0.4438,0.0797,0.0084) 
\pstThreeDLine(-0.4188,0.1322,0.0751)(-0.3813,-0.0473,0.0646) 
\pstThreeDLine(-0.4188,0.1322,0.0751)(-0.3166,-0.0438,0.1285) 
\pstThreeDLine(-0.4188,0.1322,0.0751)(-0.4188,0.0934,0.0241) 
\pstThreeDLine(-0.4573,0.0026,0.0890)(-0.3166,-0.0438,0.1285) 
\pstThreeDLine(-0.4573,0.0026,0.0890)(-0.3166,-0.0438,0.1285) 
\pstThreeDLine(-0.4573,0.0026,0.0890)(-0.3776,0.0170,0.1198) 
\pstThreeDLine(-0.4269,0.1338,0.0221)(-0.3166,-0.0438,0.1285) 
\pstThreeDLine(-0.4269,0.1338,0.0221)(-0.3813,-0.0473,0.0646) 
\pstThreeDLine(-0.3166,-0.0438,0.1285)(-0.3166,-0.0438,0.1285) 
\pstThreeDLine(-0.4068,0.0142,0.1478)(-0.3166,-0.0438,0.1285) 
\pstThreeDLine(-0.4374,0.0304,0.0586)(-0.3813,-0.0473,0.0646) 
\pstThreeDLine(-0.4374,0.0304,0.0586)(-0.3776,0.0170,0.1198) 
\pstThreeDLine(-0.2235,0.0807,0.2188)(-0.3166,-0.0438,0.1285) 
\pstThreeDLine(-0.3166,-0.0438,0.1285)(-0.4438,0.0797,0.0084) 
\pstThreeDLine(-0.3166,-0.0438,0.1285)(-0.3813,-0.0473,0.0646) 
\pstThreeDLine(-0.3166,-0.0438,0.1285)(-0.3166,-0.0438,0.1285) 
\pstThreeDLine(-0.3166,-0.0438,0.1285)(-0.4188,0.0934,0.0241) 
\pstThreeDLine(-0.3166,-0.0438,0.1285)(-0.3776,0.0170,0.1198) 
\pstThreeDLine(-0.3776,0.0170,0.1198)(-0.4188,0.0934,0.0241) 
\pstThreeDLine(-0.3776,0.0170,0.1198)(-0.3776,0.0170,0.1198) 
\pstThreeDLine(-0.4188,0.0934,0.0241)(-0.3776,0.0170,0.1198) 
\pstThreeDDot[linecolor=orange,linewidth=1.2pt](-0.3058,0.4366,-0.2130) 
\pstThreeDDot[linecolor=orange,linewidth=1.2pt](-0.2370,0.3880,-0.2352) 
\pstThreeDDot[linecolor=orange,linewidth=1.2pt](-0.3105,0.3550,-0.2430) 
\pstThreeDDot[linecolor=blue,linewidth=1.2pt](-0.4188,0.1322,0.0751) 
\pstThreeDDot[linecolor=blue,linewidth=1.2pt](-0.4573,0.0026,0.0890) 
\pstThreeDDot[linecolor=blue,linewidth=1.2pt](-0.4269,0.1338,0.0221) 
\pstThreeDDot[linecolor=blue,linewidth=1.2pt](-0.3757,0.0990,0.1322) 
\pstThreeDDot[linecolor=blue,linewidth=1.2pt](-0.4374,0.0304,0.0586) 
\pstThreeDDot[linecolor=blue,linewidth=1.2pt](-0.2235,0.0807,0.2188) 
\pstThreeDDot[linecolor=blue,linewidth=1.2pt](-0.4068,0.0142,0.1478) 
\pstThreeDDot[linecolor=blue,linewidth=1.2pt](-0.4438,0.0797,0.0084) 
\pstThreeDDot[linecolor=blue,linewidth=1.2pt](-0.3813,-0.0473,0.0646) 
\pstThreeDDot[linecolor=blue,linewidth=1.2pt](-0.3166,-0.0438,0.1285) 
\pstThreeDDot[linecolor=blue,linewidth=1.2pt](-0.4188,0.0934,0.0241) 
\pstThreeDDot[linecolor=blue,linewidth=1.2pt](-0.3776,0.0170,0.1198) 
\pstThreeDPut(-0.3058,0.4366,-0.2430){S\&P}
\pstThreeDPut(-0.2370,0.3880,-0.2652){Nasd}
\pstThreeDPut(-0.3105,0.3550,-0.2730){Cana}
\pstThreeDPut(-0.4188,0.1322,0.1051){UK}
\pstThreeDPut(-0.4573,0.0026,0.1190){Fran}
\pstThreeDPut(-0.4269,0.1338,0.0521){Germ}
\pstThreeDPut(-0.3757,0.0990,0.1622){Swit}
\pstThreeDPut(-0.4374,0.0304,0.0886){Ital}
\pstThreeDPut(-0.2235,0.0807,0.2488){Belg}
\pstThreeDPut(-0.4068,0.0142,0.1778){Neth}
\pstThreeDPut(-0.4438,0.0797,0.0384){Swed}
\pstThreeDPut(-0.3813,-0.0473,0.0946){Finl}
\pstThreeDPut(-0.3166,-0.0438,0.1585){Norw}
\pstThreeDPut(-0.4188,0.0934,0.0541){Spai}
\pstThreeDPut(-0.3776,0.0170,0.1498){Port}
\end{pspicture}

\vskip 0.2 cm

\begin{pspicture}(-3,-2.3)(1,3.9)
\rput(-2.5,2.7){T=0.5}
\psline(-3.4,-2.7)(4.5,-2.7)(4.5,3.2)(-3.4,3.2)(-3.4,-2.7)
\psset{xunit=7,yunit=7,Alpha=40,Beta=0} \scriptsize
\pstThreeDLine(-0.3058,0.4366,-0.2130)(-0.2370,0.3880,-0.2352) 
\pstThreeDLine(-0.4573,0.0026,0.0890)(-0.4374,0.0304,0.0586) 
\pstThreeDLine(-0.4573,0.0026,0.0890)(-0.4068,0.0142,0.1478) 
\pstThreeDLine(-0.3058,0.4366,-0.2130)(-0.3105,0.3550,-0.2430) 
\pstThreeDLine(-0.2370,0.3880,-0.2352)(-0.3105,0.3550,-0.2430) 
\pstThreeDLine(-0.4188,0.1322,0.0751)(-0.4573,0.0026,0.0890) 
\pstThreeDLine(-0.4573,0.0026,0.0890)(-0.4269,0.1338,0.0221) 
\pstThreeDLine(-0.4573,0.0026,0.0890)(-0.4438,0.0797,0.0084) 
\pstThreeDLine(-0.4573,0.0026,0.0890)(-0.3813,-0.0473,0.0646) 
\pstThreeDLine(-0.4573,0.0026,0.0890)(-0.4188,0.0934,0.0241) 
\pstThreeDLine(-0.4269,0.1338,0.0221)(-0.4374,0.0304,0.0586) 
\pstThreeDLine(-0.4269,0.1338,0.0221)(-0.4068,0.0142,0.1478) 
\pstThreeDLine(-0.4269,0.1338,0.0221)(-0.4438,0.0797,0.0084) 
\pstThreeDLine(-0.4269,0.1338,0.0221)(-0.4188,0.0934,0.0241) 
\pstThreeDLine(-0.4374,0.0304,0.0586)(-0.4068,0.0142,0.1478) 
\pstThreeDLine(-0.4374,0.0304,0.0586)(-0.4438,0.0797,0.0084) 
\pstThreeDLine(-0.4374,0.0304,0.0586)(-0.4188,0.0934,0.0241) 
\pstThreeDLine(-0.4438,0.0797,0.0084)(-0.3813,-0.0473,0.0646) 
\pstThreeDLine(-0.4188,0.1322,0.0751)(-0.4269,0.1338,0.0221) 
\pstThreeDLine(-0.4188,0.1322,0.0751)(-0.4374,0.0304,0.0586) 
\pstThreeDLine(-0.4188,0.1322,0.0751)(-0.4068,0.0142,0.1478) 
\pstThreeDLine(-0.4188,0.1322,0.0751)(-0.4438,0.0797,0.0084) 
\pstThreeDLine(-0.4188,0.1322,0.0751)(-0.3813,-0.0473,0.0646) 
\pstThreeDLine(-0.4188,0.1322,0.0751)(-0.3166,-0.0438,0.1285) 
\pstThreeDLine(-0.4188,0.1322,0.0751)(-0.4188,0.0934,0.0241) 
\pstThreeDLine(-0.4573,0.0026,0.0890)(-0.3166,-0.0438,0.1285) 
\pstThreeDLine(-0.4573,0.0026,0.0890)(-0.3166,-0.0438,0.1285) 
\pstThreeDLine(-0.4573,0.0026,0.0890)(-0.3776,0.0170,0.1198) 
\pstThreeDLine(-0.4269,0.1338,0.0221)(-0.3166,-0.0438,0.1285) 
\pstThreeDLine(-0.4269,0.1338,0.0221)(-0.3813,-0.0473,0.0646) 
\pstThreeDLine(-0.3166,-0.0438,0.1285)(-0.3166,-0.0438,0.1285) 
\pstThreeDLine(-0.4068,0.0142,0.1478)(-0.3166,-0.0438,0.1285) 
\pstThreeDLine(-0.4374,0.0304,0.0586)(-0.3813,-0.0473,0.0646) 
\pstThreeDLine(-0.4374,0.0304,0.0586)(-0.3776,0.0170,0.1198) 
\pstThreeDLine(-0.2235,0.0807,0.2188)(-0.3166,-0.0438,0.1285) 
\pstThreeDLine(-0.3166,-0.0438,0.1285)(-0.4438,0.0797,0.0084) 
\pstThreeDLine(-0.3166,-0.0438,0.1285)(-0.3813,-0.0473,0.0646) 
\pstThreeDLine(-0.3166,-0.0438,0.1285)(-0.3166,-0.0438,0.1285) 
\pstThreeDLine(-0.3166,-0.0438,0.1285)(-0.4188,0.0934,0.0241) 
\pstThreeDLine(-0.3166,-0.0438,0.1285)(-0.3776,0.0170,0.1198) 
\pstThreeDLine(-0.3776,0.0170,0.1198)(-0.4188,0.0934,0.0241) 
\pstThreeDLine(-0.3776,0.0170,0.1198)(-0.3776,0.0170,0.1198) 
\pstThreeDLine(-0.4188,0.0934,0.0241)(-0.3776,0.0170,0.1198) 
\pstThreeDLine(-0.3058,0.4366,-0.2130)(-0.1427,0.2351,-0.2382) 
\pstThreeDLine(-0.3058,0.4366,-0.2130)(-0.4269,0.1338,0.0221) 
\pstThreeDLine(-0.3105,0.3550,-0.2430)(-0.1427,0.2351,-0.2382) 
\pstThreeDLine(-0.3105,0.3550,-0.2430)(-0.4573,0.0026,0.0890) 
\pstThreeDLine(-0.3105,0.3550,-0.2430)(-0.4269,0.1338,0.0221) 
\pstThreeDLine(-0.1364,0.2466,-0.0671)(-0.1755,0.1852,-0.2335) 
\pstThreeDLine(-0.4188,0.1322,0.0751)(-0.4188,0.0934,0.0241) 
\pstThreeDLine(-0.4188,0.1322,0.0751)(-0.3143,-0.1175,0.0845) 
\pstThreeDLine(-0.4188,0.1322,0.0751)(-0.3776,0.0170,0.1198) 
\pstThreeDLine(-0.2113,-0.0210,0.3142)(-0.3143,-0.1175,0.0845) 
\pstThreeDLine(-0.4573,0.0026,0.0890)(-0.3143,-0.1175,0.0845) 
\pstThreeDLine(-0.4573,0.0026,0.0890)(-0.2436,-0.0325,0.1106) 
\pstThreeDLine(-0.4573,0.0026,0.0890)(-0.2283,-0.0290,0.1351) 
\pstThreeDLine(-0.4269,0.1338,0.0221)(-0.3143,-0.1175,0.0845) 
\pstThreeDLine(-0.4269,0.1338,0.0221)(-0.3166,-0.0438,0.1285) 
\pstThreeDLine(-0.4269,0.1338,0.0221)(-0.3776,0.0170,0.1198) 
\pstThreeDLine(-0.3757,0.0990,0.1322)(-0.4374,0.0304,0.0586) 
\pstThreeDLine(-0.3757,0.0990,0.1322)(-0.2235,0.0807,0.2188) 
\pstThreeDLine(-0.3757,0.0990,0.1322)(-0.4438,0.0797,0.0084) 
\pstThreeDLine(-0.3757,0.0990,0.1322)(-0.3143,-0.1175,0.0845) 
\pstThreeDLine(-0.3757,0.0990,0.1322)(-0.3813,-0.0473,0.0646) 
\pstThreeDLine(-0.3757,0.0990,0.1322)(-0.4188,0.0934,0.0241) 
\pstThreeDLine(-0.3757,0.0990,0.1322)(-0.3776,0.0170,0.1198) 
\pstThreeDLine(-0.4374,0.0304,0.0586)(-0.3166,-0.0438,0.1285) 
\pstThreeDLine(-0.4068,0.0142,0.1478)(-0.3143,-0.1175,0.0845) 
\pstThreeDLine(-0.4068,0.0142,0.1478)(-0.2436,-0.0325,0.1106) 
\pstThreeDLine(-0.4068,0.0142,0.1478)(-0.2283,-0.0290,0.1351) 
\pstThreeDLine(-0.4438,0.0797,0.0084)(-0.3166,-0.0438,0.1285) 
\pstThreeDLine(-0.4438,0.0797,0.0084)(-0.2598,-0.1947,-0.0174) 
\pstThreeDLine(-0.4438,0.0797,0.0084)(-0.2436,-0.0325,0.1106) 
\pstThreeDLine(-0.3143,-0.1175,0.0845)(-0.3166,-0.0438,0.1285) 
\pstThreeDLine(-0.3813,-0.0473,0.0646)(-0.3166,-0.0438,0.1285) 
\pstThreeDLine(-0.3813,-0.0473,0.0646)(-0.4188,0.0934,0.0241) 
\pstThreeDLine(-0.3813,-0.0473,0.0646)(-0.3776,0.0170,0.1198) 
\pstThreeDLine(-0.3813,-0.0473,0.0646)(-0.2598,-0.1947,-0.0174) 
\pstThreeDLine(-0.3813,-0.0473,0.0646)(-0.2436,-0.0325,0.1106) 
\pstThreeDLine(-0.3813,-0.0473,0.0646)(-0.2283,-0.0290,0.1351) 
\pstThreeDLine(-0.3166,-0.0438,0.1285)(-0.4188,0.0934,0.0241) 
\pstThreeDLine(-0.3059,0.0266,-0.0979)(-0.4573,0.0026,0.0890) 
\pstThreeDLine(-0.3059,0.0266,-0.0979)(-0.4269,0.1338,0.0221) 
\pstThreeDLine(-0.3059,0.0266,-0.0979)(-0.4438,0.0797,0.0084) 
\pstThreeDLine(-0.3059,0.0266,-0.0979)(-0.3813,-0.0473,0.0646) 
\pstThreeDLine(0.1027,-0.3304,-0.2516)(-0.0560,-0.3811,-0.1022) 
\pstThreeDLine(0.1027,-0.3304,-0.2516)(0.0485,-0.4309,-0.2026) 
\pstThreeDLine(-0.0560,-0.3811,-0.1022)(0.0485,-0.4309,-0.2026) 
\pstThreeDDot[linecolor=orange,linewidth=1.2pt](-0.3058,0.4366,-0.2130) 
\pstThreeDDot[linecolor=orange,linewidth=1.2pt](-0.2370,0.3880,-0.2352) 
\pstThreeDDot[linecolor=orange,linewidth=1.2pt](-0.3105,0.3550,-0.2430) 
\pstThreeDDot[linecolor=orange,linewidth=1.2pt](-0.1427,0.2351,-0.2382) 
\pstThreeDDot[linecolor=green,linewidth=1.2pt](-0.1364,0.2466,-0.0671) 
\pstThreeDDot[linecolor=green,linewidth=1.2pt](-0.1755,0.1852,-0.2335) 
\pstThreeDDot[linecolor=blue,linewidth=1.2pt](-0.4188,0.1322,0.0751) 
\pstThreeDDot[linecolor=blue,linewidth=1.2pt](-0.2113,-0.0210,0.3142) 
\pstThreeDDot[linecolor=blue,linewidth=1.2pt](-0.4573,0.0026,0.0890) 
\pstThreeDDot[linecolor=blue,linewidth=1.2pt](-0.4269,0.1338,0.0221) 
\pstThreeDDot[linecolor=blue,linewidth=1.2pt](-0.3757,0.0990,0.1322) 
\pstThreeDDot[linecolor=blue,linewidth=1.2pt](-0.4374,0.0304,0.0586) 
\pstThreeDDot[linecolor=blue,linewidth=1.2pt](-0.2235,0.0807,0.2188) 
\pstThreeDDot[linecolor=blue,linewidth=1.2pt](-0.4068,0.0142,0.1478) 
\pstThreeDDot[linecolor=blue,linewidth=1.2pt](-0.4438,0.0797,0.0084) 
\pstThreeDDot[linecolor=blue,linewidth=1.2pt](-0.3143,-0.1175,0.0845) 
\pstThreeDDot[linecolor=blue,linewidth=1.2pt](-0.3813,-0.0473,0.0646) 
\pstThreeDDot[linecolor=blue,linewidth=1.2pt](-0.3166,-0.0438,0.1285) 
\pstThreeDDot[linecolor=blue,linewidth=1.2pt](-0.4188,0.0934,0.0241) 
\pstThreeDDot[linecolor=blue,linewidth=1.2pt](-0.3776,0.0170,0.1198) 
\pstThreeDDot[linecolor=blue,linewidth=1.2pt](-0.2598,-0.1947,-0.0174) 
\pstThreeDDot[linecolor=blue,linewidth=1.2pt](-0.2436,-0.0325,0.1106) 
\pstThreeDDot[linecolor=brown,linewidth=1.2pt](-0.2283,-0.0290,0.1351) 
\pstThreeDDot[linecolor=red,linewidth=1.2pt](0.1027,-0.3304,-0.2516) 
\pstThreeDDot[linecolor=red,linewidth=1.2pt](-0.0560,-0.3811,-0.1022) 
\pstThreeDDot[linecolor=red,linewidth=1.2pt](0.0485,-0.4309,-0.2026) 
\pstThreeDDot[linecolor=magenta,linewidth=1.2pt](-0.3059,0.0266,-0.0979) 
\pstThreeDPut(-0.3058,0.4366,-0.2430){S\&P}
\pstThreeDPut(-0.2370,0.3880,-0.2652){Nasd}
\pstThreeDPut(-0.3105,0.3550,-0.2730){Cana}
\pstThreeDPut(-0.1427,0.2351,-0.2682){Mexi}
\pstThreeDPut(-0.1364,0.2466,-0.0971){Braz}
\pstThreeDPut(-0.1755,0.1852,-0.2635){Arge}
\pstThreeDPut(-0.4188,0.1322,0.1051){UK}
\pstThreeDPut(-0.2113,-0.0210,0.3442){Irel}
\pstThreeDPut(-0.4573,0.0026,0.1190){Fran}
\pstThreeDPut(-0.4269,0.1338,0.0521){Germ}
\pstThreeDPut(-0.3757,0.0990,0.1622){Swit}
\pstThreeDPut(-0.4374,0.0304,0.0886){Ital}
\pstThreeDPut(-0.2235,0.0807,0.2488){Belg}
\pstThreeDPut(-0.4068,0.0142,0.1778){Neth}
\pstThreeDPut(-0.4438,0.0797,0.0384){Swed}
\pstThreeDPut(-0.3143,-0.1175,0.1145){Denm}
\pstThreeDPut(-0.3813,-0.0473,0.0946){Finl}
\pstThreeDPut(-0.3166,-0.0438,0.1585){Norw}
\pstThreeDPut(-0.4188,0.0934,0.0541){Spai}
\pstThreeDPut(-0.3776,0.0170,0.1498){Port}
\pstThreeDPut(-0.2598,-0.1947,-0.0474){CzRe}
\pstThreeDPut(-0.2436,-0.0325,0.1406){Hung}
\pstThreeDPut(-0.2283,-0.0290,0.1651){Isra}
\pstThreeDPut(0.1027,-0.3304,-0.2816){Japa}
\pstThreeDPut(-0.0560,-0.3811,-0.1322){HoKo}
\pstThreeDPut(0.0485,-0.4309,-0.2326){SoKo}
\pstThreeDPut(-0.3059,0.0266,-0.1279){SoAf}
\end{pspicture}
\begin{pspicture}(-8,-2.3)(1,3.9)
\rput(-2.5,2.7){T=0.6}
\psline(-3.4,-2.7)(4.5,-2.7)(4.5,3.2)(-3.4,3.2)(-3.4,-2.7)
\psset{xunit=7,yunit=7,Alpha=40,Beta=0} \scriptsize
\pstThreeDLine(-0.3058,0.4366,-0.2130)(-0.2370,0.3880,-0.2352) 
\pstThreeDLine(-0.4573,0.0026,0.0890)(-0.4374,0.0304,0.0586) 
\pstThreeDLine(-0.4573,0.0026,0.0890)(-0.4068,0.0142,0.1478) 
\pstThreeDLine(-0.3058,0.4366,-0.2130)(-0.3105,0.3550,-0.2430) 
\pstThreeDLine(-0.2370,0.3880,-0.2352)(-0.3105,0.3550,-0.2430) 
\pstThreeDLine(-0.4188,0.1322,0.0751)(-0.4573,0.0026,0.0890) 
\pstThreeDLine(-0.4573,0.0026,0.0890)(-0.4269,0.1338,0.0221) 
\pstThreeDLine(-0.4573,0.0026,0.0890)(-0.4438,0.0797,0.0084) 
\pstThreeDLine(-0.4573,0.0026,0.0890)(-0.3813,-0.0473,0.0646) 
\pstThreeDLine(-0.4573,0.0026,0.0890)(-0.4188,0.0934,0.0241) 
\pstThreeDLine(-0.4269,0.1338,0.0221)(-0.4374,0.0304,0.0586) 
\pstThreeDLine(-0.4269,0.1338,0.0221)(-0.4068,0.0142,0.1478) 
\pstThreeDLine(-0.4269,0.1338,0.0221)(-0.4438,0.0797,0.0084) 
\pstThreeDLine(-0.4269,0.1338,0.0221)(-0.4188,0.0934,0.0241) 
\pstThreeDLine(-0.4374,0.0304,0.0586)(-0.4068,0.0142,0.1478) 
\pstThreeDLine(-0.4374,0.0304,0.0586)(-0.4438,0.0797,0.0084) 
\pstThreeDLine(-0.4374,0.0304,0.0586)(-0.4188,0.0934,0.0241) 
\pstThreeDLine(-0.4438,0.0797,0.0084)(-0.3813,-0.0473,0.0646) 
\pstThreeDLine(-0.4188,0.1322,0.0751)(-0.4269,0.1338,0.0221) 
\pstThreeDLine(-0.4188,0.1322,0.0751)(-0.4374,0.0304,0.0586) 
\pstThreeDLine(-0.4188,0.1322,0.0751)(-0.4068,0.0142,0.1478) 
\pstThreeDLine(-0.4188,0.1322,0.0751)(-0.4438,0.0797,0.0084) 
\pstThreeDLine(-0.4188,0.1322,0.0751)(-0.3813,-0.0473,0.0646) 
\pstThreeDLine(-0.4188,0.1322,0.0751)(-0.3166,-0.0438,0.1285) 
\pstThreeDLine(-0.4188,0.1322,0.0751)(-0.4188,0.0934,0.0241) 
\pstThreeDLine(-0.4573,0.0026,0.0890)(-0.3166,-0.0438,0.1285) 
\pstThreeDLine(-0.4573,0.0026,0.0890)(-0.3166,-0.0438,0.1285) 
\pstThreeDLine(-0.4573,0.0026,0.0890)(-0.3776,0.0170,0.1198) 
\pstThreeDLine(-0.4269,0.1338,0.0221)(-0.3166,-0.0438,0.1285) 
\pstThreeDLine(-0.4269,0.1338,0.0221)(-0.3813,-0.0473,0.0646) 
\pstThreeDLine(-0.3166,-0.0438,0.1285)(-0.3166,-0.0438,0.1285) 
\pstThreeDLine(-0.4068,0.0142,0.1478)(-0.3166,-0.0438,0.1285) 
\pstThreeDLine(-0.4374,0.0304,0.0586)(-0.3813,-0.0473,0.0646) 
\pstThreeDLine(-0.4374,0.0304,0.0586)(-0.3776,0.0170,0.1198) 
\pstThreeDLine(-0.2235,0.0807,0.2188)(-0.3166,-0.0438,0.1285) 
\pstThreeDLine(-0.3166,-0.0438,0.1285)(-0.4438,0.0797,0.0084) 
\pstThreeDLine(-0.3166,-0.0438,0.1285)(-0.3813,-0.0473,0.0646) 
\pstThreeDLine(-0.3166,-0.0438,0.1285)(-0.3166,-0.0438,0.1285) 
\pstThreeDLine(-0.3166,-0.0438,0.1285)(-0.4188,0.0934,0.0241) 
\pstThreeDLine(-0.3166,-0.0438,0.1285)(-0.3776,0.0170,0.1198) 
\pstThreeDLine(-0.3776,0.0170,0.1198)(-0.4188,0.0934,0.0241) 
\pstThreeDLine(-0.3776,0.0170,0.1198)(-0.3776,0.0170,0.1198) 
\pstThreeDLine(-0.4188,0.0934,0.0241)(-0.3776,0.0170,0.1198) 
\pstThreeDLine(-0.3058,0.4366,-0.2130)(-0.1427,0.2351,-0.2382) 
\pstThreeDLine(-0.3058,0.4366,-0.2130)(-0.4269,0.1338,0.0221) 
\pstThreeDLine(-0.3105,0.3550,-0.2430)(-0.1427,0.2351,-0.2382) 
\pstThreeDLine(-0.3105,0.3550,-0.2430)(-0.4573,0.0026,0.0890) 
\pstThreeDLine(-0.3105,0.3550,-0.2430)(-0.4269,0.1338,0.0221) 
\pstThreeDLine(-0.1364,0.2466,-0.0671)(-0.1755,0.1852,-0.2335) 
\pstThreeDLine(-0.4188,0.1322,0.0751)(-0.4188,0.0934,0.0241) 
\pstThreeDLine(-0.4188,0.1322,0.0751)(-0.3143,-0.1175,0.0845) 
\pstThreeDLine(-0.4188,0.1322,0.0751)(-0.3776,0.0170,0.1198) 
\pstThreeDLine(-0.2113,-0.0210,0.3142)(-0.3143,-0.1175,0.0845) 
\pstThreeDLine(-0.4573,0.0026,0.0890)(-0.3143,-0.1175,0.0845) 
\pstThreeDLine(-0.4573,0.0026,0.0890)(-0.2436,-0.0325,0.1106) 
\pstThreeDLine(-0.4573,0.0026,0.0890)(-0.2283,-0.0290,0.1351) 
\pstThreeDLine(-0.4269,0.1338,0.0221)(-0.3143,-0.1175,0.0845) 
\pstThreeDLine(-0.4269,0.1338,0.0221)(-0.3166,-0.0438,0.1285) 
\pstThreeDLine(-0.4269,0.1338,0.0221)(-0.3776,0.0170,0.1198) 
\pstThreeDLine(-0.3757,0.0990,0.1322)(-0.4374,0.0304,0.0586) 
\pstThreeDLine(-0.3757,0.0990,0.1322)(-0.2235,0.0807,0.2188) 
\pstThreeDLine(-0.3757,0.0990,0.1322)(-0.4438,0.0797,0.0084) 
\pstThreeDLine(-0.3757,0.0990,0.1322)(-0.3143,-0.1175,0.0845) 
\pstThreeDLine(-0.3757,0.0990,0.1322)(-0.3813,-0.0473,0.0646) 
\pstThreeDLine(-0.3757,0.0990,0.1322)(-0.4188,0.0934,0.0241) 
\pstThreeDLine(-0.3757,0.0990,0.1322)(-0.3776,0.0170,0.1198) 
\pstThreeDLine(-0.4374,0.0304,0.0586)(-0.3166,-0.0438,0.1285) 
\pstThreeDLine(-0.4068,0.0142,0.1478)(-0.3143,-0.1175,0.0845) 
\pstThreeDLine(-0.4068,0.0142,0.1478)(-0.2436,-0.0325,0.1106) 
\pstThreeDLine(-0.4068,0.0142,0.1478)(-0.2283,-0.0290,0.1351) 
\pstThreeDLine(-0.4438,0.0797,0.0084)(-0.3166,-0.0438,0.1285) 
\pstThreeDLine(-0.4438,0.0797,0.0084)(-0.2598,-0.1947,-0.0174) 
\pstThreeDLine(-0.4438,0.0797,0.0084)(-0.2436,-0.0325,0.1106) 
\pstThreeDLine(-0.3143,-0.1175,0.0845)(-0.3166,-0.0438,0.1285) 
\pstThreeDLine(-0.3813,-0.0473,0.0646)(-0.3166,-0.0438,0.1285) 
\pstThreeDLine(-0.3813,-0.0473,0.0646)(-0.4188,0.0934,0.0241) 
\pstThreeDLine(-0.3813,-0.0473,0.0646)(-0.3776,0.0170,0.1198) 
\pstThreeDLine(-0.3813,-0.0473,0.0646)(-0.2598,-0.1947,-0.0174) 
\pstThreeDLine(-0.3813,-0.0473,0.0646)(-0.2436,-0.0325,0.1106) 
\pstThreeDLine(-0.3813,-0.0473,0.0646)(-0.2283,-0.0290,0.1351) 
\pstThreeDLine(-0.3166,-0.0438,0.1285)(-0.4188,0.0934,0.0241) 
\pstThreeDLine(-0.3059,0.0266,-0.0979)(-0.4573,0.0026,0.0890) 
\pstThreeDLine(-0.3059,0.0266,-0.0979)(-0.4269,0.1338,0.0221) 
\pstThreeDLine(-0.3059,0.0266,-0.0979)(-0.4438,0.0797,0.0084) 
\pstThreeDLine(-0.3059,0.0266,-0.0979)(-0.3813,-0.0473,0.0646) 
\pstThreeDLine(0.1027,-0.3304,-0.2516)(-0.0560,-0.3811,-0.1022) 
\pstThreeDLine(0.1027,-0.3304,-0.2516)(0.0485,-0.4309,-0.2026) 
\pstThreeDLine(-0.0560,-0.3811,-0.1022)(0.0485,-0.4309,-0.2026) 
\pstThreeDLine(-0.3058,0.4366,-0.2130)(-0.1755,0.1852,-0.2335) 
\pstThreeDLine(-0.3058,0.4366,-0.2130)(-0.4188,0.1322,0.0751) 
\pstThreeDLine(-0.3058,0.4366,-0.2130)(-0.4573,0.0026,0.0890) 
\pstThreeDLine(-0.3058,0.4366,-0.2130)(-0.3757,0.0990,0.1322) 
\pstThreeDLine(-0.3058,0.4366,-0.2130)(-0.4374,0.0304,0.0586) 
\pstThreeDLine(-0.3058,0.4366,-0.2130)(-0.4438,0.0797,0.0084) 
\pstThreeDLine(-0.3058,0.4366,-0.2130)(-0.4188,0.0934,0.0241) 
\pstThreeDLine(-0.2370,0.3880,-0.2352)(-0.1427,0.2351,-0.2382) 
\pstThreeDLine(-0.2370,0.3880,-0.2352)(-0.4269,0.1338,0.0221) 
\pstThreeDLine(-0.3058,0.4366,-0.2130)(-0.1427,0.2351,-0.2382) 
\pstThreeDLine(-0.3105,0.3550,-0.2430)(-0.2426,-0.0812,-0.2983) 
\pstThreeDLine(-0.3105,0.3550,-0.2430)(-0.4188,0.1322,0.0751) 
\pstThreeDLine(-0.3105,0.3550,-0.2430)(-0.3757,0.0990,0.1322) 
\pstThreeDLine(-0.3105,0.3550,-0.2430)(-0.4374,0.0304,0.0586) 
\pstThreeDLine(-0.3105,0.3550,-0.2430)(-0.4068,0.0142,0.1478) 
\pstThreeDLine(-0.3105,0.3550,-0.2430)(-0.4438,0.0797,0.0084) 
\pstThreeDLine(-0.3105,0.3550,-0.2430)(-0.4188,0.0934,0.0241) 
\pstThreeDLine(-0.1427,0.2351,-0.2382)(-0.1364,0.2466,-0.0671) 
\pstThreeDLine(-0.1427,0.2351,-0.2382)(-0.1755,0.1852,-0.2335) 
\pstThreeDLine(-0.1364,0.2466,-0.0671)(-0.4269,0.1338,0.0221) 
\pstThreeDLine(-0.1755,0.1852,-0.2335)(-0.4269,0.1338,0.0221) 
\pstThreeDLine(-0.1755,0.1852,-0.2335)(-0.4188,0.0934,0.0241) 
\pstThreeDLine(-0.2426,-0.0812,-0.2983)(-0.4573,0.0026,0.0890) 
\pstThreeDLine(-0.2426,-0.0812,-0.2983)(-0.4269,0.1338,0.0221) 
\pstThreeDLine(-0.2426,-0.0812,-0.2983)(-0.4374,0.0304,0.0586) 
\pstThreeDLine(-0.2426,-0.0812,-0.2983)(-0.4068,0.0142,0.1478) 
\pstThreeDLine(-0.2426,-0.0812,-0.2983)(-0.4438,0.0797,0.0084) 
\pstThreeDLine(-0.2426,-0.0812,-0.2983)(-0.4188,0.0934,0.0241) 
\pstThreeDLine(-0.2426,-0.0812,-0.2983)(-0.3776,0.0170,0.1198) 
\pstThreeDLine(-0.4188,0.1322,0.0751)(-0.2113,-0.0210,0.3142) 
\pstThreeDLine(-0.4188,0.1322,0.0751)(-0.2235,0.0807,0.2188) 
\pstThreeDLine(-0.4188,0.1322,0.0751)(-0.2283,-0.0290,0.1351) 
\pstThreeDLine(-0.2113,-0.0210,0.3142)(-0.4573,0.0026,0.0890) 
\pstThreeDLine(-0.2113,-0.0210,0.3142)(-0.3757,0.0990,0.1322) 
\pstThreeDLine(-0.2113,-0.0210,0.3142)(-0.2235,0.0807,0.2188) 
\pstThreeDLine(-0.2113,-0.0210,0.3142)(-0.4068,0.0142,0.1478) 
\pstThreeDLine(-0.2113,-0.0210,0.3142)(-0.4438,0.0797,0.0084) 
\pstThreeDLine(-0.2113,-0.0210,0.3142)(-0.3166,-0.0438,0.1285) 
\pstThreeDLine(-0.4573,0.0026,0.0890)(-0.1827,0.0687,0.1471) 
\pstThreeDLine(-0.4573,0.0026,0.0890)(-0.2235,0.0807,0.2188) 
\pstThreeDLine(-0.4573,0.0026,0.0890)(-0.1431,-0.2751,-0.0514) 
\pstThreeDLine(-0.4573,0.0026,0.0890)(-0.2598,-0.1947,-0.0174) 
\pstThreeDLine(-0.4573,0.0026,0.0890)(-0.2753,-0.0948,-0.0362) 
\pstThreeDLine(-0.4269,0.1338,0.0221)(-0.2235,0.0807,0.2188) 
\pstThreeDLine(-0.4269,0.1338,0.0221)(-0.2598,-0.1947,-0.0174) 
\pstThreeDLine(-0.4269,0.1338,0.0221)(-0.2436,-0.0325,0.1106) 
\pstThreeDLine(-0.4269,0.1338,0.0221)(-0.2283,-0.0290,0.1351) 
\pstThreeDLine(-0.3757,0.0990,0.1322)(-0.2436,-0.0325,0.1106) 
\pstThreeDLine(-0.3757,0.0990,0.1322)(-0.2283,-0.0290,0.1351) 
\pstThreeDLine(-0.1827,0.0687,0.1471)(-0.4068,0.0142,0.1478) 
\pstThreeDLine(-0.1827,0.0687,0.1471)(-0.4188,0.0934,0.0241) 
\pstThreeDLine(-0.4374,0.0304,0.0586)(-0.2235,0.0807,0.2188) 
\pstThreeDLine(-0.4374,0.0304,0.0586)(-0.3143,-0.1175,0.0845) 
\pstThreeDLine(-0.4374,0.0304,0.0586)(-0.2598,-0.1947,-0.0174) 
\pstThreeDLine(-0.4374,0.0304,0.0586)(-0.2436,-0.0325,0.1106) 
\pstThreeDLine(-0.4374,0.0304,0.0586)(-0.2283,-0.0290,0.1351) 
\pstThreeDLine(-0.2235,0.0807,0.2188)(-0.4438,0.0797,0.0084) 
\pstThreeDLine(-0.2235,0.0807,0.2188)(-0.3166,-0.0438,0.1285) 
\pstThreeDLine(-0.2235,0.0807,0.2188)(-0.4188,0.0934,0.0241) 
\pstThreeDLine(-0.4068,0.0142,0.1478)(-0.1431,-0.2751,-0.0514) 
\pstThreeDLine(-0.4068,0.0142,0.1478)(-0.2598,-0.1947,-0.0174) 
\pstThreeDLine(-0.4068,0.0142,0.1478)(-0.2753,-0.0948,-0.0362) 
\pstThreeDLine(-0.4438,0.0797,0.0084)(-0.3143,-0.1175,0.0845) 
\pstThreeDLine(-0.4438,0.0797,0.0084)(-0.2753,-0.0948,-0.0362) 
\pstThreeDLine(-0.4438,0.0797,0.0084)(-0.2283,-0.0290,0.1351) 
\pstThreeDLine(-0.3143,-0.1175,0.0845)(-0.3813,-0.0473,0.0646) 
\pstThreeDLine(-0.3143,-0.1175,0.0845)(-0.4188,0.0934,0.0241) 
\pstThreeDLine(-0.3143,-0.1175,0.0845)(-0.3776,0.0170,0.1198) 
\pstThreeDLine(-0.3143,-0.1175,0.0845)(-0.2436,-0.0325,0.1106) 
\pstThreeDLine(-0.3813,-0.0473,0.0646)(-0.2753,-0.0948,-0.0362) 
\pstThreeDLine(-0.3813,-0.0473,0.0646)(-0.1353,-0.0787,0.0476) 
\pstThreeDLine(-0.3166,-0.0438,0.1285)(-0.3776,0.0170,0.1198) 
\pstThreeDLine(-0.3166,-0.0438,0.1285)(-0.1253,0.0695,0.1526) 
\pstThreeDLine(-0.3166,-0.0438,0.1285)(-0.2598,-0.1947,-0.0174) 
\pstThreeDLine(-0.4188,0.0934,0.0241)(-0.2598,-0.1947,-0.0174) 
\pstThreeDLine(-0.4188,0.0934,0.0241)(-0.2283,-0.0290,0.1351) 
\pstThreeDLine(-0.3776,0.0170,0.1198)(-0.2436,-0.0325,0.1106) 
\pstThreeDLine(-0.2598,-0.1947,-0.0174)(-0.2436,-0.0325,0.1106) 
\pstThreeDLine(-0.2598,-0.1947,-0.0174)(-0.2753,-0.0948,-0.0362) 
\pstThreeDLine(-0.2598,-0.1947,-0.0174)(-0.1353,-0.0787,0.0476) 
\pstThreeDLine(-0.2598,-0.1947,-0.0174)(-0.0560,-0.3811,-0.1022) 
\pstThreeDLine(-0.2436,-0.0325,0.1106)(-0.2753,-0.0948,-0.0362) 
\pstThreeDLine(-0.2436,-0.0325,0.1106)(-0.2283,-0.0290,0.1351) 
\pstThreeDLine(-0.2753,-0.0948,-0.0362)(-0.1353,-0.0787,0.0476) 
\pstThreeDLine(-0.0484,-0.3222,0.1614)(-0.4068,0.0142,0.1478) 
\pstThreeDLine(-0.0484,-0.3222,0.1614)(-0.0560,-0.3811,-0.1022) 
\pstThreeDLine(-0.0484,-0.3222,0.1614)(0.0485,-0.4309,-0.2026) 
\pstThreeDLine(-0.0484,-0.3222,0.1614)(-0.0481,-0.2618,-0.0828) 
\pstThreeDLine(-0.3059,0.0266,-0.0979)(-0.4188,0.1322,0.0751) 
\pstThreeDLine(-0.3059,0.0266,-0.0979)(-0.3757,0.0990,0.1322) 
\pstThreeDLine(-0.3059,0.0266,-0.0979)(-0.4374,0.0304,0.0586) 
\pstThreeDLine(-0.3059,0.0266,-0.0979)(-0.4068,0.0142,0.1478) 
\pstThreeDLine(-0.3059,0.0266,-0.0979)(-0.3166,-0.0438,0.1285) 
\pstThreeDLine(-0.3059,0.0266,-0.0979)(-0.4188,0.0934,0.0241) 
\pstThreeDLine(-0.3059,0.0266,-0.0979)(-0.3776,0.0170,0.1198) 
\pstThreeDLine(-0.3059,0.0266,-0.0979)(-0.2598,-0.1947,-0.0174) 
\pstThreeDLine(-0.3059,0.0266,-0.0979)(-0.2753,-0.0948,-0.0362) 
\pstThreeDLine(-0.3059,0.0266,-0.0979)(-0.1353,-0.0787,0.0476) 
\pstThreeDDot[linecolor=orange,linewidth=1.2pt](-0.3058,0.4366,-0.2130) 
\pstThreeDDot[linecolor=orange,linewidth=1.2pt](-0.2370,0.3880,-0.2352) 
\pstThreeDDot[linecolor=orange,linewidth=1.2pt](-0.3105,0.3550,-0.2430) 
\pstThreeDDot[linecolor=orange,linewidth=1.2pt](-0.1427,0.2351,-0.2382) 
\pstThreeDDot[linecolor=green,linewidth=1.2pt](-0.1364,0.2466,-0.0671) 
\pstThreeDDot[linecolor=green,linewidth=1.2pt](-0.1755,0.1852,-0.2335) 
\pstThreeDDot[linecolor=green,linewidth=1.2pt](-0.2426,-0.0812,-0.2983) 
\pstThreeDDot[linecolor=blue,linewidth=1.2pt](-0.4188,0.1322,0.0751) 
\pstThreeDDot[linecolor=blue,linewidth=1.2pt](-0.2113,-0.0210,0.3142) 
\pstThreeDDot[linecolor=blue,linewidth=1.2pt](-0.4573,0.0026,0.0890) 
\pstThreeDDot[linecolor=blue,linewidth=1.2pt](-0.4269,0.1338,0.0221) 
\pstThreeDDot[linecolor=blue,linewidth=1.2pt](-0.3757,0.0990,0.1322) 
\pstThreeDDot[linecolor=blue,linewidth=1.2pt](-0.1827,0.0687,0.1471) 
\pstThreeDDot[linecolor=blue,linewidth=1.2pt](-0.4374,0.0304,0.0586) 
\pstThreeDDot[linecolor=blue,linewidth=1.2pt](-0.2235,0.0807,0.2188) 
\pstThreeDDot[linecolor=blue,linewidth=1.2pt](-0.4068,0.0142,0.1478) 
\pstThreeDDot[linecolor=blue,linewidth=1.2pt](-0.1431,-0.2751,-0.0514) 
\pstThreeDDot[linecolor=blue,linewidth=1.2pt](-0.4438,0.0797,0.0084) 
\pstThreeDDot[linecolor=blue,linewidth=1.2pt](-0.3143,-0.1175,0.0845) 
\pstThreeDDot[linecolor=blue,linewidth=1.2pt](-0.3813,-0.0473,0.0646) 
\pstThreeDDot[linecolor=blue,linewidth=1.2pt](-0.3166,-0.0438,0.1285) 
\pstThreeDDot[linecolor=blue,linewidth=1.2pt](-0.4188,0.0934,0.0241) 
\pstThreeDDot[linecolor=blue,linewidth=1.2pt](-0.3776,0.0170,0.1198) 
\pstThreeDDot[linecolor=blue,linewidth=1.2pt](-0.1253,0.0695,0.1526) 
\pstThreeDDot[linecolor=blue,linewidth=1.2pt](-0.2598,-0.1947,-0.0174) 
\pstThreeDDot[linecolor=blue,linewidth=1.2pt](-0.2436,-0.0325,0.1106) 
\pstThreeDDot[linecolor=blue,linewidth=1.2pt](-0.2753,-0.0948,-0.0362) 
\pstThreeDDot[linecolor=bluish,linewidth=1.2pt](-0.1353,-0.0787,0.0476) 
\pstThreeDDot[linecolor=brown,linewidth=1.2pt](-0.2283,-0.0290,0.1351) 
\pstThreeDDot[linecolor=red,linewidth=1.2pt](0.1027,-0.3304,-0.2516) 
\pstThreeDDot[linecolor=red,linewidth=1.2pt](-0.0560,-0.3811,-0.1022) 
\pstThreeDDot[linecolor=red,linewidth=1.2pt](0.0485,-0.4309,-0.2026) 
\pstThreeDDot[linecolor=red,linewidth=1.2pt](-0.0484,-0.3222,0.1614) 
\pstThreeDDot[linecolor=black,linewidth=1.2pt](-0.0481,-0.2618,-0.0828) 
\pstThreeDDot[linecolor=magenta,linewidth=1.2pt](-0.3059,0.0266,-0.0979) 
\pstThreeDPut(-0.3058,0.4366,-0.2430){S\&P}
\pstThreeDPut(-0.2370,0.3880,-0.2652){Nasd}
\pstThreeDPut(-0.3105,0.3550,-0.2730){Cana}
\pstThreeDPut(-0.1427,0.2351,-0.2682){Mexi}
\pstThreeDPut(-0.1364,0.2466,-0.0971){Braz}
\pstThreeDPut(-0.1755,0.1852,-0.2635){Arge}
\pstThreeDPut(-0.2426,-0.0812,-0.3283){Chil}
\pstThreeDPut(-0.4188,0.1322,0.1051){UK}
\pstThreeDPut(-0.2113,-0.0210,0.3442){Irel}
\pstThreeDPut(-0.4573,0.0026,0.1190){Fran}
\pstThreeDPut(-0.4269,0.1338,0.0521){Germ}
\pstThreeDPut(-0.3757,0.0990,0.1622){Swit}
\pstThreeDPut(-0.1827,0.0687,0.1771){Autr}
\pstThreeDPut(-0.4374,0.0304,0.0886){Ital}
\pstThreeDPut(-0.2235,0.0807,0.2488){Belg}
\pstThreeDPut(-0.4068,0.0142,0.1778){Neth}
\pstThreeDPut(-0.1431,-0.2751,-0.0814){Luxe}
\pstThreeDPut(-0.4438,0.0797,0.0384){Swed}
\pstThreeDPut(-0.3143,-0.1175,0.1145){Denm}
\pstThreeDPut(-0.3813,-0.0473,0.0946){Finl}
\pstThreeDPut(-0.3166,-0.0438,0.1585){Norw}
\pstThreeDPut(-0.4188,0.0934,0.0541){Spai}
\pstThreeDPut(-0.3776,0.0170,0.1498){Port}
\pstThreeDPut(-0.1253,0.0695,0.1826){Gree}
\pstThreeDPut(-0.2598,-0.1947,-0.0474){CzRe}
\pstThreeDPut(-0.2436,-0.0325,0.1406){Hung}
\pstThreeDPut(-0.2753,-0.0948,-0.0662){Pola}
\pstThreeDPut(-0.1353,-0.0787,0.0776){Russ}
\pstThreeDPut(-0.2283,-0.0290,0.1651){Isra}
\pstThreeDPut(0.1027,-0.3304,-0.2816){Japa}
\pstThreeDPut(-0.0560,-0.3811,-0.1322){HoKo}
\pstThreeDPut(0.0485,-0.4309,-0.2326){SoKo}
\pstThreeDPut(-0.0484,-0.3222,0.1914){Sing}
\pstThreeDPut(-0.0481,-0.2618,-0.1128){NZ}
\pstThreeDPut(-0.3059,0.0266,-0.1279){SoAf}
\end{pspicture}

\vskip 0.7 cm

\noindent Fig. 11. Three dimensional view of the asset trees for the first semester of 2001, with threshold ranging from $T=0.3$ to $T=0.6$.

\vskip 0.3 cm

\subsubsection{Second semester, 2001}

Figure 12 shows the three dimensional view of the asset trees for the second semester of 2001, with threshold ranging from $T=0.3$ to $T=0.6$.

For the data concerning the second semester of 2001, for $T=0.1$, there is already a connection between France and Netherlands. At $T=0.2$, two small clusters are formed: S\&P and Nasdaq (North American cluster) and an European cluster comprised of the UK, France, Germany, Switzerland, Italy, Netherlands, Sweden, and Spain. At $T=0.3$, Canada joins the North American cluster, the European cluster becomes more densely knit, and Belgium and Finland join the fold. At $T=0.4$, the North American cluster joins with the European one via Canada, and the European cluster receives the addition of Portugal.

At $T=0.5$, there are many more connections formed: Mexico joins the North American cluster, and more connections are made between this cluster and the European one. Europe receives Ireland, Denmark, Norway, Greece, the Czech Republic, and Estonia. Strange connections form, like one between Sweden and Peru, and between Estonia and Hong Kong, what makes us think that random noise already has some effects here. At this same threshold, a Pacific Asian cluster is formed, already linked with Europe, comprised of Hong Kong, Taiwan, South Korea, and Singapore. Australia connects both with the Asian cluster and with the European cluster. For $T=0.6$, the threshold that we are considering as the limit to the region dominated by noise, Brazil connects with the North American cluster via Canada, and also connects itself with Chile, which establishes connections with Europe, while Peru connects with many European indices. More connections are formed between America and Europe, and the European cluster now adds to itself Luxembourg, Hungary, and Poland. Russia, Turkey, Israel, and South Africa connect with the European cluster, and Japan and India join the Pacific Asian one. At higher thresholds, an explosion of connections occur, some of them reinforcing already made clusters, some establishing more connections between them, and others connecting more indices to those clusters, many of them apparently at random.



\vskip 0.7 cm

\noindent Fig. 12. Three dimensional view of the asset trees for the second semester of 2001, with threshold ranging from $T=0.3$ to $T=0.6$.

\vskip 0.3 cm

\subsection{Overview}

By considering all the time periods being studied, the first fact that stands out is the presence of two cluster throughout all periods: the American and the European clusters. The connections between the S\&P 500 and Nasdaq are present at all times and at the smallest thresholds. What we call an European cluster also appears early on, but its core varies with time, beginning with the pair Germany-Netherlands and then evolving to a new core, consisting on the UK, France, Germany, Switzerland, Italy, Netherlands, Sweden, and Spain. It is interesting to notice that the UK, which was more connected to the American cluster in the 80's, moves to the European cluster later on.

At higher thresholds, the American cluster is joined by Canada and some South American indices, and the European cluster grows with the addition of some more indices of Scandinavian countries and of other Western European indices. As the thresholds grow, Europe is joined by Eastern European countries and only then we witness the formation of a Pacific Asian cluster, which starts to solidify after 1997, which coincides with the Asian Financial Crisis. South Africa and, in a lesser proportion, Israel, connect with Europe rather than with their neighbouring countries. Australia and New Zealand are more connected with Pacific Asia than with Europe. Indices from the Caribbean, of most islands, of the majority of Africa, and of the Arab countries, connect only at much higher values of the threshold, where noise reigns.

What is also clear from the graphics is that the financial indices tend to group according to the geographic proximity of their countries, and also according to cultural similarities, as are the case of South Africa and Israel with Europe. Other striking feature is that the American and European clusters are fairly independent, and only connect at higher thresholds, when both cluster have already grown larger and denser. It is good to remind ourselves that networks built from correlation matrices are not directed networks, so that we can deduct no effects of causality between the indices using just the correlation matrix (an alternative would be parcial correlation \cite{Dror1}-\cite{Dror4}).

Also to be noticed is that networks shrink in size in times of crises, reflecting the growth in correlation between markets in those times \cite{leocorr}.

\section{Information in the second largest eigenvalue}

It is well documented in the literature that the eigenvalues of the correlation matrix of the time series of assets may be used to identify part of the information carried by those time series separating it from noise. This is usually done using Random Matrix Theory \cite{rmt1}, that is based on the analysis of the eigenvalue frequency distribution of a matrix which is obtained from the correlation between $N$ time series of random numbers generated over a Gaussian frequency distribution with zero mean and standard deviation $\sigma $. If $L$ is the number of elements of each time series, then, in the limit $L\to \infty $ and $N\to \infty $ such that $Q=L/N$ is constant, finite and greater than one, then the probability distribution function of the eigenvalues of such a matrix is given by the expression
\begin{equation}
\label{dist}
\rho(\lambda )=\frac{Q}{2\pi \sigma ^2}\frac{\sqrt{(\lambda_+-\lambda )(\lambda -\lambda_-)}}{\lambda }\ ,
\end{equation}
where
\begin{equation}
\lambda_-=\sigma ^2\left( 1+\frac{1}{Q}-2\sqrt{\frac{1}{Q}}\right) \ \ ,\ \ \lambda_+=\sigma ^2\left( 1+\frac{1}{Q}+2\sqrt{\frac{1}{Q}}\right) \ ,
\end{equation}
and $\lambda $ is restricted to the interval $\left[ \lambda_-,\lambda_+\right] $.

This probability distribution function is called a Mar\v{c}enku-Pastur distribution \cite{rmt2}, and it establishes limits for the eigenvalues generated from random data. So, in theory, any eigenvalue falling out of the interval $\left[ \lambda_-,\lambda_+\right] $ has a good chance of representing true information about the system.

As stated before, the Mar\v{c}enku-Pastur distribution is valid only for the limit of an infinite amount of available data, and also for random time series generated over a Gaussian. Since this is not the case for real time series of financial data, for they are finite and usually their probability frequency distributions are not Gaussian, an alternative would be to use randomized data for a collection of time series and then analyze the minimum and maximum values of eigenvalues found for the correlation matrix thus generated. The result is similar to the one obtained for theoretical distributions, but not quite the same.

In this section, we use the results obtained from randomized data in order to establish regions where noise can be a major concern and then analize the eigenvectors of some eigenvalues that are outside those regions. Figures 13 to 15 show the eigenvalues (represented as lines) and the values associated with noise (shaded region) in the graphs for the years being studied in this article.

The first feature which is clearly visible is that the highest eigenvalue always stands out from the others. Each eigenvalue has a correponding eigenvector with the same number of indices used in creating the correlation matrix. Each entry in an eigenvector corresponds to the participation of that index in the building of a ``portfolio'' of indices. It is well known that the eigenvectors of the highest eigenvalues are usually built in such a way that every index has a similar participation in such a portfolio, and that this particular combination emulates the general behavior of a world market index \cite{leocorr}.



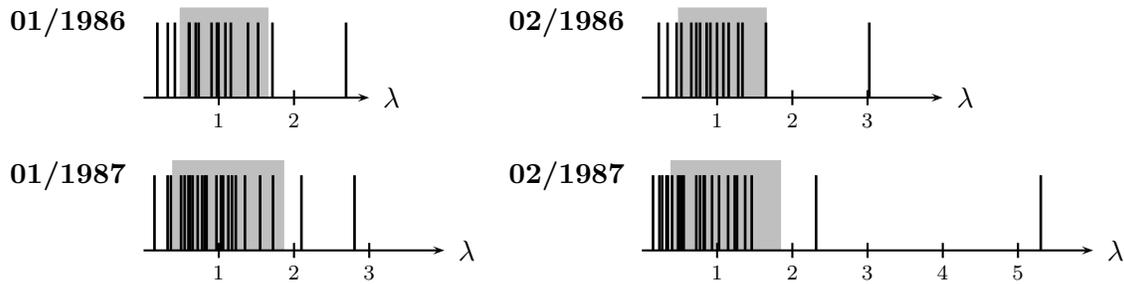
\noindent Fig. 13. Eigenvalues for the first and second semesters of 1986 and 1987 in order of magnitude. The shaded area corresponds to the eigenvalues predicted for randomized data.

\vskip 0.4 cm

%

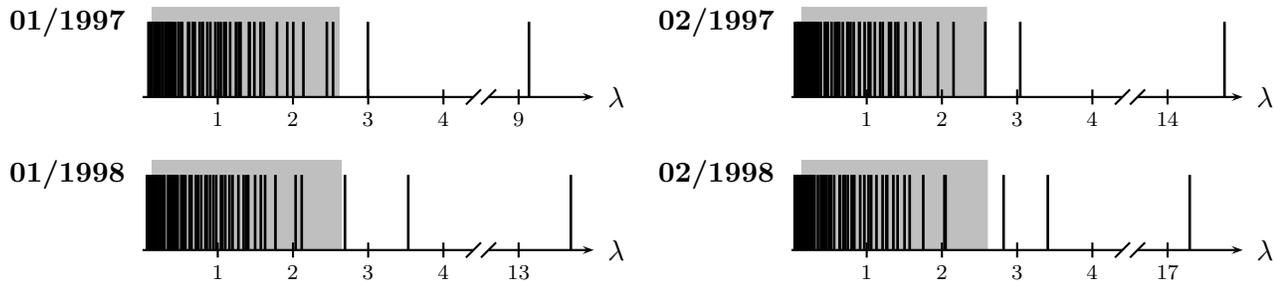
\noindent Fig. 14. Eigenvalues for the first and second semesters of 1997 and 1998 in order of magnitude. The shaded area corresponds to the eigenvalues predicted for randomized data.

\vskip 0.4 cm

%

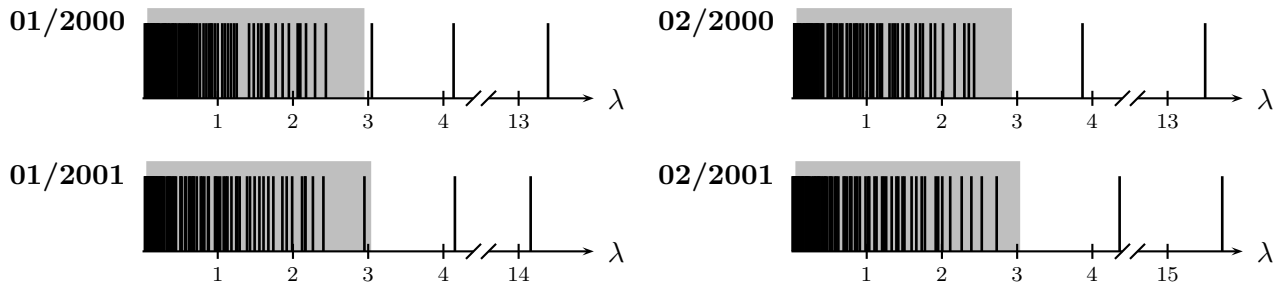
\noindent Fig. 15. Eigenvalues for the first and second semesters of 2000 and 2001 in order of magnitude. The shaded area corresponds to the eigenvalues predicted for randomized data.

\vskip 0.3 cm

To illustrate this, if one compares the log-returns of a portfolio consisting on the indices as expressed by the eingenvectors corresponding to the highest eigenvalues with the log-returns of the world index calculated by the MSCI (Morgan Stanley Capital International), then one obtains the correlations between those indices as stated in Table 1.

\[ \begin{array}{|c|cccc|cccc|} \hline \text{Semester/Year} & 01/1986 & 02/1986 & 01/1987 & 02/1987 & 01/1997 & 02/1997 & 01/1998 & 02/1998 \\ \hline \text{Correlation} & 0.69 & 0.60 & 0.40 & 0.87 & 0.64 & 0.81 & 0.83 & 0.82 \\ \hline \end{array} \]
\[ \begin{array}{|c|cccc|} \hline \text{Semester/Year} & 01/2000 & 02/2000 & 01/2001 & 02/2001 \\ \hline \text{Correlation} & 0.71 & 0.83 & 0.77 & 0.82 \\ \hline \end{array} \]

\noindent Table 1. Correlations between indices built on the eigenvector corresponding to the highest eigenvalue and the MSCI world index according to semester and year.

\vskip 0.3 cm

Although the correlation is weak for early years, probably due to the small number of indices that were considered for them, correlation is strong for subsequent years. So, the eigenvector corresponding to the highest eigenvalue is associated with a ``market mode'', or the general oscillations common to all indices.

Now, we move our attention to the second highest eigenvalue. Figures 13 to 15 show that such eigenvalue is also detached from the noisy region, although less so than the highest eigenvalue. This second highest eigenvalue has been connected with some internal structures of markets when dealing with assets from particular markets. For stock market indices, it has a different meaning that is peculiar to systems that operate at different times. Figures 16 to 21 show representations of the eigenvectors corresponding to the second highest eigenvalues for each of the intervals of time we are studying in this article. White rectangles correspond to positive values and dark rectangles correspond to negative values of the elements of the eigenvectors.



\noindent Fig. 16. Contributions of the stock market indices to eigenvector $e_{2}$, corresponding to the second largest eigenvalue of the correlation matrix. White bars indicate positive values, and gray bars indicate negative values, corresponding to the first and second semesters of 1986.

%

\noindent Fig. 17. Contributions of the stock market indices to eigenvector $e_{2}$, corresponding to the second largest eigenvalue of the correlation matrix. White bars indicate positive values, and gray bars indicate negative values, corresponding to the first and second semesters of 1987.

%

\vskip 0.4 cm

\noindent Fig. 18: contributions of the stock market indices to eigenvector $e_{2}$, corresponding to the second largest eigenvalue of the correlation matrix. White bars indicate positive values, and gray bars indicate negative values, corresponding to the first and second semesters of 1997. The indices are aligned in the following way: {\bf S\&P}, Nasd, Cana, Mexi, {\bf CoRi}, Berm, Jama, Bra, Arg, {\bf Chil}, Ven, Peru, UK, Irel, {\bf Fran}, Germ, Swit, Autr, Belg, {\bf Neth}, Swed, Denm, Finl, Norw, {\bf Icel}, Spai, Port, Gree, CzRe, {\bf Slok}, Hung, Pola, Esto, Russ, {\bf Turk}, Isra, Leba, SaAr, Ohma, {\bf Paki}, Indi, SrLa, Bang, Japa, {\bf HoKo}, Chin, Taiw, SoKo, Thai, {\bf Mala}, Indo, Phil, Aust, Moro, {\bf Ghan}, Keny, SoAf, {\bf Maur}.

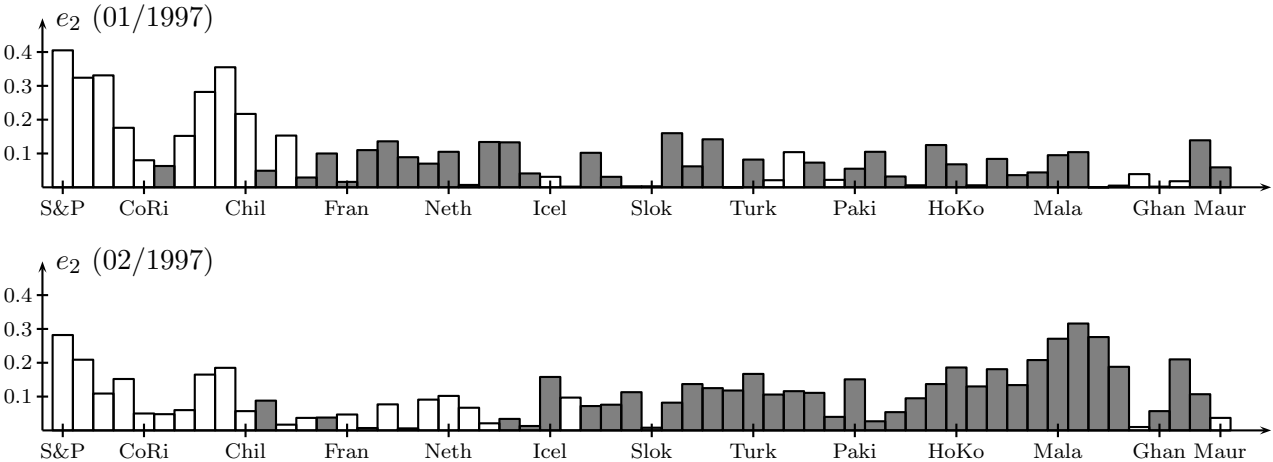
\begin{pspicture}(-0.1,-0.3)(3.5,2.9)
\psset{xunit=0.27,yunit=4.5}
\pspolygon*[linecolor=white](0.5,0)(0.5,0.002)(1.5,0.002)(1.5,0)
\pspolygon*[linecolor=white](1.5,0)(1.5,0.039)(2.5,0.039)(2.5,0)
\pspolygon*[linecolor=gray](2.5,0)(2.5,0.026)(3.5,0.026)(3.5,0)
\pspolygon*[linecolor=white](3.5,0)(3.5,0.148)(4.5,0.148)(4.5,0)
\pspolygon*[linecolor=white](4.5,0)(4.5,0.116)(5.5,0.116)(5.5,0)
\pspolygon*[linecolor=gray](5.5,0)(5.5,0.013)(6.5,0.013)(6.5,0)
\pspolygon*[linecolor=gray](6.5,0)(6.5,0.079)(7.5,0.079)(7.5,0)
\pspolygon*[linecolor=white](7.5,0)(7.5,0.076)(8.5,0.076)(8.5,0)
\pspolygon*[linecolor=white](8.5,0)(8.5,0.185)(9.5,0.185)(9.5,0)
\pspolygon*[linecolor=white](9.5,0)(9.5,0.079)(10.5,0.079)(10.5,0)
\pspolygon*[linecolor=white](10.5,0)(10.5,0.065)(11.5,0.065)(11.5,0)
\pspolygon*[linecolor=gray](11.5,0)(11.5,0.001)(12.5,0.001)(12.5,0)
\pspolygon*[linecolor=gray](12.5,0)(12.5,0.155)(13.5,0.155)(13.5,0)
\pspolygon*[linecolor=gray](13.5,0)(13.5,0.060)(14.5,0.060)(14.5,0)
\pspolygon*[linecolor=gray](14.5,0)(14.5,0.192)(15.5,0.192)(15.5,0)
\pspolygon*[linecolor=gray](15.5,0)(15.5,0.114)(16.5,0.114)(16.5,0)
\pspolygon*[linecolor=gray](16.5,0)(16.5,0.168)(17.5,0.168)(17.5,0)
\pspolygon*[linecolor=gray](17.5,0)(17.5,0.105)(18.5,0.105)(18.5,0)
\pspolygon*[linecolor=gray](18.5,0)(18.5,0.142)(19.5,0.142)(19.5,0)
\pspolygon*[linecolor=gray](19.5,0)(19.5,0.170)(20.5,0.170)(20.5,0)
\pspolygon*[linecolor=gray](20.5,0)(20.5,0.181)(21.5,0.181)(21.5,0)
\pspolygon*[linecolor=gray](21.5,0)(21.5,0.158)(22.5,0.158)(22.5,0)
\pspolygon*[linecolor=gray](22.5,0)(22.5,0.126)(23.5,0.126)(23.5,0)
\pspolygon*[linecolor=gray](23.5,0)(23.5,0.065)(24.5,0.065)(24.5,0)
\pspolygon*[linecolor=gray](24.5,0)(24.5,0.058)(25.5,0.058)(25.5,0)
\pspolygon*[linecolor=gray](25.5,0)(25.5,0.218)(26.5,0.218)(26.5,0)
\pspolygon*[linecolor=gray](26.5,0)(26.5,0.157)(27.5,0.157)(27.5,0)
\pspolygon*[linecolor=white](27.5,0)(27.5,0.080)(28.5,0.080)(28.5,0)
\pspolygon*[linecolor=white](28.5,0)(28.5,0.177)(29.5,0.177)(29.5,0)
\pspolygon*[linecolor=white](29.5,0)(29.5,0.000)(30.5,0.000)(30.5,0)
\pspolygon*[linecolor=white](30.5,0)(30.5,0.115)(31.5,0.115)(31.5,0)
\pspolygon*[linecolor=white](31.5,0)(31.5,0.264)(32.5,0.264)(32.5,0)
\pspolygon*[linecolor=white](32.5,0)(32.5,0.088)(33.5,0.088)(33.5,0)
\pspolygon*[linecolor=white](33.5,0)(33.5,0.239)(34.5,0.239)(34.5,0)
\pspolygon*[linecolor=gray](34.5,0)(34.5,0.044)(35.5,0.044)(35.5,0)
\pspolygon*[linecolor=white](35.5,0)(35.5,0.036)(36.5,0.036)(36.5,0)
\pspolygon*[linecolor=white](36.5,0)(36.5,0.094)(37.5,0.094)(37.5,0)
\pspolygon*[linecolor=white](37.5,0)(37.5,0.136)(38.5,0.136)(38.5,0)
\pspolygon*[linecolor=gray](38.5,0)(38.5,0.038)(39.5,0.038)(39.5,0)
\pspolygon*[linecolor=white](39.5,0)(39.5,0.195)(40.5,0.195)(40.5,0)
\pspolygon*[linecolor=gray](40.5,0)(40.5,0.051)(41.5,0.051)(41.5,0)
\pspolygon*[linecolor=white](41.5,0)(41.5,0.087)(42.5,0.087)(42.5,0)
\pspolygon*[linecolor=white](42.5,0)(42.5,0.042)(43.5,0.042)(43.5,0)
\pspolygon*[linecolor=white](43.5,0)(43.5,0.169)(44.5,0.169)(44.5,0)
\pspolygon*[linecolor=white](44.5,0)(44.5,0.151)(45.5,0.151)(45.5,0)
\pspolygon*[linecolor=gray](45.5,0)(45.5,0.049)(46.5,0.049)(46.5,0)
\pspolygon*[linecolor=white](46.5,0)(46.5,0.076)(47.5,0.076)(47.5,0)
\pspolygon*[linecolor=white](47.5,0)(47.5,0.166)(48.5,0.166)(48.5,0)
\pspolygon*[linecolor=white](48.5,0)(48.5,0.231)(49.5,0.231)(49.5,0)
\pspolygon*[linecolor=white](49.5,0)(49.5,0.267)(50.5,0.267)(50.5,0)
\pspolygon*[linecolor=white](50.5,0)(50.5,0.149)(51.5,0.149)(51.5,0)
\pspolygon*[linecolor=white](51.5,0)(51.5,0.214)(52.5,0.214)(52.5,0)
\pspolygon*[linecolor=white](52.5,0)(52.5,0.100)(53.5,0.100)(53.5,0)
\pspolygon*[linecolor=white](53.5,0)(53.5,0.058)(54.5,0.058)(54.5,0)
\pspolygon*[linecolor=white](54.5,0)(54.5,0.073)(55.5,0.073)(55.5,0)
\pspolygon*[linecolor=gray](55.5,0)(55.5,0.003)(56.5,0.003)(56.5,0)
\pspolygon*[linecolor=white](56.5,0)(56.5,0.033)(57.5,0.033)(57.5,0)
\pspolygon*[linecolor=white](57.5,0)(57.5,0.137)(58.5,0.137)(58.5,0)
\pspolygon(0.5,0)(0.5,0.002)(1.5,0.002)(1.5,0)
\pspolygon(1.5,0)(1.5,0.039)(2.5,0.039)(2.5,0)
\pspolygon(2.5,0)(2.5,0.026)(3.5,0.026)(3.5,0)
\pspolygon(3.5,0)(3.5,0.148)(4.5,0.148)(4.5,0)
\pspolygon(4.5,0)(4.5,0.116)(5.5,0.116)(5.5,0)
\pspolygon(5.5,0)(5.5,0.013)(6.5,0.013)(6.5,0)
\pspolygon(6.5,0)(6.5,0.079)(7.5,0.079)(7.5,0)
\pspolygon(7.5,0)(7.5,0.076)(8.5,0.076)(8.5,0)
\pspolygon(8.5,0)(8.5,0.185)(9.5,0.185)(9.5,0)
\pspolygon(9.5,0)(9.5,0.079)(10.5,0.079)(10.5,0)
\pspolygon(10.5,0)(10.5,0.065)(11.5,0.065)(11.5,0)
\pspolygon(11.5,0)(11.5,0.001)(12.5,0.001)(12.5,0)
\pspolygon(12.5,0)(12.5,0.155)(13.5,0.155)(13.5,0)
\pspolygon(13.5,0)(13.5,0.060)(14.5,0.060)(14.5,0)
\pspolygon(14.5,0)(14.5,0.192)(15.5,0.192)(15.5,0)
\pspolygon(15.5,0)(15.5,0.114)(16.5,0.114)(16.5,0)
\pspolygon(16.5,0)(16.5,0.168)(17.5,0.168)(17.5,0)
\pspolygon(17.5,0)(17.5,0.105)(18.5,0.105)(18.5,0)
\pspolygon(18.5,0)(18.5,0.142)(19.5,0.142)(19.5,0)
\pspolygon(19.5,0)(19.5,0.170)(20.5,0.170)(20.5,0)
\pspolygon(20.5,0)(20.5,0.181)(21.5,0.181)(21.5,0)
\pspolygon(21.5,0)(21.5,0.158)(22.5,0.158)(22.5,0)
\pspolygon(22.5,0)(22.5,0.126)(23.5,0.126)(23.5,0)
\pspolygon(23.5,0)(23.5,0.065)(24.5,0.065)(24.5,0)
\pspolygon(24.5,0)(24.5,0.058)(25.5,0.058)(25.5,0)
\pspolygon(25.5,0)(25.5,0.218)(26.5,0.218)(26.5,0)
\pspolygon(26.5,0)(26.5,0.157)(27.5,0.157)(27.5,0)
\pspolygon(27.5,0)(27.5,0.080)(28.5,0.080)(28.5,0)
\pspolygon(28.5,0)(28.5,0.177)(29.5,0.177)(29.5,0)
\pspolygon(29.5,0)(29.5,0.000)(30.5,0.000)(30.5,0)
\pspolygon(30.5,0)(30.5,0.115)(31.5,0.115)(31.5,0)
\pspolygon(31.5,0)(31.5,0.264)(32.5,0.264)(32.5,0)
\pspolygon(32.5,0)(32.5,0.088)(33.5,0.088)(33.5,0)
\pspolygon(33.5,0)(33.5,0.239)(34.5,0.239)(34.5,0)
\pspolygon(34.5,0)(34.5,0.044)(35.5,0.044)(35.5,0)
\pspolygon(35.5,0)(35.5,0.036)(36.5,0.036)(36.5,0)
\pspolygon(36.5,0)(36.5,0.094)(37.5,0.094)(37.5,0)
\pspolygon(37.5,0)(37.5,0.136)(38.5,0.136)(38.5,0)
\pspolygon(38.5,0)(38.5,0.038)(39.5,0.038)(39.5,0)
\pspolygon(39.5,0)(39.5,0.195)(40.5,0.195)(40.5,0)
\pspolygon(40.5,0)(40.5,0.051)(41.5,0.051)(41.5,0)
\pspolygon(41.5,0)(41.5,0.087)(42.5,0.087)(42.5,0)
\pspolygon(42.5,0)(42.5,0.042)(43.5,0.042)(43.5,0)
\pspolygon(43.5,0)(43.5,0.169)(44.5,0.169)(44.5,0)
\pspolygon(44.5,0)(44.5,0.151)(45.5,0.151)(45.5,0)
\pspolygon(45.5,0)(45.5,0.049)(46.5,0.049)(46.5,0)
\pspolygon(46.5,0)(46.5,0.076)(47.5,0.076)(47.5,0)
\pspolygon(47.5,0)(47.5,0.166)(48.5,0.166)(48.5,0)
\pspolygon(48.5,0)(48.5,0.231)(49.5,0.231)(49.5,0)
\pspolygon(49.5,0)(49.5,0.267)(50.5,0.267)(50.5,0)
\pspolygon(50.5,0)(50.5,0.149)(51.5,0.149)(51.5,0)
\pspolygon(51.5,0)(51.5,0.214)(52.5,0.214)(52.5,0)
\pspolygon(52.5,0)(52.5,0.100)(53.5,0.100)(53.5,0)
\pspolygon(53.5,0)(53.5,0.058)(54.5,0.058)(54.5,0)
\pspolygon(54.5,0)(54.5,0.073)(55.5,0.073)(55.5,0)
\pspolygon(55.5,0)(55.5,0.003)(56.5,0.003)(56.5,0)
\pspolygon(56.5,0)(56.5,0.033)(57.5,0.033)(57.5,0)
\pspolygon(57.5,0)(57.5,0.137)(58.5,0.137)(58.5,0)
\psline{->}(0,0)(60.5,0) \psline{->}(0,0)(0,0.5) \rput(4.6,0.5){$e_{2}\ (01/1998)$} \scriptsize \psline(1,-0.02)(1,0.02) \rput(1,-0.06){S\&P} \psline(5,-0.02)(5,0.02) \rput(5,-0.06){CoRi} \psline(10,-0.02)(10,0.02) \rput(10,-0.06){Chil} \psline(15,-0.02)(15,0.02) \rput(15,-0.06){Fran} \psline(20,-0.02)(20,0.02) \rput(20,-0.06){Neth} \psline(25,-0.02)(25,0.02) \rput(25,-0.06){Icel} \psline(30,-0.02)(30,0.02) \rput(30,-0.06){Slok} \psline(35,-0.02)(35,0.02) \rput(35,-0.06){Turk} \psline(40,-0.02)(40,0.02) \rput(40,-0.06){Paki} \psline(45,-0.02)(45,0.02) \rput(45,-0.06){HoKo} \psline(50,-0.02)(50,0.02) \rput(50,-0.06){Mala} \psline(55,-0.02)(55,0.02) \rput(55,-0.06){Ghan} \psline(58,-0.02)(58,0.02) \rput(58,-0.06){Maur} \scriptsize \psline(-0.28,0.1)(0.28,0.1) \rput(-1.2,0.1){$0.1$} \psline(-0.28,0.2)(0.28,0.2) \rput(-1.2,0.2){$0.2$} \psline(-0.28,0.3)(0.28,0.3) \rput(-1.2,0.3){$0.3$} \psline(-0.28,0.4)(0.28,0.4) \rput(-1.2,0.4){$0.4$}
\end{pspicture}

\begin{pspicture}(-0.1,-0.3)(3.5,2.9)
\psset{xunit=0.27,yunit=4.5}
\pspolygon*[linecolor=white](0.5,0)(0.5,0.341)(1.5,0.341)(1.5,0)
\pspolygon*[linecolor=white](1.5,0)(1.5,0.309)(2.5,0.309)(2.5,0)
\pspolygon*[linecolor=white](2.5,0)(2.5,0.194)(3.5,0.194)(3.5,0)
\pspolygon*[linecolor=white](3.5,0)(3.5,0.252)(4.5,0.252)(4.5,0)
\pspolygon*[linecolor=gray](4.5,0)(4.5,0.009)(5.5,0.009)(5.5,0)
\pspolygon*[linecolor=gray](5.5,0)(5.5,0.014)(6.5,0.014)(6.5,0)
\pspolygon*[linecolor=gray](6.5,0)(6.5,0.011)(7.5,0.011)(7.5,0)
\pspolygon*[linecolor=white](7.5,0)(7.5,0.288)(8.5,0.288)(8.5,0)
\pspolygon*[linecolor=white](8.5,0)(8.5,0.277)(9.5,0.277)(9.5,0)
\pspolygon*[linecolor=white](9.5,0)(9.5,0.258)(10.5,0.258)(10.5,0)
\pspolygon*[linecolor=white](10.5,0)(10.5,0.084)(11.5,0.084)(11.5,0)
\pspolygon*[linecolor=gray](11.5,0)(11.5,0.029)(12.5,0.029)(12.5,0)
\pspolygon*[linecolor=white](12.5,0)(12.5,0.019)(13.5,0.019)(13.5,0)
\pspolygon*[linecolor=gray](13.5,0)(13.5,0.069)(14.5,0.069)(14.5,0)
\pspolygon*[linecolor=white](14.5,0)(14.5,0.059)(15.5,0.059)(15.5,0)
\pspolygon*[linecolor=white](15.5,0)(15.5,0.023)(16.5,0.023)(16.5,0)
\pspolygon*[linecolor=white](16.5,0)(16.5,0.038)(17.5,0.038)(17.5,0)
\pspolygon*[linecolor=gray](17.5,0)(17.5,0.042)(18.5,0.042)(18.5,0)
\pspolygon*[linecolor=gray](18.5,0)(18.5,0.030)(19.5,0.030)(19.5,0)
\pspolygon*[linecolor=gray](19.5,0)(19.5,0.001)(20.5,0.001)(20.5,0)
\pspolygon*[linecolor=white](20.5,0)(20.5,0.063)(21.5,0.063)(21.5,0)
\pspolygon*[linecolor=gray](21.5,0)(21.5,0.011)(22.5,0.011)(22.5,0)
\pspolygon*[linecolor=white](22.5,0)(22.5,0.012)(23.5,0.012)(23.5,0)
\pspolygon*[linecolor=gray](23.5,0)(23.5,0.025)(24.5,0.025)(24.5,0)
\pspolygon*[linecolor=gray](24.5,0)(24.5,0.100)(25.5,0.100)(25.5,0)
\pspolygon*[linecolor=white](25.5,0)(25.5,0.054)(26.5,0.054)(26.5,0)
\pspolygon*[linecolor=gray](26.5,0)(26.5,0.037)(27.5,0.037)(27.5,0)
\pspolygon*[linecolor=gray](27.5,0)(27.5,0.157)(28.5,0.157)(28.5,0)
\pspolygon*[linecolor=gray](28.5,0)(28.5,0.103)(29.5,0.103)(29.5,0)
\pspolygon*[linecolor=white](29.5,0)(29.5,0.012)(30.5,0.012)(30.5,0)
\pspolygon*[linecolor=gray](30.5,0)(30.5,0.103)(31.5,0.103)(31.5,0)
\pspolygon*[linecolor=gray](31.5,0)(31.5,0.179)(32.5,0.179)(32.5,0)
\pspolygon*[linecolor=white](32.5,0)(32.5,0.017)(33.5,0.017)(33.5,0)
\pspolygon*[linecolor=gray](33.5,0)(33.5,0.061)(34.5,0.061)(34.5,0)
\pspolygon*[linecolor=gray](34.5,0)(34.5,0.033)(35.5,0.033)(35.5,0)
\pspolygon*[linecolor=white](35.5,0)(35.5,0.037)(36.5,0.037)(36.5,0)
\pspolygon*[linecolor=white](36.5,0)(36.5,0.035)(37.5,0.035)(37.5,0)
\pspolygon*[linecolor=gray](37.5,0)(37.5,0.102)(38.5,0.102)(38.5,0)
\pspolygon*[linecolor=gray](38.5,0)(38.5,0.023)(39.5,0.023)(39.5,0)
\pspolygon*[linecolor=white](39.5,0)(39.5,0.067)(40.5,0.067)(40.5,0)
\pspolygon*[linecolor=gray](40.5,0)(40.5,0.175)(41.5,0.175)(41.5,0)
\pspolygon*[linecolor=gray](41.5,0)(41.5,0.027)(42.5,0.027)(42.5,0)
\pspolygon*[linecolor=white](42.5,0)(42.5,0.015)(43.5,0.015)(43.5,0)
\pspolygon*[linecolor=gray](43.5,0)(43.5,0.090)(44.5,0.090)(44.5,0)
\pspolygon*[linecolor=gray](44.5,0)(44.5,0.139)(45.5,0.139)(45.5,0)
\pspolygon*[linecolor=gray](45.5,0)(45.5,0.064)(46.5,0.064)(46.5,0)
\pspolygon*[linecolor=gray](46.5,0)(46.5,0.119)(47.5,0.119)(47.5,0)
\pspolygon*[linecolor=gray](47.5,0)(47.5,0.221)(48.5,0.221)(48.5,0)
\pspolygon*[linecolor=gray](48.5,0)(48.5,0.141)(49.5,0.141)(49.5,0)
\pspolygon*[linecolor=gray](49.5,0)(49.5,0.181)(50.5,0.181)(50.5,0)
\pspolygon*[linecolor=gray](50.5,0)(50.5,0.241)(51.5,0.241)(51.5,0)
\pspolygon*[linecolor=gray](51.5,0)(51.5,0.117)(52.5,0.117)(52.5,0)
\pspolygon*[linecolor=gray](52.5,0)(52.5,0.189)(53.5,0.189)(53.5,0)
\pspolygon*[linecolor=white](53.5,0)(53.5,0.083)(54.5,0.083)(54.5,0)
\pspolygon*[linecolor=gray](54.5,0)(54.5,0.097)(55.5,0.097)(55.5,0)
\pspolygon*[linecolor=gray](55.5,0)(55.5,0.020)(56.5,0.020)(56.5,0)
\pspolygon*[linecolor=gray](56.5,0)(56.5,0.075)(57.5,0.075)(57.5,0)
\pspolygon*[linecolor=gray](57.5,0)(57.5,0.100)(58.5,0.100)(58.5,0)
\pspolygon(0.5,0)(0.5,0.341)(1.5,0.341)(1.5,0)
\pspolygon(1.5,0)(1.5,0.309)(2.5,0.309)(2.5,0)
\pspolygon(2.5,0)(2.5,0.194)(3.5,0.194)(3.5,0)
\pspolygon(3.5,0)(3.5,0.252)(4.5,0.252)(4.5,0)
\pspolygon(4.5,0)(4.5,0.009)(5.5,0.009)(5.5,0)
\pspolygon(5.5,0)(5.5,0.014)(6.5,0.014)(6.5,0)
\pspolygon(6.5,0)(6.5,0.011)(7.5,0.011)(7.5,0)
\pspolygon(7.5,0)(7.5,0.288)(8.5,0.288)(8.5,0)
\pspolygon(8.5,0)(8.5,0.277)(9.5,0.277)(9.5,0)
\pspolygon(9.5,0)(9.5,0.258)(10.5,0.258)(10.5,0)
\pspolygon(10.5,0)(10.5,0.084)(11.5,0.084)(11.5,0)
\pspolygon(11.5,0)(11.5,0.029)(12.5,0.029)(12.5,0)
\pspolygon(12.5,0)(12.5,0.019)(13.5,0.019)(13.5,0)
\pspolygon(13.5,0)(13.5,0.069)(14.5,0.069)(14.5,0)
\pspolygon(14.5,0)(14.5,0.059)(15.5,0.059)(15.5,0)
\pspolygon(15.5,0)(15.5,0.023)(16.5,0.023)(16.5,0)
\pspolygon(16.5,0)(16.5,0.038)(17.5,0.038)(17.5,0)
\pspolygon(17.5,0)(17.5,0.042)(18.5,0.042)(18.5,0)
\pspolygon(18.5,0)(18.5,0.030)(19.5,0.030)(19.5,0)
\pspolygon(19.5,0)(19.5,0.001)(20.5,0.001)(20.5,0)
\pspolygon(20.5,0)(20.5,0.063)(21.5,0.063)(21.5,0)
\pspolygon(21.5,0)(21.5,0.011)(22.5,0.011)(22.5,0)
\pspolygon(22.5,0)(22.5,0.012)(23.5,0.012)(23.5,0)
\pspolygon(23.5,0)(23.5,0.025)(24.5,0.025)(24.5,0)
\pspolygon(24.5,0)(24.5,0.100)(25.5,0.100)(25.5,0)
\pspolygon(25.5,0)(25.5,0.054)(26.5,0.054)(26.5,0)
\pspolygon(26.5,0)(26.5,0.037)(27.5,0.037)(27.5,0)
\pspolygon(27.5,0)(27.5,0.157)(28.5,0.157)(28.5,0)
\pspolygon(28.5,0)(28.5,0.103)(29.5,0.103)(29.5,0)
\pspolygon(29.5,0)(29.5,0.012)(30.5,0.012)(30.5,0)
\pspolygon(30.5,0)(30.5,0.103)(31.5,0.103)(31.5,0)
\pspolygon(31.5,0)(31.5,0.179)(32.5,0.179)(32.5,0)
\pspolygon(32.5,0)(32.5,0.017)(33.5,0.017)(33.5,0)
\pspolygon(33.5,0)(33.5,0.061)(34.5,0.061)(34.5,0)
\pspolygon(34.5,0)(34.5,0.033)(35.5,0.033)(35.5,0)
\pspolygon(35.5,0)(35.5,0.037)(36.5,0.037)(36.5,0)
\pspolygon(36.5,0)(36.5,0.035)(37.5,0.035)(37.5,0)
\pspolygon(37.5,0)(37.5,0.102)(38.5,0.102)(38.5,0)
\pspolygon(38.5,0)(38.5,0.023)(39.5,0.023)(39.5,0)
\pspolygon(39.5,0)(39.5,0.067)(40.5,0.067)(40.5,0)
\pspolygon(40.5,0)(40.5,0.175)(41.5,0.175)(41.5,0)
\pspolygon(41.5,0)(41.5,0.027)(42.5,0.027)(42.5,0)
\pspolygon(42.5,0)(42.5,0.015)(43.5,0.015)(43.5,0)
\pspolygon(43.5,0)(43.5,0.090)(44.5,0.090)(44.5,0)
\pspolygon(44.5,0)(44.5,0.139)(45.5,0.139)(45.5,0)
\pspolygon(45.5,0)(45.5,0.064)(46.5,0.064)(46.5,0)
\pspolygon(46.5,0)(46.5,0.119)(47.5,0.119)(47.5,0)
\pspolygon(47.5,0)(47.5,0.221)(48.5,0.221)(48.5,0)
\pspolygon(48.5,0)(48.5,0.141)(49.5,0.141)(49.5,0)
\pspolygon(49.5,0)(49.5,0.181)(50.5,0.181)(50.5,0)
\pspolygon(50.5,0)(50.5,0.241)(51.5,0.241)(51.5,0)
\pspolygon(51.5,0)(51.5,0.117)(52.5,0.117)(52.5,0)
\pspolygon(52.5,0)(52.5,0.189)(53.5,0.189)(53.5,0)
\pspolygon(53.5,0)(53.5,0.083)(54.5,0.083)(54.5,0)
\pspolygon(54.5,0)(54.5,0.097)(55.5,0.097)(55.5,0)
\pspolygon(55.5,0)(55.5,0.020)(56.5,0.020)(56.5,0)
\pspolygon(56.5,0)(56.5,0.075)(57.5,0.075)(57.5,0)
\pspolygon(57.5,0)(57.5,0.100)(58.5,0.100)(58.5,0)
\psline{->}(0,0)(60.5,0) \psline{->}(0,0)(0,0.5) \rput(4.6,0.5){$e_{2}\ (02/1998)$} \scriptsize \psline(1,-0.02)(1,0.02) \rput(1,-0.06){S\&P} \psline(5,-0.02)(5,0.02) \rput(5,-0.06){CoRi} \psline(10,-0.02)(10,0.02) \rput(10,-0.06){Chil} \psline(15,-0.02)(15,0.02) \rput(15,-0.06){Fran} \psline(20,-0.02)(20,0.02) \rput(20,-0.06){Neth} \psline(25,-0.02)(25,0.02) \rput(25,-0.06){Icel} \psline(30,-0.02)(30,0.02) \rput(30,-0.06){Slok} \psline(35,-0.02)(35,0.02) \rput(35,-0.06){Turk} \psline(40,-0.02)(40,0.02) \rput(40,-0.06){Paki} \psline(45,-0.02)(45,0.02) \rput(45,-0.06){HoKo} \psline(50,-0.02)(50,0.02) \rput(50,-0.06){Mala} \psline(55,-0.02)(55,0.02) \rput(55,-0.06){Ghan} \psline(58,-0.02)(58,0.02) \rput(58,-0.06){Maur} \scriptsize \psline(-0.28,0.1)(0.28,0.1) \rput(-1.2,0.1){$0.1$} \psline(-0.28,0.2)(0.28,0.2) \rput(-1.2,0.2){$0.2$} \psline(-0.28,0.3)(0.28,0.3) \rput(-1.2,0.3){$0.3$} \psline(-0.28,0.4)(0.28,0.4) \rput(-1.2,0.4){$0.4$}
\end{pspicture}

\vskip 0.4 cm

\noindent Fig. 19. Contributions of the stock market indices to eigenvector $e_{2}$, corresponding to the second largest eigenvalue of the correlation matrix. White bars indicate positive values, and gray bars indicate negative values, corresponding to the first and second semesters of 1997. The indices are aligned in the following way: {\bf S\&P}, Nasd, Cana, Mexi, {\bf CoRi}, Berm, Jama, Bra, Arg, {\bf Chil}, Ven, Peru, UK, Irel, {\bf Fran}, Germ, Swit, Autr, Belg, {\bf Neth}, Swed, Denm, Finl, Norw, {\bf Icel}, Spai, Port, Gree, CzRe, {\bf Slok}, Hung, Pola, Esto, Russ, {\bf Turk}, Isra, Leba, SaAr, Ohma, {\bf Paki}, Indi, SrLa, Bang, Japa, {\bf HoKo}, Chin, Taiw, SoKo, Thai, {\bf Mala}, Indo, Phil, Aust, Moro, {\bf Ghan}, Keny, SoAf, {\bf Maur}.

\vskip 0.3 cm

\begin{pspicture}(-0.1,0)(3.5,2.7)
\psset{xunit=0.215,yunit=4.5}
\pspolygon*[linecolor=white](0.5,0)(0.5,0.362)(1.5,0.362)(1.5,0)
\pspolygon*[linecolor=white](1.5,0)(1.5,0.293)(2.5,0.293)(2.5,0)
\pspolygon*[linecolor=white](2.5,0)(2.5,0.226)(3.5,0.226)(3.5,0)
\pspolygon*[linecolor=white](3.5,0)(3.5,0.278)(4.5,0.278)(4.5,0)
\pspolygon*[linecolor=white](4.5,0)(4.5,0.024)(5.5,0.024)(5.5,0)
\pspolygon*[linecolor=gray](5.5,0)(5.5,0.033)(6.5,0.033)(6.5,0)
\pspolygon*[linecolor=gray](6.5,0)(6.5,0.017)(7.5,0.017)(7.5,0)
\pspolygon*[linecolor=white](7.5,0)(7.5,0.112)(8.5,0.112)(8.5,0)
\pspolygon*[linecolor=white](8.5,0)(8.5,0.265)(9.5,0.265)(9.5,0)
\pspolygon*[linecolor=white](9.5,0)(9.5,0.270)(10.5,0.270)(10.5,0)
\pspolygon*[linecolor=white](10.5,0)(10.5,0.161)(11.5,0.161)(11.5,0)
\pspolygon*[linecolor=white](11.5,0)(11.5,0.011)(12.5,0.011)(12.5,0)
\pspolygon*[linecolor=white](12.5,0)(12.5,0.057)(13.5,0.057)(13.5,0)
\pspolygon*[linecolor=white](13.5,0)(13.5,0.049)(14.5,0.049)(14.5,0)
\pspolygon*[linecolor=gray](14.5,0)(14.5,0.100)(15.5,0.100)(15.5,0)
\pspolygon*[linecolor=white](15.5,0)(15.5,0.000)(16.5,0.000)(16.5,0)
\pspolygon*[linecolor=white](16.5,0)(16.5,0.043)(17.5,0.043)(17.5,0)
\pspolygon*[linecolor=gray](17.5,0)(17.5,0.006)(18.5,0.006)(18.5,0)
\pspolygon*[linecolor=gray](18.5,0)(18.5,0.094)(19.5,0.094)(19.5,0)
\pspolygon*[linecolor=white](19.5,0)(19.5,0.025)(20.5,0.025)(20.5,0)
\pspolygon*[linecolor=gray](20.5,0)(20.5,0.104)(21.5,0.104)(21.5,0)
\pspolygon*[linecolor=white](21.5,0)(21.5,0.043)(22.5,0.043)(22.5,0)
\pspolygon*[linecolor=gray](22.5,0)(22.5,0.024)(23.5,0.024)(23.5,0)
\pspolygon*[linecolor=gray](23.5,0)(23.5,0.091)(24.5,0.091)(24.5,0)
\pspolygon*[linecolor=gray](24.5,0)(24.5,0.083)(25.5,0.083)(25.5,0)
\pspolygon*[linecolor=gray](25.5,0)(25.5,0.092)(26.5,0.092)(26.5,0)
\pspolygon*[linecolor=gray](26.5,0)(26.5,0.113)(27.5,0.113)(27.5,0)
\pspolygon*[linecolor=gray](27.5,0)(27.5,0.017)(28.5,0.017)(28.5,0)
\pspolygon*[linecolor=gray](28.5,0)(28.5,0.117)(29.5,0.117)(29.5,0)
\pspolygon*[linecolor=gray](29.5,0)(29.5,0.005)(30.5,0.005)(30.5,0)
\pspolygon*[linecolor=white](30.5,0)(30.5,0.053)(31.5,0.053)(31.5,0)
\pspolygon*[linecolor=gray](31.5,0)(31.5,0.127)(32.5,0.127)(32.5,0)
\pspolygon*[linecolor=gray](32.5,0)(32.5,0.023)(33.5,0.023)(33.5,0)
\pspolygon*[linecolor=gray](33.5,0)(33.5,0.051)(34.5,0.051)(34.5,0)
\pspolygon*[linecolor=gray](34.5,0)(34.5,0.003)(35.5,0.003)(35.5,0)
\pspolygon*[linecolor=gray](35.5,0)(35.5,0.178)(36.5,0.178)(36.5,0)
\pspolygon*[linecolor=gray](36.5,0)(36.5,0.011)(37.5,0.011)(37.5,0)
\pspolygon*[linecolor=gray](37.5,0)(37.5,0.166)(38.5,0.166)(38.5,0)
\pspolygon*[linecolor=white](38.5,0)(38.5,0.030)(39.5,0.030)(39.5,0)
\pspolygon*[linecolor=gray](39.5,0)(39.5,0.014)(40.5,0.014)(40.5,0)
\pspolygon*[linecolor=gray](40.5,0)(40.5,0.036)(41.5,0.036)(41.5,0)
\pspolygon*[linecolor=gray](41.5,0)(41.5,0.103)(42.5,0.103)(42.5,0)
\pspolygon*[linecolor=gray](42.5,0)(42.5,0.026)(43.5,0.026)(43.5,0)
\pspolygon*[linecolor=white](43.5,0)(43.5,0.111)(44.5,0.111)(44.5,0)
\pspolygon*[linecolor=gray](44.5,0)(44.5,0.070)(45.5,0.070)(45.5,0)
\pspolygon*[linecolor=gray](45.5,0)(45.5,0.097)(46.5,0.097)(46.5,0)
\pspolygon*[linecolor=gray](46.5,0)(46.5,0.017)(47.5,0.017)(47.5,0)
\pspolygon*[linecolor=white](47.5,0)(47.5,0.033)(48.5,0.033)(48.5,0)
\pspolygon*[linecolor=white](48.5,0)(48.5,0.062)(49.5,0.062)(49.5,0)
\pspolygon*[linecolor=gray](49.5,0)(49.5,0.038)(50.5,0.038)(50.5,0)
\pspolygon*[linecolor=gray](50.5,0)(50.5,0.080)(51.5,0.080)(51.5,0)
\pspolygon*[linecolor=gray](51.5,0)(51.5,0.061)(52.5,0.061)(52.5,0)
\pspolygon*[linecolor=gray](52.5,0)(52.5,0.012)(53.5,0.012)(53.5,0)
\pspolygon*[linecolor=white](53.5,0)(53.5,0.030)(54.5,0.030)(54.5,0)
\pspolygon*[linecolor=gray](54.5,0)(54.5,0.200)(55.5,0.200)(55.5,0)
\pspolygon*[linecolor=gray](55.5,0)(55.5,0.152)(56.5,0.152)(56.5,0)
\pspolygon*[linecolor=gray](56.5,0)(56.5,0.067)(57.5,0.067)(57.5,0)
\pspolygon*[linecolor=gray](57.5,0)(57.5,0.021)(58.5,0.021)(58.5,0)
\pspolygon*[linecolor=gray](58.5,0)(58.5,0.061)(59.5,0.061)(59.5,0)
\pspolygon*[linecolor=gray](59.5,0)(59.5,0.167)(60.5,0.167)(60.5,0)
\pspolygon*[linecolor=gray](60.5,0)(60.5,0.137)(61.5,0.137)(61.5,0)
\pspolygon*[linecolor=gray](61.5,0)(61.5,0.128)(62.5,0.128)(62.5,0)
\pspolygon*[linecolor=gray](62.5,0)(62.5,0.122)(63.5,0.122)(63.5,0)
\pspolygon*[linecolor=gray](63.5,0)(63.5,0.111)(64.5,0.111)(64.5,0)
\pspolygon*[linecolor=gray](64.5,0)(64.5,0.157)(65.5,0.157)(65.5,0)
\pspolygon*[linecolor=gray](65.5,0)(65.5,0.093)(66.5,0.093)(66.5,0)
\pspolygon*[linecolor=white](66.5,0)(66.5,0.067)(67.5,0.067)(67.5,0)
\pspolygon*[linecolor=gray](67.5,0)(67.5,0.062)(68.5,0.062)(68.5,0)
\pspolygon*[linecolor=gray](68.5,0)(68.5,0.041)(69.5,0.041)(69.5,0)
\pspolygon*[linecolor=gray](69.5,0)(69.5,0.162)(70.5,0.162)(70.5,0)
\pspolygon*[linecolor=gray](70.5,0)(70.5,0.058)(71.5,0.058)(71.5,0)
\pspolygon*[linecolor=gray](71.5,0)(71.5,0.012)(72.5,0.012)(72.5,0)
\pspolygon*[linecolor=gray](72.5,0)(72.5,0.067)(73.5,0.067)(73.5,0)
\pspolygon*[linecolor=white](73.5,0)(73.5,0.033)(74.5,0.033)(74.5,0)
\pspolygon*[linecolor=white](74.5,0)(74.5,0.000)(75.5,0.000)(75.5,0)
\pspolygon(0.5,0)(0.5,0.362)(1.5,0.362)(1.5,0)
\pspolygon(1.5,0)(1.5,0.293)(2.5,0.293)(2.5,0)
\pspolygon(2.5,0)(2.5,0.226)(3.5,0.226)(3.5,0)
\pspolygon(3.5,0)(3.5,0.278)(4.5,0.278)(4.5,0)
\pspolygon(4.5,0)(4.5,0.024)(5.5,0.024)(5.5,0)
\pspolygon(5.5,0)(5.5,0.033)(6.5,0.033)(6.5,0)
\pspolygon(6.5,0)(6.5,0.017)(7.5,0.017)(7.5,0)
\pspolygon(7.5,0)(7.5,0.112)(8.5,0.112)(8.5,0)
\pspolygon(8.5,0)(8.5,0.265)(9.5,0.265)(9.5,0)
\pspolygon(9.5,0)(9.5,0.270)(10.5,0.270)(10.5,0)
\pspolygon(10.5,0)(10.5,0.161)(11.5,0.161)(11.5,0)
\pspolygon(11.5,0)(11.5,0.011)(12.5,0.011)(12.5,0)
\pspolygon(12.5,0)(12.5,0.057)(13.5,0.057)(13.5,0)
\pspolygon(13.5,0)(13.5,0.049)(14.5,0.049)(14.5,0)
\pspolygon(14.5,0)(14.5,0.100)(15.5,0.100)(15.5,0)
\pspolygon(15.5,0)(15.5,0.000)(16.5,0.000)(16.5,0)
\pspolygon(16.5,0)(16.5,0.043)(17.5,0.043)(17.5,0)
\pspolygon(17.5,0)(17.5,0.006)(18.5,0.006)(18.5,0)
\pspolygon(18.5,0)(18.5,0.094)(19.5,0.094)(19.5,0)
\pspolygon(19.5,0)(19.5,0.025)(20.5,0.025)(20.5,0)
\pspolygon(20.5,0)(20.5,0.104)(21.5,0.104)(21.5,0)
\pspolygon(21.5,0)(21.5,0.043)(22.5,0.043)(22.5,0)
\pspolygon(22.5,0)(22.5,0.024)(23.5,0.024)(23.5,0)
\pspolygon(23.5,0)(23.5,0.091)(24.5,0.091)(24.5,0)
\pspolygon(24.5,0)(24.5,0.083)(25.5,0.083)(25.5,0)
\pspolygon(25.5,0)(25.5,0.092)(26.5,0.092)(26.5,0)
\pspolygon(26.5,0)(26.5,0.113)(27.5,0.113)(27.5,0)
\pspolygon(27.5,0)(27.5,0.017)(28.5,0.017)(28.5,0)
\pspolygon(28.5,0)(28.5,0.117)(29.5,0.117)(29.5,0)
\pspolygon(29.5,0)(29.5,0.005)(30.5,0.005)(30.5,0)
\pspolygon(30.5,0)(30.5,0.053)(31.5,0.053)(31.5,0)
\pspolygon(31.5,0)(31.5,0.127)(32.5,0.127)(32.5,0)
\pspolygon(32.5,0)(32.5,0.023)(33.5,0.023)(33.5,0)
\pspolygon(33.5,0)(33.5,0.051)(34.5,0.051)(34.5,0)
\pspolygon(34.5,0)(34.5,0.003)(35.5,0.003)(35.5,0)
\pspolygon(35.5,0)(35.5,0.178)(36.5,0.178)(36.5,0)
\pspolygon(36.5,0)(36.5,0.011)(37.5,0.011)(37.5,0)
\pspolygon(37.5,0)(37.5,0.166)(38.5,0.166)(38.5,0)
\pspolygon(38.5,0)(38.5,0.030)(39.5,0.030)(39.5,0)
\pspolygon(39.5,0)(39.5,0.014)(40.5,0.014)(40.5,0)
\pspolygon(40.5,0)(40.5,0.036)(41.5,0.036)(41.5,0)
\pspolygon(41.5,0)(41.5,0.103)(42.5,0.103)(42.5,0)
\pspolygon(42.5,0)(42.5,0.026)(43.5,0.026)(43.5,0)
\pspolygon(43.5,0)(43.5,0.111)(44.5,0.111)(44.5,0)
\pspolygon(44.5,0)(44.5,0.070)(45.5,0.070)(45.5,0)
\pspolygon(45.5,0)(45.5,0.097)(46.5,0.097)(46.5,0)
\pspolygon(46.5,0)(46.5,0.017)(47.5,0.017)(47.5,0)
\pspolygon(47.5,0)(47.5,0.033)(48.5,0.033)(48.5,0)
\pspolygon(48.5,0)(48.5,0.062)(49.5,0.062)(49.5,0)
\pspolygon(49.5,0)(49.5,0.038)(50.5,0.038)(50.5,0)
\pspolygon(50.5,0)(50.5,0.080)(51.5,0.080)(51.5,0)
\pspolygon(51.5,0)(51.5,0.061)(52.5,0.061)(52.5,0)
\pspolygon(52.5,0)(52.5,0.012)(53.5,0.012)(53.5,0)
\pspolygon(53.5,0)(53.5,0.030)(54.5,0.030)(54.5,0)
\pspolygon(54.5,0)(54.5,0.200)(55.5,0.200)(55.5,0)
\pspolygon(55.5,0)(55.5,0.152)(56.5,0.152)(56.5,0)
\pspolygon(56.5,0)(56.5,0.067)(57.5,0.067)(57.5,0)
\pspolygon(57.5,0)(57.5,0.021)(58.5,0.021)(58.5,0)
\pspolygon(58.5,0)(58.5,0.061)(59.5,0.061)(59.5,0)
\pspolygon(59.5,0)(59.5,0.167)(60.5,0.167)(60.5,0)
\pspolygon(60.5,0)(60.5,0.137)(61.5,0.137)(61.5,0)
\pspolygon(61.5,0)(61.5,0.128)(62.5,0.128)(62.5,0)
\pspolygon(62.5,0)(62.5,0.122)(63.5,0.122)(63.5,0)
\pspolygon(63.5,0)(63.5,0.111)(64.5,0.111)(64.5,0)
\pspolygon(64.5,0)(64.5,0.157)(65.5,0.157)(65.5,0)
\pspolygon(65.5,0)(65.5,0.093)(66.5,0.093)(66.5,0)
\pspolygon(66.5,0)(66.5,0.067)(67.5,0.067)(67.5,0)
\pspolygon(67.5,0)(67.5,0.062)(68.5,0.062)(68.5,0)
\pspolygon(68.5,0)(68.5,0.041)(69.5,0.041)(69.5,0)
\pspolygon(69.5,0)(69.5,0.162)(70.5,0.162)(70.5,0)
\pspolygon(70.5,0)(70.5,0.058)(71.5,0.058)(71.5,0)
\pspolygon(71.5,0)(71.5,0.012)(72.5,0.012)(72.5,0)
\pspolygon(72.5,0)(72.5,0.067)(73.5,0.067)(73.5,0)
\pspolygon(73.5,0)(73.5,0.033)(74.5,0.033)(74.5,0)
\pspolygon(74.5,0)(74.5,0.000)(75.5,0.000)(75.5,0)
\psline{->}(0,0)(77,0) \psline{->}(0,0)(0,0.5) \rput(5.8,0.5){$e_{2}\ (01/2000)$} \scriptsize \psline(1,-0.02)(1,0.02) \rput(1,-0.06){S\&P} \psline(5,-0.02)(5,0.02) \rput(5,-0.06){Pana} \psline(10,-0.02)(10,0.02) \rput(10,-0.06){Arge} \psline(15,-0.02)(15,0.02) \rput(15,-0.06){Irel} \psline(20,-0.02)(20,0.02) \rput(20,-0.06){Ital} \psline(25,-0.02)(25,0.02) \rput(25,-0.06){Swed} \psline(30,-0.02)(30,0.02) \rput(30,-0.06){Spai} \psline(35,-0.02)(35,0.02) \rput(35,-0.06){Hung} \psline(40,-0.02)(40,0.02) \rput(40,-0.06){Lith} \psline(45,-0.02)(45,0.02) \rput(45,-0.06){Pale} \psline(50,-0.02)(50,0.02) \rput(49,-0.06){Ohma} \psline(55,-0.02)(55,0.02) \rput(55,-0.06){Japa} \psline(60,-0.02)(60,0.02) \rput(60,-0.06){SoKo} \psline(65,-0.02)(65,0.02) \rput(65,-0.06){Phil} \psline(70,-0.02)(70,0.02) \rput(70,-0.06){Ghan} \psline(74,-0.02)(74,0.02) \rput(74,-0.06){Maur} \scriptsize \psline(-0.28,0.1)(0.28,0.1) \rput(-1.38,0.1){$0.1$} \psline(-0.28,0.2)(0.28,0.2) \rput(-1.38,0.2){$0.2$} \psline(-0.28,0.3)(0.28,0.3) \rput(-1.38,0.3){$0.3$} \psline(-0.28,0.4)(0.28,0.4) \rput(-1.38,0.4){$0.4$}
\end{pspicture}

\begin{pspicture}(-0.1,0)(3.5,3.2)
\psset{xunit=0.215,yunit=4.5}
\pspolygon*[linecolor=white](0.5,0)(0.5,0.362)(1.5,0.362)(1.5,0)
\pspolygon*[linecolor=white](1.5,0)(1.5,0.216)(2.5,0.216)(2.5,0)
\pspolygon*[linecolor=white](2.5,0)(2.5,0.194)(3.5,0.194)(3.5,0)
\pspolygon*[linecolor=white](3.5,0)(3.5,0.091)(4.5,0.091)(4.5,0)
\pspolygon*[linecolor=white](4.5,0)(4.5,0.157)(5.5,0.157)(5.5,0)
\pspolygon*[linecolor=gray](5.5,0)(5.5,0.006)(6.5,0.006)(6.5,0)
\pspolygon*[linecolor=gray](6.5,0)(6.5,0.023)(7.5,0.023)(7.5,0)
\pspolygon*[linecolor=gray](7.5,0)(7.5,0.089)(8.5,0.089)(8.5,0)
\pspolygon*[linecolor=white](8.5,0)(8.5,0.016)(9.5,0.016)(9.5,0)
\pspolygon*[linecolor=white](9.5,0)(9.5,0.150)(10.5,0.150)(10.5,0)
\pspolygon*[linecolor=white](10.5,0)(10.5,0.281)(11.5,0.281)(11.5,0)
\pspolygon*[linecolor=white](11.5,0)(11.5,0.026)(12.5,0.026)(12.5,0)
\pspolygon*[linecolor=white](12.5,0)(12.5,0.039)(13.5,0.039)(13.5,0)
\pspolygon*[linecolor=white](13.5,0)(13.5,0.025)(14.5,0.025)(14.5,0)
\pspolygon*[linecolor=white](14.5,0)(14.5,0.019)(15.5,0.019)(15.5,0)
\pspolygon*[linecolor=gray](15.5,0)(15.5,0.147)(16.5,0.147)(16.5,0)
\pspolygon*[linecolor=white](16.5,0)(16.5,0.121)(17.5,0.121)(17.5,0)
\pspolygon*[linecolor=white](17.5,0)(17.5,0.144)(18.5,0.144)(18.5,0)
\pspolygon*[linecolor=white](18.5,0)(18.5,0.090)(19.5,0.090)(19.5,0)
\pspolygon*[linecolor=white](19.5,0)(19.5,0.011)(20.5,0.011)(20.5,0)
\pspolygon*[linecolor=white](20.5,0)(20.5,0.147)(21.5,0.147)(21.5,0)
\pspolygon*[linecolor=gray](21.5,0)(21.5,0.183)(22.5,0.183)(22.5,0)
\pspolygon*[linecolor=white](22.5,0)(22.5,0.002)(23.5,0.002)(23.5,0)
\pspolygon*[linecolor=white](23.5,0)(23.5,0.031)(24.5,0.031)(24.5,0)
\pspolygon*[linecolor=gray](24.5,0)(24.5,0.057)(25.5,0.057)(25.5,0)
\pspolygon*[linecolor=white](25.5,0)(25.5,0.116)(26.5,0.116)(26.5,0)
\pspolygon*[linecolor=gray](26.5,0)(26.5,0.056)(27.5,0.056)(27.5,0)
\pspolygon*[linecolor=white](27.5,0)(27.5,0.037)(28.5,0.037)(28.5,0)
\pspolygon*[linecolor=gray](28.5,0)(28.5,0.019)(29.5,0.019)(29.5,0)
\pspolygon*[linecolor=gray](29.5,0)(29.5,0.007)(30.5,0.007)(30.5,0)
\pspolygon*[linecolor=white](30.5,0)(30.5,0.116)(31.5,0.116)(31.5,0)
\pspolygon*[linecolor=white](31.5,0)(31.5,0.097)(32.5,0.097)(32.5,0)
\pspolygon*[linecolor=gray](32.5,0)(32.5,0.015)(33.5,0.015)(33.5,0)
\pspolygon*[linecolor=gray](33.5,0)(33.5,0.080)(34.5,0.080)(34.5,0)
\pspolygon*[linecolor=gray](34.5,0)(34.5,0.032)(35.5,0.032)(35.5,0)
\pspolygon*[linecolor=gray](35.5,0)(35.5,0.116)(36.5,0.116)(36.5,0)
\pspolygon*[linecolor=gray](36.5,0)(36.5,0.168)(37.5,0.168)(37.5,0)
\pspolygon*[linecolor=white](37.5,0)(37.5,0.073)(38.5,0.073)(38.5,0)
\pspolygon*[linecolor=gray](38.5,0)(38.5,0.074)(39.5,0.074)(39.5,0)
\pspolygon*[linecolor=gray](39.5,0)(39.5,0.023)(40.5,0.023)(40.5,0)
\pspolygon*[linecolor=gray](40.5,0)(40.5,0.147)(41.5,0.147)(41.5,0)
\pspolygon*[linecolor=gray](41.5,0)(41.5,0.021)(42.5,0.021)(42.5,0)
\pspolygon*[linecolor=gray](42.5,0)(42.5,0.200)(43.5,0.200)(43.5,0)
\pspolygon*[linecolor=gray](43.5,0)(43.5,0.178)(44.5,0.178)(44.5,0)
\pspolygon*[linecolor=gray](44.5,0)(44.5,0.015)(45.5,0.015)(45.5,0)
\pspolygon*[linecolor=gray](45.5,0)(45.5,0.036)(46.5,0.036)(46.5,0)
\pspolygon*[linecolor=white](46.5,0)(46.5,0.076)(47.5,0.076)(47.5,0)
\pspolygon*[linecolor=white](47.5,0)(47.5,0.059)(48.5,0.059)(48.5,0)
\pspolygon*[linecolor=gray](48.5,0)(48.5,0.103)(49.5,0.103)(49.5,0)
\pspolygon*[linecolor=white](49.5,0)(49.5,0.095)(50.5,0.095)(50.5,0)
\pspolygon*[linecolor=white](50.5,0)(50.5,0.018)(51.5,0.018)(51.5,0)
\pspolygon*[linecolor=gray](51.5,0)(51.5,0.030)(52.5,0.030)(52.5,0)
\pspolygon*[linecolor=gray](52.5,0)(52.5,0.127)(53.5,0.127)(53.5,0)
\pspolygon*[linecolor=gray](53.5,0)(53.5,0.045)(54.5,0.045)(54.5,0)
\pspolygon*[linecolor=gray](54.5,0)(54.5,0.052)(55.5,0.052)(55.5,0)
\pspolygon*[linecolor=gray](55.5,0)(55.5,0.190)(56.5,0.190)(56.5,0)
\pspolygon*[linecolor=gray](56.5,0)(56.5,0.255)(57.5,0.255)(57.5,0)
\pspolygon*[linecolor=gray](57.5,0)(57.5,0.117)(58.5,0.117)(58.5,0)
\pspolygon*[linecolor=white](58.5,0)(58.5,0.080)(59.5,0.080)(59.5,0)
\pspolygon*[linecolor=gray](59.5,0)(59.5,0.153)(60.5,0.153)(60.5,0)
\pspolygon*[linecolor=gray](60.5,0)(60.5,0.209)(61.5,0.209)(61.5,0)
\pspolygon*[linecolor=gray](61.5,0)(61.5,0.209)(62.5,0.209)(62.5,0)
\pspolygon*[linecolor=gray](62.5,0)(62.5,0.176)(63.5,0.176)(63.5,0)
\pspolygon*[linecolor=gray](63.5,0)(63.5,0.105)(64.5,0.105)(64.5,0)
\pspolygon*[linecolor=gray](64.5,0)(64.5,0.110)(65.5,0.110)(65.5,0)
\pspolygon*[linecolor=gray](65.5,0)(65.5,0.143)(66.5,0.143)(66.5,0)
\pspolygon*[linecolor=gray](66.5,0)(66.5,0.223)(67.5,0.223)(67.5,0)
\pspolygon*[linecolor=gray](67.5,0)(67.5,0.024)(68.5,0.024)(68.5,0)
\pspolygon*[linecolor=gray](68.5,0)(68.5,0.067)(69.5,0.067)(69.5,0)
\pspolygon*[linecolor=gray](69.5,0)(69.5,0.051)(70.5,0.051)(70.5,0)
\pspolygon*[linecolor=white](70.5,0)(70.5,0.115)(71.5,0.115)(71.5,0)
\pspolygon*[linecolor=white](71.5,0)(71.5,0.047)(72.5,0.047)(72.5,0)
\pspolygon*[linecolor=white](72.5,0)(72.5,0.082)(73.5,0.082)(73.5,0)
\pspolygon*[linecolor=gray](73.5,0)(73.5,0.068)(74.5,0.068)(74.5,0)
\pspolygon*[linecolor=gray](74.5,0)(74.5,0.080)(75.5,0.080)(75.5,0)
\pspolygon(0.5,0)(0.5,0.362)(1.5,0.362)(1.5,0)
\pspolygon(1.5,0)(1.5,0.216)(2.5,0.216)(2.5,0)
\pspolygon(2.5,0)(2.5,0.194)(3.5,0.194)(3.5,0)
\pspolygon(3.5,0)(3.5,0.091)(4.5,0.091)(4.5,0)
\pspolygon(4.5,0)(4.5,0.157)(5.5,0.157)(5.5,0)
\pspolygon(5.5,0)(5.5,0.006)(6.5,0.006)(6.5,0)
\pspolygon(6.5,0)(6.5,0.023)(7.5,0.023)(7.5,0)
\pspolygon(7.5,0)(7.5,0.089)(8.5,0.089)(8.5,0)
\pspolygon(8.5,0)(8.5,0.016)(9.5,0.016)(9.5,0)
\pspolygon(9.5,0)(9.5,0.150)(10.5,0.150)(10.5,0)
\pspolygon(10.5,0)(10.5,0.281)(11.5,0.281)(11.5,0)
\pspolygon(11.5,0)(11.5,0.026)(12.5,0.026)(12.5,0)
\pspolygon(12.5,0)(12.5,0.039)(13.5,0.039)(13.5,0)
\pspolygon(13.5,0)(13.5,0.025)(14.5,0.025)(14.5,0)
\pspolygon(14.5,0)(14.5,0.019)(15.5,0.019)(15.5,0)
\pspolygon(15.5,0)(15.5,0.147)(16.5,0.147)(16.5,0)
\pspolygon(16.5,0)(16.5,0.121)(17.5,0.121)(17.5,0)
\pspolygon(17.5,0)(17.5,0.144)(18.5,0.144)(18.5,0)
\pspolygon(18.5,0)(18.5,0.090)(19.5,0.090)(19.5,0)
\pspolygon(19.5,0)(19.5,0.011)(20.5,0.011)(20.5,0)
\pspolygon(20.5,0)(20.5,0.147)(21.5,0.147)(21.5,0)
\pspolygon(21.5,0)(21.5,0.183)(22.5,0.183)(22.5,0)
\pspolygon(22.5,0)(22.5,0.002)(23.5,0.002)(23.5,0)
\pspolygon(23.5,0)(23.5,0.031)(24.5,0.031)(24.5,0)
\pspolygon(24.5,0)(24.5,0.057)(25.5,0.057)(25.5,0)
\pspolygon(25.5,0)(25.5,0.116)(26.5,0.116)(26.5,0)
\pspolygon(26.5,0)(26.5,0.056)(27.5,0.056)(27.5,0)
\pspolygon(27.5,0)(27.5,0.037)(28.5,0.037)(28.5,0)
\pspolygon(28.5,0)(28.5,0.019)(29.5,0.019)(29.5,0)
\pspolygon(29.5,0)(29.5,0.007)(30.5,0.007)(30.5,0)
\pspolygon(30.5,0)(30.5,0.116)(31.5,0.116)(31.5,0)
\pspolygon(31.5,0)(31.5,0.097)(32.5,0.097)(32.5,0)
\pspolygon(32.5,0)(32.5,0.015)(33.5,0.015)(33.5,0)
\pspolygon(33.5,0)(33.5,0.080)(34.5,0.080)(34.5,0)
\pspolygon(34.5,0)(34.5,0.032)(35.5,0.032)(35.5,0)
\pspolygon(35.5,0)(35.5,0.116)(36.5,0.116)(36.5,0)
\pspolygon(36.5,0)(36.5,0.168)(37.5,0.168)(37.5,0)
\pspolygon(37.5,0)(37.5,0.073)(38.5,0.073)(38.5,0)
\pspolygon(38.5,0)(38.5,0.074)(39.5,0.074)(39.5,0)
\pspolygon(39.5,0)(39.5,0.023)(40.5,0.023)(40.5,0)
\pspolygon(40.5,0)(40.5,0.147)(41.5,0.147)(41.5,0)
\pspolygon(41.5,0)(41.5,0.021)(42.5,0.021)(42.5,0)
\pspolygon(42.5,0)(42.5,0.200)(43.5,0.200)(43.5,0)
\pspolygon(43.5,0)(43.5,0.178)(44.5,0.178)(44.5,0)
\pspolygon(44.5,0)(44.5,0.015)(45.5,0.015)(45.5,0)
\pspolygon(45.5,0)(45.5,0.036)(46.5,0.036)(46.5,0)
\pspolygon(46.5,0)(46.5,0.076)(47.5,0.076)(47.5,0)
\pspolygon(47.5,0)(47.5,0.059)(48.5,0.059)(48.5,0)
\pspolygon(48.5,0)(48.5,0.103)(49.5,0.103)(49.5,0)
\pspolygon(49.5,0)(49.5,0.095)(50.5,0.095)(50.5,0)
\pspolygon(50.5,0)(50.5,0.018)(51.5,0.018)(51.5,0)
\pspolygon(51.5,0)(51.5,0.030)(52.5,0.030)(52.5,0)
\pspolygon(52.5,0)(52.5,0.127)(53.5,0.127)(53.5,0)
\pspolygon(53.5,0)(53.5,0.045)(54.5,0.045)(54.5,0)
\pspolygon(54.5,0)(54.5,0.052)(55.5,0.052)(55.5,0)
\pspolygon(55.5,0)(55.5,0.190)(56.5,0.190)(56.5,0)
\pspolygon(56.5,0)(56.5,0.255)(57.5,0.255)(57.5,0)
\pspolygon(57.5,0)(57.5,0.117)(58.5,0.117)(58.5,0)
\pspolygon(58.5,0)(58.5,0.080)(59.5,0.080)(59.5,0)
\pspolygon(59.5,0)(59.5,0.153)(60.5,0.153)(60.5,0)
\pspolygon(60.5,0)(60.5,0.209)(61.5,0.209)(61.5,0)
\pspolygon(61.5,0)(61.5,0.209)(62.5,0.209)(62.5,0)
\pspolygon(62.5,0)(62.5,0.176)(63.5,0.176)(63.5,0)
\pspolygon(63.5,0)(63.5,0.105)(64.5,0.105)(64.5,0)
\pspolygon(64.5,0)(64.5,0.110)(65.5,0.110)(65.5,0)
\pspolygon(65.5,0)(65.5,0.143)(66.5,0.143)(66.5,0)
\pspolygon(66.5,0)(66.5,0.223)(67.5,0.223)(67.5,0)
\pspolygon(67.5,0)(67.5,0.024)(68.5,0.024)(68.5,0)
\pspolygon(68.5,0)(68.5,0.067)(69.5,0.067)(69.5,0)
\pspolygon(69.5,0)(69.5,0.051)(70.5,0.051)(70.5,0)
\pspolygon(70.5,0)(70.5,0.115)(71.5,0.115)(71.5,0)
\pspolygon(71.5,0)(71.5,0.047)(72.5,0.047)(72.5,0)
\pspolygon(72.5,0)(72.5,0.082)(73.5,0.082)(73.5,0)
\pspolygon(73.5,0)(73.5,0.068)(74.5,0.068)(74.5,0)
\pspolygon(74.5,0)(74.5,0.080)(75.5,0.080)(75.5,0)
\psline{->}(0,0)(77,0) \psline{->}(0,0)(0,0.5) \rput(5.8,0.5){$e_{2}\ (02/2000)$} \scriptsize \psline(1,-0.02)(1,0.02) \rput(1,-0.06){S\&P} \psline(5,-0.02)(5,0.02) \rput(5,-0.06){Pana} \psline(10,-0.02)(10,0.02) \rput(10,-0.06){Arge} \psline(15,-0.02)(15,0.02) \rput(15,-0.06){Irel} \psline(20,-0.02)(20,0.02) \rput(20,-0.06){Ital} \psline(25,-0.02)(25,0.02) \rput(25,-0.06){Swed} \psline(30,-0.02)(30,0.02) \rput(30,-0.06){Spai} \psline(35,-0.02)(35,0.02) \rput(35,-0.06){Hung} \psline(40,-0.02)(40,0.02) \rput(40,-0.06){Lith} \psline(45,-0.02)(45,0.02) \rput(45,-0.06){Pale} \psline(50,-0.02)(50,0.02) \rput(49,-0.06){Ohma} \psline(55,-0.02)(55,0.02) \rput(55,-0.06){Japa} \psline(60,-0.02)(60,0.02) \rput(60,-0.06){SoKo} \psline(65,-0.02)(65,0.02) \rput(65,-0.06){Phil} \psline(70,-0.02)(70,0.02) \rput(70,-0.06){Ghan} \psline(74,-0.02)(74,0.02) \rput(74,-0.06){Maur} \scriptsize \psline(-0.28,0.1)(0.28,0.1) \rput(-1.38,0.1){$0.1$} \psline(-0.28,0.2)(0.28,0.2) \rput(-1.38,0.2){$0.2$} \psline(-0.28,0.3)(0.28,0.3) \rput(-1.38,0.3){$0.3$} \psline(-0.28,0.4)(0.28,0.4) \rput(-1.38,0.4){$0.4$}
\end{pspicture}

\vskip 0.6 cm

\noindent Fig. 20. Contributions of the stock market indices to eigenvector $e_{2}$, corresponding to the second largest eigenvalue of the correlation matrix. White bars indicate positive values, and gray bars indicate negative values, corresponding to the first and second semesters of 2000. The indices are aligned in the following way: {\bf S\&P}, Nasd, Cana, Mexi, {\bf Pana}, CoRi, Berm, Jama, Braz, {\bf Arge}, Chil, Vene, Peru, UK, {\bf Irel}, Fran, Germ, Swit, Autr, {\bf Ita}, Malt, Belg, Neth, Luxe, {\bf Swed}, Denm, Finl, Norw, Icel, {\bf Spai}, Port, Gree, CzRe, Slok, {\bf Hung}, Pola, Roma, Esto, Latv, {\bf Lith}, Ukra, Russ, Turk, Isra, {\bf Pale}, Leba, Jord, SaAr, Qata, {\bf Ohma}, Paki, Indi, SrLa, Bang, {\bf Japa}, HoKo, Chin, Mong, Taiw, {\bf SoKo}, Thai, Mala, Sing, Indo, {\bf Phil}, Aust, Moro, Tuni, Egyp, {\bf Ghan}, Nige, Keny, SoAf, {\bf Maur}.

\begin{pspicture}(-0.1,0)(3.5,3.1)
\psset{xunit=0.21,yunit=4.5}
\pspolygon*[linecolor=white](0.5,0)(0.5,0.362)(1.5,0.362)(1.5,0)
\pspolygon*[linecolor=white](1.5,0)(1.5,0.280)(2.5,0.280)(2.5,0)
\pspolygon*[linecolor=white](2.5,0)(2.5,0.238)(3.5,0.238)(3.5,0)
\pspolygon*[linecolor=white](3.5,0)(3.5,0.233)(4.5,0.233)(4.5,0)
\pspolygon*[linecolor=white](4.5,0)(4.5,0.126)(5.5,0.126)(5.5,0)
\pspolygon*[linecolor=gray](5.5,0)(5.5,0.063)(6.5,0.063)(6.5,0)
\pspolygon*[linecolor=white](6.5,0)(6.5,0.019)(7.5,0.019)(7.5,0)
\pspolygon*[linecolor=white](7.5,0)(7.5,0.040)(8.5,0.040)(8.5,0)
\pspolygon*[linecolor=white](8.5,0)(8.5,0.024)(9.5,0.024)(9.5,0)
\pspolygon*[linecolor=white](9.5,0)(9.5,0.100)(10.5,0.100)(10.5,0)
\pspolygon*[linecolor=white](10.5,0)(10.5,0.098)(11.5,0.098)(11.5,0)
\pspolygon*[linecolor=gray](11.5,0)(11.5,0.032)(12.5,0.032)(12.5,0)
\pspolygon*[linecolor=white](12.5,0)(12.5,0.001)(13.5,0.001)(13.5,0)
\pspolygon*[linecolor=gray](13.5,0)(13.5,0.018)(14.5,0.018)(14.5,0)
\pspolygon*[linecolor=white](14.5,0)(14.5,0.121)(15.5,0.121)(15.5,0)
\pspolygon*[linecolor=gray](15.5,0)(15.5,0.026)(16.5,0.026)(16.5,0)
\pspolygon*[linecolor=white](16.5,0)(16.5,0.063)(17.5,0.063)(17.5,0)
\pspolygon*[linecolor=white](17.5,0)(17.5,0.130)(18.5,0.130)(18.5,0)
\pspolygon*[linecolor=white](18.5,0)(18.5,0.093)(19.5,0.093)(19.5,0)
\pspolygon*[linecolor=white](19.5,0)(19.5,0.036)(20.5,0.036)(20.5,0)
\pspolygon*[linecolor=white](20.5,0)(20.5,0.083)(21.5,0.083)(21.5,0)
\pspolygon*[linecolor=white](21.5,0)(21.5,0.093)(22.5,0.093)(22.5,0)
\pspolygon*[linecolor=white](22.5,0)(22.5,0.043)(23.5,0.043)(23.5,0)
\pspolygon*[linecolor=white](23.5,0)(23.5,0.037)(24.5,0.037)(24.5,0)
\pspolygon*[linecolor=gray](24.5,0)(24.5,0.161)(25.5,0.161)(25.5,0)
\pspolygon*[linecolor=white](25.5,0)(25.5,0.089)(26.5,0.089)(26.5,0)
\pspolygon*[linecolor=gray](26.5,0)(26.5,0.047)(27.5,0.047)(27.5,0)
\pspolygon*[linecolor=gray](27.5,0)(27.5,0.000)(28.5,0.000)(28.5,0)
\pspolygon*[linecolor=gray](28.5,0)(28.5,0.024)(29.5,0.024)(29.5,0)
\pspolygon*[linecolor=gray](29.5,0)(29.5,0.124)(30.5,0.124)(30.5,0)
\pspolygon*[linecolor=white](30.5,0)(30.5,0.106)(31.5,0.106)(31.5,0)
\pspolygon*[linecolor=white](31.5,0)(31.5,0.048)(32.5,0.048)(32.5,0)
\pspolygon*[linecolor=white](32.5,0)(32.5,0.003)(33.5,0.003)(33.5,0)
\pspolygon*[linecolor=gray](33.5,0)(33.5,0.102)(34.5,0.102)(34.5,0)
\pspolygon*[linecolor=gray](34.5,0)(34.5,0.010)(35.5,0.010)(35.5,0)
\pspolygon*[linecolor=gray](35.5,0)(35.5,0.046)(36.5,0.046)(36.5,0)
\pspolygon*[linecolor=gray](36.5,0)(36.5,0.044)(37.5,0.044)(37.5,0)
\pspolygon*[linecolor=white](37.5,0)(37.5,0.013)(38.5,0.013)(38.5,0)
\pspolygon*[linecolor=gray](38.5,0)(38.5,0.007)(39.5,0.007)(39.5,0)
\pspolygon*[linecolor=gray](39.5,0)(39.5,0.142)(40.5,0.142)(40.5,0)
\pspolygon*[linecolor=gray](40.5,0)(40.5,0.048)(41.5,0.048)(41.5,0)
\pspolygon*[linecolor=gray](41.5,0)(41.5,0.176)(42.5,0.176)(42.5,0)
\pspolygon*[linecolor=gray](42.5,0)(42.5,0.096)(43.5,0.096)(43.5,0)
\pspolygon*[linecolor=gray](43.5,0)(43.5,0.087)(44.5,0.087)(44.5,0)
\pspolygon*[linecolor=white](44.5,0)(44.5,0.037)(45.5,0.037)(45.5,0)
\pspolygon*[linecolor=gray](45.5,0)(45.5,0.022)(46.5,0.022)(46.5,0)
\pspolygon*[linecolor=gray](46.5,0)(46.5,0.033)(47.5,0.033)(47.5,0)
\pspolygon*[linecolor=white](47.5,0)(47.5,0.017)(48.5,0.017)(48.5,0)
\pspolygon*[linecolor=gray](48.5,0)(48.5,0.063)(49.5,0.063)(49.5,0)
\pspolygon*[linecolor=white](49.5,0)(49.5,0.052)(50.5,0.052)(50.5,0)
\pspolygon*[linecolor=gray](50.5,0)(50.5,0.052)(51.5,0.052)(51.5,0)
\pspolygon*[linecolor=gray](51.5,0)(51.5,0.007)(52.5,0.007)(52.5,0)
\pspolygon*[linecolor=gray](52.5,0)(52.5,0.032)(53.5,0.032)(53.5,0)
\pspolygon*[linecolor=white](53.5,0)(53.5,0.076)(54.5,0.076)(54.5,0)
\pspolygon*[linecolor=gray](54.5,0)(54.5,0.121)(55.5,0.121)(55.5,0)
\pspolygon*[linecolor=gray](55.5,0)(55.5,0.124)(56.5,0.124)(56.5,0)
\pspolygon*[linecolor=gray](56.5,0)(56.5,0.096)(57.5,0.096)(57.5,0)
\pspolygon*[linecolor=gray](57.5,0)(57.5,0.229)(58.5,0.229)(58.5,0)
\pspolygon*[linecolor=gray](58.5,0)(58.5,0.260)(59.5,0.260)(59.5,0)
\pspolygon*[linecolor=white](59.5,0)(59.5,0.086)(60.5,0.086)(60.5,0)
\pspolygon*[linecolor=gray](60.5,0)(60.5,0.069)(61.5,0.069)(61.5,0)
\pspolygon*[linecolor=gray](61.5,0)(61.5,0.199)(62.5,0.199)(62.5,0)
\pspolygon*[linecolor=gray](62.5,0)(62.5,0.286)(63.5,0.286)(63.5,0)
\pspolygon*[linecolor=gray](63.5,0)(63.5,0.183)(64.5,0.183)(64.5,0)
\pspolygon*[linecolor=gray](64.5,0)(64.5,0.039)(65.5,0.039)(65.5,0)
\pspolygon*[linecolor=gray](65.5,0)(65.5,0.183)(66.5,0.183)(66.5,0)
\pspolygon*[linecolor=gray](66.5,0)(66.5,0.235)(67.5,0.235)(67.5,0)
\pspolygon*[linecolor=gray](67.5,0)(67.5,0.122)(68.5,0.122)(68.5,0)
\pspolygon*[linecolor=gray](68.5,0)(68.5,0.105)(69.5,0.105)(69.5,0)
\pspolygon*[linecolor=gray](69.5,0)(69.5,0.188)(70.5,0.188)(70.5,0)
\pspolygon*[linecolor=gray](70.5,0)(70.5,0.166)(71.5,0.166)(71.5,0)
\pspolygon*[linecolor=white](71.5,0)(71.5,0.027)(72.5,0.027)(72.5,0)
\pspolygon*[linecolor=gray](72.5,0)(72.5,0.063)(73.5,0.063)(73.5,0)
\pspolygon*[linecolor=gray](73.5,0)(73.5,0.099)(74.5,0.099)(74.5,0)
\pspolygon*[linecolor=white](74.5,0)(74.5,0.077)(75.5,0.077)(75.5,0)
\pspolygon*[linecolor=gray](75.5,0)(75.5,0.031)(76.5,0.031)(76.5,0)
\pspolygon*[linecolor=white](76.5,0)(76.5,0.005)(77.5,0.005)(77.5,0)
\pspolygon*[linecolor=white](77.5,0)(77.5,0.018)(78.5,0.018)(78.5,0)
\pspolygon*[linecolor=white](78.5,0)(78.5,0.036)(79.5,0.036)(79.5,0)
\pspolygon*[linecolor=gray](79.5,0)(79.5,0.072)(80.5,0.072)(80.5,0)
\pspolygon(0.5,0)(0.5,0.362)(1.5,0.362)(1.5,0)
\pspolygon(1.5,0)(1.5,0.280)(2.5,0.280)(2.5,0)
\pspolygon(2.5,0)(2.5,0.238)(3.5,0.238)(3.5,0)
\pspolygon(3.5,0)(3.5,0.233)(4.5,0.233)(4.5,0)
\pspolygon(4.5,0)(4.5,0.126)(5.5,0.126)(5.5,0)
\pspolygon(5.5,0)(5.5,0.063)(6.5,0.063)(6.5,0)
\pspolygon(6.5,0)(6.5,0.019)(7.5,0.019)(7.5,0)
\pspolygon(7.5,0)(7.5,0.040)(8.5,0.040)(8.5,0)
\pspolygon(8.5,0)(8.5,0.024)(9.5,0.024)(9.5,0)
\pspolygon(9.5,0)(9.5,0.100)(10.5,0.100)(10.5,0)
\pspolygon(10.5,0)(10.5,0.098)(11.5,0.098)(11.5,0)
\pspolygon(11.5,0)(11.5,0.032)(12.5,0.032)(12.5,0)
\pspolygon(12.5,0)(12.5,0.001)(13.5,0.001)(13.5,0)
\pspolygon(13.5,0)(13.5,0.018)(14.5,0.018)(14.5,0)
\pspolygon(14.5,0)(14.5,0.121)(15.5,0.121)(15.5,0)
\pspolygon(15.5,0)(15.5,0.026)(16.5,0.026)(16.5,0)
\pspolygon(16.5,0)(16.5,0.063)(17.5,0.063)(17.5,0)
\pspolygon(17.5,0)(17.5,0.130)(18.5,0.130)(18.5,0)
\pspolygon(18.5,0)(18.5,0.093)(19.5,0.093)(19.5,0)
\pspolygon(19.5,0)(19.5,0.036)(20.5,0.036)(20.5,0)
\pspolygon(20.5,0)(20.5,0.083)(21.5,0.083)(21.5,0)
\pspolygon(21.5,0)(21.5,0.093)(22.5,0.093)(22.5,0)
\pspolygon(22.5,0)(22.5,0.043)(23.5,0.043)(23.5,0)
\pspolygon(23.5,0)(23.5,0.037)(24.5,0.037)(24.5,0)
\pspolygon(24.5,0)(24.5,0.161)(25.5,0.161)(25.5,0)
\pspolygon(25.5,0)(25.5,0.089)(26.5,0.089)(26.5,0)
\pspolygon(26.5,0)(26.5,0.047)(27.5,0.047)(27.5,0)
\pspolygon(27.5,0)(27.5,0.000)(28.5,0.000)(28.5,0)
\pspolygon(28.5,0)(28.5,0.024)(29.5,0.024)(29.5,0)
\pspolygon(29.5,0)(29.5,0.124)(30.5,0.124)(30.5,0)
\pspolygon(30.5,0)(30.5,0.106)(31.5,0.106)(31.5,0)
\pspolygon(31.5,0)(31.5,0.048)(32.5,0.048)(32.5,0)
\pspolygon(32.5,0)(32.5,0.003)(33.5,0.003)(33.5,0)
\pspolygon(33.5,0)(33.5,0.102)(34.5,0.102)(34.5,0)
\pspolygon(34.5,0)(34.5,0.010)(35.5,0.010)(35.5,0)
\pspolygon(35.5,0)(35.5,0.046)(36.5,0.046)(36.5,0)
\pspolygon(36.5,0)(36.5,0.044)(37.5,0.044)(37.5,0)
\pspolygon(37.5,0)(37.5,0.013)(38.5,0.013)(38.5,0)
\pspolygon(38.5,0)(38.5,0.007)(39.5,0.007)(39.5,0)
\pspolygon(39.5,0)(39.5,0.142)(40.5,0.142)(40.5,0)
\pspolygon(40.5,0)(40.5,0.048)(41.5,0.048)(41.5,0)
\pspolygon(41.5,0)(41.5,0.176)(42.5,0.176)(42.5,0)
\pspolygon(42.5,0)(42.5,0.096)(43.5,0.096)(43.5,0)
\pspolygon(43.5,0)(43.5,0.087)(44.5,0.087)(44.5,0)
\pspolygon(44.5,0)(44.5,0.037)(45.5,0.037)(45.5,0)
\pspolygon(45.5,0)(45.5,0.022)(46.5,0.022)(46.5,0)
\pspolygon(46.5,0)(46.5,0.033)(47.5,0.033)(47.5,0)
\pspolygon(47.5,0)(47.5,0.017)(48.5,0.017)(48.5,0)
\pspolygon(48.5,0)(48.5,0.063)(49.5,0.063)(49.5,0)
\pspolygon(49.5,0)(49.5,0.052)(50.5,0.052)(50.5,0)
\pspolygon(50.5,0)(50.5,0.052)(51.5,0.052)(51.5,0)
\pspolygon(51.5,0)(51.5,0.007)(52.5,0.007)(52.5,0)
\pspolygon(52.5,0)(52.5,0.032)(53.5,0.032)(53.5,0)
\pspolygon(53.5,0)(53.5,0.076)(54.5,0.076)(54.5,0)
\pspolygon(54.5,0)(54.5,0.121)(55.5,0.121)(55.5,0)
\pspolygon(55.5,0)(55.5,0.124)(56.5,0.124)(56.5,0)
\pspolygon(56.5,0)(56.5,0.096)(57.5,0.096)(57.5,0)
\pspolygon(57.5,0)(57.5,0.229)(58.5,0.229)(58.5,0)
\pspolygon(58.5,0)(58.5,0.260)(59.5,0.260)(59.5,0)
\pspolygon(59.5,0)(59.5,0.086)(60.5,0.086)(60.5,0)
\pspolygon(60.5,0)(60.5,0.069)(61.5,0.069)(61.5,0)
\pspolygon(61.5,0)(61.5,0.199)(62.5,0.199)(62.5,0)
\pspolygon(62.5,0)(62.5,0.286)(63.5,0.286)(63.5,0)
\pspolygon(63.5,0)(63.5,0.183)(64.5,0.183)(64.5,0)
\pspolygon(64.5,0)(64.5,0.039)(65.5,0.039)(65.5,0)
\pspolygon(65.5,0)(65.5,0.183)(66.5,0.183)(66.5,0)
\pspolygon(66.5,0)(66.5,0.235)(67.5,0.235)(67.5,0)
\pspolygon(67.5,0)(67.5,0.122)(68.5,0.122)(68.5,0)
\pspolygon(68.5,0)(68.5,0.105)(69.5,0.105)(69.5,0)
\pspolygon(69.5,0)(69.5,0.188)(70.5,0.188)(70.5,0)
\pspolygon(70.5,0)(70.5,0.166)(71.5,0.166)(71.5,0)
\pspolygon(71.5,0)(71.5,0.027)(72.5,0.027)(72.5,0)
\pspolygon(72.5,0)(72.5,0.063)(73.5,0.063)(73.5,0)
\pspolygon(73.5,0)(73.5,0.099)(74.5,0.099)(74.5,0)
\pspolygon(74.5,0)(74.5,0.077)(75.5,0.077)(75.5,0)
\pspolygon(75.5,0)(75.5,0.031)(76.5,0.031)(76.5,0)
\pspolygon(76.5,0)(76.5,0.005)(77.5,0.005)(77.5,0)
\pspolygon(77.5,0)(77.5,0.018)(78.5,0.018)(78.5,0)
\pspolygon(78.5,0)(78.5,0.036)(79.5,0.036)(79.5,0)
\pspolygon(79.5,0)(79.5,0.072)(80.5,0.072)(80.5,0)
\psline{->}(0,0)(83,0) \psline{->}(0,0)(0,0.5) \rput(5.8,0.5){$e_{2}\ (01/2001)$} \scriptsize \psline(1,-0.02)(1,0.02) \rput(1,-0.06){S\&P} \psline(5,-0.02)(5,0.02) \rput(5,-0.06){Pana} \psline(10,-0.02)(10,0.02) \rput(10,-0.06){Arge} \psline(15,-0.02)(15,0.02) \rput(15,-0.06){Irel} \psline(20,-0.02)(20,0.02) \rput(20,-0.06){Ital} \psline(25,-0.02)(25,0.02) \rput(25,-0.06){Swed} \psline(30,-0.02)(30,0.02) \rput(30,-0.06){Spai} \psline(35,-0.02)(35,0.02) \rput(35,-0.06){Hung} \psline(40,-0.02)(40,0.02) \rput(40,-0.06){Latv} \psline(45,-0.02)(45,0.02) \rput(45,-0.06){Turk} \psline(50,-0.02)(50,0.02) \rput(49,-0.06){SaAr} \psline(55,-0.02)(55,0.02) \rput(55,-0.06){SrLa} \psline(60,-0.02)(60,0.02) \rput(60,-0.06){Mongo} \psline(65,-0.02)(65,0.02) \rput(65,-0.06){Mala} \psline(70,-0.02)(70,0.02) \rput(70,-0.06){NeZe} \psline(75,-0.02)(75,0.02) \rput(75,-0.06){Nige} \psline(79,-0.02)(79,0.02) \rput(79,-0.06){Maur} \scriptsize \psline(-0.28,0.1)(0.28,0.1) \rput(-1.38,0.1){$0.1$} \psline(-0.28,0.2)(0.28,0.2) \rput(-1.38,0.2){$0.2$} \psline(-0.28,0.3)(0.28,0.3) \rput(-1.38,0.3){$0.3$} \psline(-0.28,0.4)(0.28,0.4) \rput(-1.38,0.4){$0.4$}
\end{pspicture}

\begin{pspicture}(-0.1,0)(3.5,3.1)
\psset{xunit=0.21,yunit=4.5}
\pspolygon*[linecolor=white](0.5,0)(0.5,0.083)(1.5,0.083)(1.5,0)
\pspolygon*[linecolor=white](1.5,0)(1.5,0.089)(2.5,0.089)(2.5,0)
\pspolygon*[linecolor=white](2.5,0)(2.5,0.128)(3.5,0.128)(3.5,0)
\pspolygon*[linecolor=white](3.5,0)(3.5,0.048)(4.5,0.048)(4.5,0)
\pspolygon*[linecolor=gray](4.5,0)(4.5,0.053)(5.5,0.053)(5.5,0)
\pspolygon*[linecolor=white](5.5,0)(5.5,0.051)(6.5,0.051)(6.5,0)
\pspolygon*[linecolor=gray](6.5,0)(6.5,0.052)(7.5,0.052)(7.5,0)
\pspolygon*[linecolor=gray](7.5,0)(7.5,0.036)(8.5,0.036)(8.5,0)
\pspolygon*[linecolor=white](8.5,0)(8.5,0.087)(9.5,0.087)(9.5,0)
\pspolygon*[linecolor=white](9.5,0)(9.5,0.068)(10.5,0.068)(10.5,0)
\pspolygon*[linecolor=white](10.5,0)(10.5,0.131)(11.5,0.131)(11.5,0)
\pspolygon*[linecolor=white](11.5,0)(11.5,0.028)(12.5,0.028)(12.5,0)
\pspolygon*[linecolor=white](12.5,0)(12.5,0.076)(13.5,0.076)(13.5,0)
\pspolygon*[linecolor=white](13.5,0)(13.5,0.152)(14.5,0.152)(14.5,0)
\pspolygon*[linecolor=gray](14.5,0)(14.5,0.060)(15.5,0.060)(15.5,0)
\pspolygon*[linecolor=white](15.5,0)(15.5,0.162)(16.5,0.162)(16.5,0)
\pspolygon*[linecolor=white](16.5,0)(16.5,0.180)(17.5,0.180)(17.5,0)
\pspolygon*[linecolor=white](17.5,0)(17.5,0.122)(18.5,0.122)(18.5,0)
\pspolygon*[linecolor=gray](18.5,0)(18.5,0.061)(19.5,0.061)(19.5,0)
\pspolygon*[linecolor=white](19.5,0)(19.5,0.155)(20.5,0.155)(20.5,0)
\pspolygon*[linecolor=white](20.5,0)(20.5,0.013)(21.5,0.013)(21.5,0)
\pspolygon*[linecolor=white](21.5,0)(21.5,0.085)(22.5,0.085)(22.5,0)
\pspolygon*[linecolor=white](22.5,0)(22.5,0.135)(23.5,0.135)(23.5,0)
\pspolygon*[linecolor=gray](23.5,0)(23.5,0.111)(24.5,0.111)(24.5,0)
\pspolygon*[linecolor=white](24.5,0)(24.5,0.162)(25.5,0.162)(25.5,0)
\pspolygon*[linecolor=gray](25.5,0)(25.5,0.009)(26.5,0.009)(26.5,0)
\pspolygon*[linecolor=white](26.5,0)(26.5,0.088)(27.5,0.088)(27.5,0)
\pspolygon*[linecolor=gray](27.5,0)(27.5,0.053)(28.5,0.053)(28.5,0)
\pspolygon*[linecolor=gray](28.5,0)(28.5,0.070)(29.5,0.070)(29.5,0)
\pspolygon*[linecolor=white](29.5,0)(29.5,0.146)(30.5,0.146)(30.5,0)
\pspolygon*[linecolor=gray](30.5,0)(30.5,0.017)(31.5,0.017)(31.5,0)
\pspolygon*[linecolor=gray](31.5,0)(31.5,0.086)(32.5,0.086)(32.5,0)
\pspolygon*[linecolor=gray](32.5,0)(32.5,0.063)(33.5,0.063)(33.5,0)
\pspolygon*[linecolor=white](33.5,0)(33.5,0.040)(34.5,0.040)(34.5,0)
\pspolygon*[linecolor=gray](34.5,0)(34.5,0.083)(35.5,0.083)(35.5,0)
\pspolygon*[linecolor=gray](35.5,0)(35.5,0.106)(36.5,0.106)(36.5,0)
\pspolygon*[linecolor=gray](36.5,0)(36.5,0.006)(37.5,0.006)(37.5,0)
\pspolygon*[linecolor=gray](37.5,0)(37.5,0.054)(38.5,0.054)(38.5,0)
\pspolygon*[linecolor=gray](38.5,0)(38.5,0.196)(39.5,0.196)(39.5,0)
\pspolygon*[linecolor=gray](39.5,0)(39.5,0.070)(40.5,0.070)(40.5,0)
\pspolygon*[linecolor=gray](40.5,0)(40.5,0.156)(41.5,0.156)(41.5,0)
\pspolygon*[linecolor=gray](41.5,0)(41.5,0.095)(42.5,0.095)(42.5,0)
\pspolygon*[linecolor=gray](42.5,0)(42.5,0.187)(43.5,0.187)(43.5,0)
\pspolygon*[linecolor=gray](43.5,0)(43.5,0.061)(44.5,0.061)(44.5,0)
\pspolygon*[linecolor=gray](44.5,0)(44.5,0.081)(45.5,0.081)(45.5,0)
\pspolygon*[linecolor=white](45.5,0)(45.5,0.000)(46.5,0.000)(46.5,0)
\pspolygon*[linecolor=gray](46.5,0)(46.5,0.087)(47.5,0.087)(47.5,0)
\pspolygon*[linecolor=gray](47.5,0)(47.5,0.017)(48.5,0.017)(48.5,0)
\pspolygon*[linecolor=gray](48.5,0)(48.5,0.179)(49.5,0.179)(49.5,0)
\pspolygon*[linecolor=white](49.5,0)(49.5,0.008)(50.5,0.008)(50.5,0)
\pspolygon*[linecolor=gray](50.5,0)(50.5,0.057)(51.5,0.057)(51.5,0)
\pspolygon*[linecolor=gray](51.5,0)(51.5,0.066)(52.5,0.066)(52.5,0)
\pspolygon*[linecolor=gray](52.5,0)(52.5,0.118)(53.5,0.118)(53.5,0)
\pspolygon*[linecolor=gray](53.5,0)(53.5,0.179)(54.5,0.179)(54.5,0)
\pspolygon*[linecolor=gray](54.5,0)(54.5,0.076)(55.5,0.076)(55.5,0)
\pspolygon*[linecolor=white](55.5,0)(55.5,0.012)(56.5,0.012)(56.5,0)
\pspolygon*[linecolor=gray](56.5,0)(56.5,0.198)(57.5,0.198)(57.5,0)
\pspolygon*[linecolor=gray](57.5,0)(57.5,0.211)(58.5,0.211)(58.5,0)
\pspolygon*[linecolor=gray](58.5,0)(58.5,0.161)(59.5,0.161)(59.5,0)
\pspolygon*[linecolor=white](59.5,0)(59.5,0.034)(60.5,0.034)(60.5,0)
\pspolygon*[linecolor=gray](60.5,0)(60.5,0.116)(61.5,0.116)(61.5,0)
\pspolygon*[linecolor=gray](61.5,0)(61.5,0.190)(62.5,0.190)(62.5,0)
\pspolygon*[linecolor=gray](62.5,0)(62.5,0.137)(63.5,0.137)(63.5,0)
\pspolygon*[linecolor=gray](63.5,0)(63.5,0.090)(64.5,0.090)(64.5,0)
\pspolygon*[linecolor=gray](64.5,0)(64.5,0.116)(65.5,0.116)(65.5,0)
\pspolygon*[linecolor=gray](65.5,0)(65.5,0.241)(66.5,0.241)(66.5,0)
\pspolygon*[linecolor=gray](66.5,0)(66.5,0.148)(67.5,0.148)(67.5,0)
\pspolygon*[linecolor=gray](67.5,0)(67.5,0.155)(68.5,0.155)(68.5,0)
\pspolygon*[linecolor=gray](68.5,0)(68.5,0.238)(69.5,0.238)(69.5,0)
\pspolygon*[linecolor=gray](69.5,0)(69.5,0.139)(70.5,0.139)(70.5,0)
\pspolygon*[linecolor=gray](70.5,0)(70.5,0.052)(71.5,0.052)(71.5,0)
\pspolygon*[linecolor=gray](71.5,0)(71.5,0.047)(72.5,0.047)(72.5,0)
\pspolygon*[linecolor=gray](72.5,0)(72.5,0.111)(73.5,0.111)(73.5,0)
\pspolygon*[linecolor=gray](73.5,0)(73.5,0.078)(74.5,0.078)(74.5,0)
\pspolygon*[linecolor=gray](74.5,0)(74.5,0.005)(75.5,0.005)(75.5,0)
\pspolygon*[linecolor=gray](75.5,0)(75.5,0.048)(76.5,0.048)(76.5,0)
\pspolygon*[linecolor=gray](76.5,0)(76.5,0.102)(77.5,0.102)(77.5,0)
\pspolygon*[linecolor=gray](77.5,0)(77.5,0.127)(78.5,0.127)(78.5,0)
\pspolygon*[linecolor=gray](78.5,0)(78.5,0.069)(79.5,0.069)(79.5,0)
\pspolygon*[linecolor=gray](79.5,0)(79.5,0.072)(80.5,0.072)(80.5,0)
\pspolygon(0.5,0)(0.5,0.083)(1.5,0.083)(1.5,0)
\pspolygon(1.5,0)(1.5,0.089)(2.5,0.089)(2.5,0)
\pspolygon(2.5,0)(2.5,0.128)(3.5,0.128)(3.5,0)
\pspolygon(3.5,0)(3.5,0.048)(4.5,0.048)(4.5,0)
\pspolygon(4.5,0)(4.5,0.053)(5.5,0.053)(5.5,0)
\pspolygon(5.5,0)(5.5,0.051)(6.5,0.051)(6.5,0)
\pspolygon(6.5,0)(6.5,0.052)(7.5,0.052)(7.5,0)
\pspolygon(7.5,0)(7.5,0.036)(8.5,0.036)(8.5,0)
\pspolygon(8.5,0)(8.5,0.087)(9.5,0.087)(9.5,0)
\pspolygon(9.5,0)(9.5,0.068)(10.5,0.068)(10.5,0)
\pspolygon(10.5,0)(10.5,0.131)(11.5,0.131)(11.5,0)
\pspolygon(11.5,0)(11.5,0.028)(12.5,0.028)(12.5,0)
\pspolygon(12.5,0)(12.5,0.076)(13.5,0.076)(13.5,0)
\pspolygon(13.5,0)(13.5,0.152)(14.5,0.152)(14.5,0)
\pspolygon(14.5,0)(14.5,0.060)(15.5,0.060)(15.5,0)
\pspolygon(15.5,0)(15.5,0.162)(16.5,0.162)(16.5,0)
\pspolygon(16.5,0)(16.5,0.180)(17.5,0.180)(17.5,0)
\pspolygon(17.5,0)(17.5,0.122)(18.5,0.122)(18.5,0)
\pspolygon(18.5,0)(18.5,0.061)(19.5,0.061)(19.5,0)
\pspolygon(19.5,0)(19.5,0.155)(20.5,0.155)(20.5,0)
\pspolygon(20.5,0)(20.5,0.013)(21.5,0.013)(21.5,0)
\pspolygon(21.5,0)(21.5,0.085)(22.5,0.085)(22.5,0)
\pspolygon(22.5,0)(22.5,0.135)(23.5,0.135)(23.5,0)
\pspolygon(23.5,0)(23.5,0.111)(24.5,0.111)(24.5,0)
\pspolygon(24.5,0)(24.5,0.162)(25.5,0.162)(25.5,0)
\pspolygon(25.5,0)(25.5,0.009)(26.5,0.009)(26.5,0)
\pspolygon(26.5,0)(26.5,0.088)(27.5,0.088)(27.5,0)
\pspolygon(27.5,0)(27.5,0.053)(28.5,0.053)(28.5,0)
\pspolygon(28.5,0)(28.5,0.070)(29.5,0.070)(29.5,0)
\pspolygon(29.5,0)(29.5,0.146)(30.5,0.146)(30.5,0)
\pspolygon(30.5,0)(30.5,0.017)(31.5,0.017)(31.5,0)
\pspolygon(31.5,0)(31.5,0.086)(32.5,0.086)(32.5,0)
\pspolygon(32.5,0)(32.5,0.063)(33.5,0.063)(33.5,0)
\pspolygon(33.5,0)(33.5,0.040)(34.5,0.040)(34.5,0)
\pspolygon(34.5,0)(34.5,0.083)(35.5,0.083)(35.5,0)
\pspolygon(35.5,0)(35.5,0.106)(36.5,0.106)(36.5,0)
\pspolygon(36.5,0)(36.5,0.006)(37.5,0.006)(37.5,0)
\pspolygon(37.5,0)(37.5,0.054)(38.5,0.054)(38.5,0)
\pspolygon(38.5,0)(38.5,0.196)(39.5,0.196)(39.5,0)
\pspolygon(39.5,0)(39.5,0.070)(40.5,0.070)(40.5,0)
\pspolygon(40.5,0)(40.5,0.156)(41.5,0.156)(41.5,0)
\pspolygon(41.5,0)(41.5,0.095)(42.5,0.095)(42.5,0)
\pspolygon(42.5,0)(42.5,0.187)(43.5,0.187)(43.5,0)
\pspolygon(43.5,0)(43.5,0.061)(44.5,0.061)(44.5,0)
\pspolygon(44.5,0)(44.5,0.081)(45.5,0.081)(45.5,0)
\pspolygon(45.5,0)(45.5,0.000)(46.5,0.000)(46.5,0)
\pspolygon(46.5,0)(46.5,0.087)(47.5,0.087)(47.5,0)
\pspolygon(47.5,0)(47.5,0.017)(48.5,0.017)(48.5,0)
\pspolygon(48.5,0)(48.5,0.179)(49.5,0.179)(49.5,0)
\pspolygon(49.5,0)(49.5,0.008)(50.5,0.008)(50.5,0)
\pspolygon(50.5,0)(50.5,0.057)(51.5,0.057)(51.5,0)
\pspolygon(51.5,0)(51.5,0.066)(52.5,0.066)(52.5,0)
\pspolygon(52.5,0)(52.5,0.118)(53.5,0.118)(53.5,0)
\pspolygon(53.5,0)(53.5,0.179)(54.5,0.179)(54.5,0)
\pspolygon(54.5,0)(54.5,0.076)(55.5,0.076)(55.5,0)
\pspolygon(55.5,0)(55.5,0.012)(56.5,0.012)(56.5,0)
\pspolygon(56.5,0)(56.5,0.198)(57.5,0.198)(57.5,0)
\pspolygon(57.5,0)(57.5,0.211)(58.5,0.211)(58.5,0)
\pspolygon(58.5,0)(58.5,0.161)(59.5,0.161)(59.5,0)
\pspolygon(59.5,0)(59.5,0.034)(60.5,0.034)(60.5,0)
\pspolygon(60.5,0)(60.5,0.116)(61.5,0.116)(61.5,0)
\pspolygon(61.5,0)(61.5,0.190)(62.5,0.190)(62.5,0)
\pspolygon(62.5,0)(62.5,0.137)(63.5,0.137)(63.5,0)
\pspolygon(63.5,0)(63.5,0.090)(64.5,0.090)(64.5,0)
\pspolygon(64.5,0)(64.5,0.116)(65.5,0.116)(65.5,0)
\pspolygon(65.5,0)(65.5,0.241)(66.5,0.241)(66.5,0)
\pspolygon(66.5,0)(66.5,0.148)(67.5,0.148)(67.5,0)
\pspolygon(67.5,0)(67.5,0.155)(68.5,0.155)(68.5,0)
\pspolygon(68.5,0)(68.5,0.238)(69.5,0.238)(69.5,0)
\pspolygon(69.5,0)(69.5,0.139)(70.5,0.139)(70.5,0)
\pspolygon(70.5,0)(70.5,0.052)(71.5,0.052)(71.5,0)
\pspolygon(71.5,0)(71.5,0.047)(72.5,0.047)(72.5,0)
\pspolygon(72.5,0)(72.5,0.111)(73.5,0.111)(73.5,0)
\pspolygon(73.5,0)(73.5,0.078)(74.5,0.078)(74.5,0)
\pspolygon(74.5,0)(74.5,0.005)(75.5,0.005)(75.5,0)
\pspolygon(75.5,0)(75.5,0.048)(76.5,0.048)(76.5,0)
\pspolygon(76.5,0)(76.5,0.102)(77.5,0.102)(77.5,0)
\pspolygon(77.5,0)(77.5,0.127)(78.5,0.127)(78.5,0)
\pspolygon(78.5,0)(78.5,0.069)(79.5,0.069)(79.5,0)
\pspolygon(79.5,0)(79.5,0.072)(80.5,0.072)(80.5,0)
\psline{->}(0,0)(83,0) \psline{->}(0,0)(0,0.5) \rput(5.8,0.5){$e_{2}\ (02/2001)$} \scriptsize \psline(1,-0.02)(1,0.02) \rput(1,-0.06){S\&P} \psline(5,-0.02)(5,0.02) \rput(5,-0.06){Pana} \psline(10,-0.02)(10,0.02) \rput(10,-0.06){Arge} \psline(15,-0.02)(15,0.02) \rput(15,-0.06){Irel} \psline(20,-0.02)(20,0.02) \rput(20,-0.06){Ital} \psline(25,-0.02)(25,0.02) \rput(25,-0.06){Swed} \psline(30,-0.02)(30,0.02) \rput(30,-0.06){Spai} \psline(35,-0.02)(35,0.02) \rput(35,-0.06){Hung} \psline(40,-0.02)(40,0.02) \rput(40,-0.06){Latv} \psline(45,-0.02)(45,0.02) \rput(45,-0.06){Turk} \psline(50,-0.02)(50,0.02) \rput(49,-0.06){SaAr} \psline(55,-0.02)(55,0.02) \rput(55,-0.06){SrLa} \psline(60,-0.02)(60,0.02) \rput(60,-0.06){Mongo} \psline(65,-0.02)(65,0.02) \rput(65,-0.06){Mala} \psline(70,-0.02)(70,0.02) \rput(70,-0.06){NeZe} \psline(75,-0.02)(75,0.02) \rput(75,-0.06){Nige} \psline(79,-0.02)(79,0.02) \rput(79,-0.06){Maur} \scriptsize \psline(-0.28,0.1)(0.28,0.1) \rput(-1.38,0.1){$0.1$} \psline(-0.28,0.2)(0.28,0.2) \rput(-1.38,0.2){$0.2$} \psline(-0.28,0.3)(0.28,0.3) \rput(-1.38,0.3){$0.3$} \psline(-0.28,0.4)(0.28,0.4) \rput(-1.38,0.4){$0.4$}
\end{pspicture}

\vskip 0.6 cm

\noindent Fig. 21. Contributions of the stock market indices to eigenvector $e_{2}$, corresponding to the second largest eigenvalue of the correlation matrix. White bars indicate positive values, and gray bars indicate negative values, corresponding to the first and second semesters of 2001. The indices are aligned in the following way: {\bf S\&P}, Nasd, Cana, Mexi, {\bf Pana}, CoRi, Berm, Jama, Braz, {\bf Arge}, Chil, Vene, Peru, UK, {\bf Irel}, Fran, Germ, Swit, Autr, {\bf Ital}, Malt, Belg, Neth, Luxe, {\bf Swed}, Denm, Finl, Norw, Icel, {\bf Spai}, Port, Gree, CzRe, Slok, {\bf Hung}, Pola, Roma, Bulg, Esto, {\bf Latv}, Lith, Ukra, Russ, Kaza, {\bf Turk}, Isra, Pale, Leba, Jord, {\bf SaAr}, Qata, Ohma, Paki, Indi, {\bf SrLa}, Bang, Japa, HoKo, Chin, {\bf Mong}, Taiw, SoKo, Thai, Viet, {\bf Mala}, Sing, Indo, Phil, Aust, {\bf NeZe}, Moro, Tuni, Egyp, Ghan, {\bf Nige}, Keny, Bots, SoAf, {\bf Maur}.

\vskip 0.4 cm

What is not clear at the beginning, but becomes more evident in later times and with more data, is that there are two main blocks that move together as a second approximation to the market movement, and that those blocks are related with time zones, which reflect the operation hours of the stock exchanges. Usually, from Eastern Europe to Pacific Asia, indices belong to the second group, which appears with negative values in the eigenvectors. This is a characteristic peculiar to data related with international financial indices, and does not appear in ordinary correlations between assets in the same stock exchange. A certain contamination exists from other internal structures, like an European Union, quite clear in the data prior to the first semester of 1998, but further studies into the future (2007 to 2010) reveal that this separation in two blocks persists and even increases in time. A third highest eigenvalue sometimes also stand out of the noisy region, but contamination by random noise makes it very difficult to analize any internal structure revealed by its corresponding eigenvector.

\section{Conclusion}

As we have seen in this article, correlation matrices between international stock market indices may be used in the construction of networks based on distance thresholds between the indices. By varying the thresholds, one may contemplate the cluster structure of those networks at different levels, revealing two strong and persistent clusters, one of American and the other of European indices, the formation of a Pacific Asian cluster in the 90's, and the slow integration of some of the indices. One could also see that those networks tend to shrink in size in times of crises, and, studying randomized data, that there are values of thresholds above which noise takes over. It was also shown that the eigenvectors of the correlation matrix hold some important information about the networks formed from it. In particular, the one corresponding to the second highest eigenvalue shows a structure based on the difference in operation hours between markets. Still, none of this study gives us information about causality between networks, and this shall indeed be a very interesting topic for future research.

\vskip 0.6 cm

\noindent{\bf \large Acknowledgements}

\vskip 0.4 cm

The author acknowledges the support of this work by a grant from Insper, Instituto de Ensino e Pesquisa. I am also grateful to Siew Ann Cheong, for useful discussions, and to the attenders and organizers of the Econophysics Colloquium 2010. This article was written using \LaTeX, all figures were made using PSTricks, and the calculations were made using Matlab and Excel. All data are freely available upon request on leonidassj@insper.edu.br.

\end{document}